\documentclass[12pt]{iopart-modified}

\usepackage{graphicx}
\usepackage{iopams}  
\usepackage{amsmath}
\def\la{\leftarrow}
\def\ra{\rightarrow}
\def\lan{\langle}
\def\ran{\rangle}
\def\pom{I\hspace{-0.1cm}P}
\def\reg{I\hspace{-0.1cm}R}
\def\Re{{\rm Re}}
\def\keV{{\rm keV}}
\def\Im{{\rm Im}}
\def\eV{{\rm eV}}
\def\MeV{{\rm MeV}}
\def\GeV{{\rm GeV}}
\def\pipluspiminus{\pi^+\pi^-}
\def\pinullpinull{\pi^0\pi^0}
\def\mpipi{m_{\pi\pi}}
\def\nb{{\rm nb}}
\def\ofsp{\hspace{-1.0mm}}

\def\pom{I\hspace{-0.1cm}P}
\def\reg{I\hspace{-0.1cm}R}
\def\Gammatot{\Gamma_{{\rm tot}}}
\def\gaga{{\gamma\gamma}}

\def\beq{\begin{equation}}
\def\eeq{\end{equation}}
\def\bea{\begin{eqnarray}}
\def\eea{\eea{eqnarray}}
\begin{document}
\title[The Status of Glueballs]{The Status of Glueballs}
\author{Wolfgang Ochs}
\address{Max-Planck-Institut f\"ur Physik, Werner-Heisenberg-Institut,\\
F\"ohringer Ring 6, D-80805 M\"unchen, Germany} 
\ead{ochs@mpp.mpg.de}

\begin{abstract}
Calculations within QCD (lattice and sum rules)
find the lightest glueball 
to be a scalar and with mass in the range of about 1000-1700 MeV.
Several phenomenological investigations are discussed which 
aim at the identification of the scalar meson 
nonets of lowest mass
and the super-numerous states if any. 
Results on the 
flavour structure of the light scalars $f_0(500),\ f_0(980)$ and
$f_0(1500)$ are presented; the
evidence for $f_0(1370)$ is scrutinized.
A significant surplus of leading clusters of neutral charge in 
gluon jets is found 
at LEP in comparison with MC´s, possibly a direct signal for glueball
production; further
studies with more energetic jets at LHC are suggested. 
As a powerful tool in the identification of 
the scalar nonets or other multiplets, along with signals
from glueballs we propose the exploration of symmetry
relations for decay rates of $C=+1$ heavy quark states like 
$\chi_c$ or $\chi_b$.
Results from $\chi_c$ decays are discussed; they are 
not in support of a tetra-quark substructure of $f_0(980)$.  
A minimal scenario for scalar quarkonium-glueball spectroscopy is presented.
\end{abstract}


\renewcommand{\contentsname}{}
\tableofcontents

\section{Introduction}
The existence of ``glueballs'', bound states of gluons, is a consequence of
the self-interaction of gluons within Quantumchromodynamics. Such states have been considered
to appear in analogy to $q\bar q$ states in a  quark-gluon
theory of hadrons already in 1972 during the early development of 
QCD~\cite{Fritzsch:1972jv}.
First specific scenarios for glueball spectroscopy on the basis of a
quark-gluon field theory with hadrons as colour singlets have been developed
by Fritzsch and Minkowski \cite{Fritzsch:1975tx} assuming an analogy between
massless gluons and photons. 
Much effort has been devoted during the last 40 years 
into the theoretical analysis 
as well as the experimental searches of this new type of hadrons. 

While a consensus is reached
about the existence and some properties of glueballs in a world without
quarks,
the details in the full theory are still controversial. 
Lattice QCD is a formulation of the quark-gluon gauge
theory on a space-time lattice and it is suitable for describing hadronic
phenomena in the non-perturbative regime with confined 
quarks. 
The theory starts directly from the QCD
Lagrangian with no other parameters than quark masses and a mass scale.
The lightest glueball has quantum numbers $J^{PC}=0^{++}$ and a mass
around 1700 MeV in the theory without quarks but could become smaller in the
full theory. Alternatively, information on the hadronic spectrum has been
derived successfully
from the ``QCD sum rules''. 
In this approach some
phenomenological input parameters (``condensates'') have to be determined
from experiment but a wide spectrum of quantitative results in hadron physics
has been established.
The predictions from these two theoretical approaches on glueballs will 
be reported and summarized in section 2.

Most of our attention concerning glueballs is focussed on the scalar sector
with $J^{PC}=0^{++}$, 
as the lightest glueball is expected with these quantum numbers. The first aim of the experimental search is
the identification of scalar nonets of $q\bar q$ bound states ($q=u,d,s$)
or other multiplets like those built from ``tetraquarks'', according to the
classification within flavour $SU(3)_{fl}$ symmetry.
Glueballs will then show up as super-numerous states. Unfortunately, it is
just this scalar sector whose identification is still ``tentative'' 
while the other multiplets are quite well established \cite{Amslerquark}.
Therefore, additional strategies, like the study of ``gluon rich'' processes
have attracted attention. We present strategies following such
investigations and describe different
scenarios for a scalar quark-gluon spectroscopy in sections 3 and 4.

It turns out that the validity of such scenarios depends crucially on 
the existence  of certain scalar states and their specific properties.
Therefore, in the subsequent sections 5 - 7 we analyze in greater detail 
the experimental data on production and decay of the isoscalar states 
which could have a gluonic contribution mixed in and analyze their flavour
structure. The 
Particle Data Group (PDG \cite{beringer2012pdg}) has listed the following
isoscalar states with $J^{PC}=0^{++}$
below a mass of 2 GeV as "established":
\begin{equation}
 f_0(500),\ f_0(980),\ f_0(1370),\ f_0(1500),\ f_0(1710).
\label{isoscalars}
\end{equation}
Special attention is paid to the reported evidence for $f_0(1370)$ which is
the cornerstone for a particular scenario.  
This state comes along with $f_0(1500)$ nearby in mass, so they are
discussed together. While attention is paid to data taken already long ago,
we are witnessing now the advent of high precision
results from $B$ factories as well as from the Large Hadron Collider and fixed target
experiments at CERN with remarkable future potential.

There are some pros and cons for the different scenarios but finally 
it is hard to
claim an ultimate evidence for the existence of a glueball. In the search
for a direct evidence of gluonic mesons 
the leading clusters in gluon jets have been
scrutinized in several LEP experiments. As a result, gluon jets are found
different indeed from the expectations of
Monte Carlo programs without glueballs, otherwise very successful (section 9).
 
Finally in section 10, we propose a new tool for identifying 
flavour multiplets based on symmetry relations for decay rates 
in heavy quarkonium decays, in
particular in states $\chi_{c0},\chi_{c2}$. Violation of such symmetries are
indicative of glueball production.
First available results are promising and should allow for further tests of
the existing scenarios. 
The last section contains a summary and our preferred solution for scalar
spectroscopy with glueball.

We focus here on questions related to glueball spectroscopy, so
there is a large variety of experimental and theoretical works on the
scalar sector which we do not address. There are recent reviews which include
complementary material on various phenomenological and experimental issues
\cite{Klempt:2007cp} and on other theoretical \cite{Mathieu:2008me} 
as well as experimental
results \cite{Crede:2008vw}. 

\section{QCD expectations for glueballs}
\subsection{Quantum numbers}
In the first specific scenarios for glueball 
spectroscopy~\cite{Fritzsch:1975tx} it has been assumed that the
counting of gluonic states is that obtained for free, massless gluons 
which have only two polarization states, so $n$ gluons can form $2^n$
different states. 
A simple case is met for the two-gluon channel. We consider the colour
singlet states which correspond to the observable hadronic states. The
colour averaged two-gluon systems carry the same quantum numbers as the
two-photon systems. They have been studied by Landau \cite{Landau1948} and Yang
\cite{Yang:1950rg}; the general case is also addressed in a group theoretical 
analysis by Minkowski \cite{Minkowski2005}. The results can be summarized as
follows: 

For a single photon there are two helicity states
$|\lambda_a\ran$ with $\lambda_a=\pm1$ corresponding to the right and left
circular polarization states. For two photons in the cms system we consider
one photon in $+z$ direction, the other one in $-z$ direction. 
The states with two photons (likewise two gluons in a colour singlet state)
have spin $J\neq 1$ (``Landau-Yang theorem'') and  
$C$-parity $C=+1$. 
Four helicity states $|\lambda_a\lambda_b\ran$ are available
from which states of definite parity can be formed

\begin{tabular}{rclclrcl}
($\ |11\ran$ & +& $ |-1-1\ran$\ )\ ;&\quad  $|1-1\ran$ \ ;
       & $ |-11\ran $\ ; &\quad  (\ $ |11\ran$ &$ -$& $ |-1-1\ran $\ )\\
$\la\ra$  & +& $ \hspace{0.3cm}\ra\la$ &\ $\ra\ra $& 
    \hspace{0.2cm}$\la\la$&$ \hspace{0.5cm} 
            \la\ra$ &$ -$ & $ \hspace{0.5cm} \ra\la$
\end{tabular}
\\
\noindent
the first three states have parity $P=+1$ the last one has $P=-1$.
They form states
of definite $J^{PC}$ as can be found by studying the behaviour of these
states under rotation and parity $P$ \cite{Yang:1950rg}
\begin{equation}
\fbox{$ 0^{++}$},\ 2^{++},\ 4^{++} \ldots); \ \fbox{$ 2^{++}$},\ 
4^{++} \ldots; 
\ 3^{++},\ 5^{++}\ldots;\ \fbox{$ 0^{-+}$},\ (2^{-+},\ 4^{-+} \ldots\ 
\label{jpc}
\end{equation}
\noindent Here the states with $J^{PC}=2^{++}$  and 
descendants correspond to the states $J_z=\pm2$ above.
Those  states which are formed by ``constituents'' in a relative 
$S$ wave are expected to be lowest in mass and
are enclosed by a box in \eref{jpc}; the higher spin excitations correspond
to even orbital wave functions.

Three gluon systems can have both $C=+1$ and $C=-1$. For the
ground states with relative  $S$ waves one finds \cite{Fritzsch:1975tx}

\begin{tabular}{rl}
\quad $C=+1:$ &\quad $J^{PC}=1^{++}$\\
\quad $C=-1:$ &\quad $J^{PC}=1^{+-},\ 1^{--},\ 3^{--}$.
\end{tabular}
\subsection{Lattice QCD.\label{lattice}}
Because of ``asymptotic freedom'' QCD interactions become weak for short
distances where they can be analyzed within 
perturbation theory at small coupling between quarks and gluons.
This method fails at large distances
where the interactions become strong. Lattice QCD is a formulation of the
theory on a space-time lattice and it is suitable for describing hadronic
phenomena in the non-perturbative regime \cite{Wilson:1974sk}
(see
\cite{Weisz2012} for an introduction 
and \cite{Kronfeld:2012uk} for a recent review). Physical quantities
can be computed from the QCD gauge and fermion actions $S_G$ and $S_F$
with the quark masses $m_q$ as parameters and a suitable mass scale.

The theory can be defined on a Euclidean 
space-time lattice with fermions on the lattice sites and gauge fields
connecting the sites. The lattice spacing $a$ acts as
an ultraviolet cutoff which provides a gauge invariant regularization of
QCD. Physical observables can be obtained from an average over the relevant
configurations of quark and gluon fields 
according to the Boltzmann weight
$e^{-S_F-S_G}$ within a space time volume $L^3T$ 
and the physical results are obtained ultimately by extrapolation 
to the continuum limit $a\to 0$.

Hadron masses are calculated from 2-point correlation functions $C_{ij}(t)$ 
for operators $O_i$ relevant for the hadrons considered
\begin{equation}
C_{ij}(t)=\frac{1}{Z}\int d\psi\int d\bar{\psi}\int dU
e^{-S_F-S_G}\lan0| O_i(t)^\dagger O_j(0)|0\ran \label{correlationf}
\end{equation}
in terms of a path integral over fermionic and gauge field variables  $\psi,\ \bar
\psi$ and $U$ with normalization $Z$. The correlation functions 
decrease exponentially for large Euclidean times $t$ 
\begin{equation}
C_{ij}(t)=c_i^mc_j^m e^{-E_m t}
\end{equation} 
where the term with the lowest energy $E_0$ dominates and determines the
mass of the ground state.
All masses are obtained in units of lattice spacing $a$ so that only 
ratios of masses
are predicted. In order to relate to absolute scales 
a suitable physical mass
scale has to be taken as input, such as the ``string tension'' or the 
``Sommer scale'' $1/r_0\sim400$ MeV \cite{Sommer:1993ce}.

As examples of recent lattice calculations from first principles, 
we mention the results in full QCD on the conventional light hadron spectrum
by the Budapest-Marseilles-Wuppertal Collaboration \cite{Durr:2008zz} 
who has calculated the masses of the 
baryon octet and decuplet states as well as the masses of
some light mesons within a few percent of accuracy. Here the masses of $\pi$, $K$
and $\Xi$ particles have been used to fix the 
masses of light and strange quarks at their physical values 
as well as the overall mass scale.
Another result, obtained by the ``Hadron Spectrum Collaboration'' 
\cite{Dudek:2011tt}  
concerns the spectrum of lightest and the first excited 
isoscalar meson states
which includes quark-annihilation contributions. Remarkably, the mixing
pattern of these mesons is reproduced close to observations.

More difficult to compute is the spectrum of glueballs in full QCD, 
as these states are heavier and therefore need higher statistics, in
particular the scalar states with vacuum quantum numbers have extra
contributions difficult to disentangle. In full QCD there is a mixing of
gluonic and fermionic degrees of freedom, correspondingly one inserts
gluonic and fermionic
operators for the relevant correlation functions.
For sufficiently light quark (pion) masses 
the glueball can decay into a meson pair which has to be included in the
consideration as well.

\begin{figure}[t]
\begin{center}
\includegraphics[width=8.0cm]{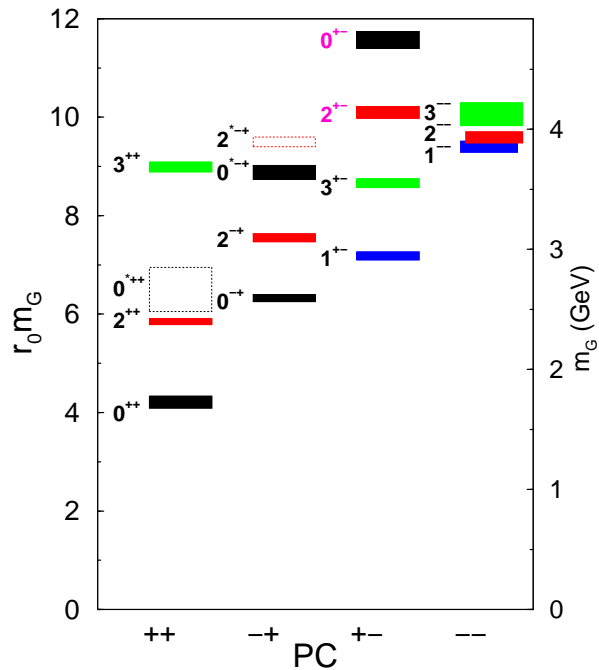}
\caption{\label{fig:massglue}Mass spectrum of glueballs (in GeV on r.h.s.) 
for different quantum numbers $PC$ according to the quenched lattice
calculations (figure from \cite{Morningstar:1999rf}).}
\end{center}
\end{figure}

The spectrum of glueballs has been calculated at first within the pure 
(Yang-Mills) 
gluon theory without quarks (``quenched approximation''). The lightest
glueballs are found for $J^{PC}=0^{++}$, $2^{++}$ and
$0^{-+}$ which correspond to the S wave ground states of the three 
different spin configurations listed in \eref{jpc}. 
Masses have been obtained by several groups in good agreement within around 10\% 
by Bali et al.
\cite{Bali:1993fb}, by Morningstar and Peardon \cite{Morningstar:1999rf}
and with a similar calculation but for larger lattices and volumes
by Chen et al.  \cite{Chen:2005mg} 
at a lattice spacing of  around $a=0.1-0.2$ fm  (see Table \ref{tab:quench}).
%
The results by Morningstar and Peardon \cite{Morningstar:1999rf} for states
below $\sim 5$ GeV  
are shown in figure~\ref{fig:massglue}. The three $S$ wave states
above are indeed the states of lowest mass, the lightest 
states formed by
three gluons with  $J^{PC}=1^{+-}$ and $1^{--}$ for $C=-1$ are found at around
3 GeV or higher.
\begin{table}[h]
\caption{\label{tab:quench} 
Glueball masses (in MeV) in quenched lattice approximation; an additional
error of $\sim 5$ \% should be added for the 
scale error of $1/r_0=410 (20)$ MeV
 \cite{Morningstar:1999rf}.}
\begin{indented}
\item[]\begin{tabular}{@{}llll}
\br
$J^{PC}$ & Bali et al. \cite{Bali:1993fb}& Morningstar et al. 
\cite{Morningstar:1999rf} & Chen et al. \cite{Chen:2005mg} \\
\mr
$0^{++}$ & 1550 (50)    &1730  (50)   &1709  (49)   \\
$2^{++}$ & 2270 (100)   &2400  (25)   & 2388 (23)   \\
$0^{-+}$ & 2330 (260)   &2590  (40)  &  2557  (25) \\
\br
\end{tabular}
\end{indented}
\end{table}

The influence of dynamical $q\bar
q$ contributions in full QCD  
on the mass of flavour singlet 
gluonic mesons has been studied by Hart and Teper
(UKQCD collaboration \cite{Hart:2001fp}). 
In this exploratory investigation the lattice spacing
was around $a\sim 0.1$ fm and the light quark masses about one half the
strange quark mass; an extrapolation to physical limits was not possible.
A significant 
suppression of the scalar glueball mass by the amount $\sim 0.84\pm0.03$ with respect to
the quenched value was found. With increasing quark mass this suppression
should disappear and pure gluo-dynamics be restored. As this effect was not
observed, not even for quark masses twice the strange quark mass, 
no real significance has been attributed to this suppression effect.
Furthermore, the spin 2 glueball did not show any 
substantial deviation from the quenched result.

A different conclusion has been reached by Hart et al. 
(UKQCD \cite{Hart:2006ps}) within a study using both gluonic and fermionic
operators to create the flavour singlet states. 
There have been $N_f=2$ degenerate sea-quarks using a sample of a total of
$\sim$1500 configurations.
At comparable lattice spacing the mass of the lightest
gluonic meson is reduced considerably with respect to the unquenched result
from 1600 MeV to about 1000 MeV. The results on the two
lightest $0^{++}$ mesons are interpreted in terms of a
maximal mixing of the $\bar q q$ and $gg$ states into the physical
flavour singlet mesons around 1000 and 1600 MeV.  

More recently, a study with higher statistics 
and including strange quarks
has been presented by Richards et al. (UKQCD \cite{Richards:2010ck}). They
use measurements at pion masses of 280 and 360 MeV and a spacing of $a=0.123$
fm and $a=0.092$ fm with 3000 and 5000 configurations respectively. 
In their analysis gluonic and $\pi\pi$ operators have been included but no
fermionic operators. 
This approach did not show any essential
difference between unquenched and quenched results for the
lightest $0^{++}$ state.
A mixing of the glueball with the $\bar q q$ state as found in
\cite{Hart:2006ps} has not been considered explicitly.

Similarly, for the $2^{++}$ and $0^{-+}$ states the results on the masses 
show little difference between quenched and unquenched calculations
and the same conclusion has been drawn by the same collaboration (UKQCD)
for glueballs of yet higher mass \cite{Gregory:2012hu}.
A summary of the 
unquenched calculations is presented in Table \ref{tab:unq}. It should be
noted that the scale parameter $r_0$ is different in quenched and unquenched
calculations, but the values actually 
used in  \cite{Morningstar:1999rf} and \cite{Richards:2010ck} differ by
$<3$\%. 
\begin{table}[h]
\caption{\label{tab:unq} 
Glueball masses from lattice QCD: listed are the ratios of masses from
unquenched over quenched calculations.}
\begin{indented}
\item[]\begin{tabular}{@{}llll}
\br
$J^{PC}$ & Hart and Teper \cite{Hart:2001fp}& {Hart\ et\ al.}
\cite{Hart:2006ps} & Richards et al.  \cite{Richards:2010ck}\\
\mr
$0^{++}$ & 0.84 (3)    &   0.63   & 1.03 (3)   \\
$2^{++}$ & 1.03 (3)    &          & 0.98 (10)   \\
$0^{-+}$ &             &          &  0.97 (2) \\
\br
\end{tabular}
\end{indented}
\end{table}

For the future studies a yet higher accuracy of the simulations 
is suggested in \cite{Richards:2010ck}, also 
smaller quark masses should be used to allow for
a more realistic decay of the glueball  into two pions.
Furthermore it is argued that a definitive calculation required a continuum
extrapolation, and the inclusion of fermionic operators
\cite{Gregory:2012hu}.
In that way the mixing problem between $gg$, $\bar q q$ and $\pi\pi$
states could ultimately be clarified.


\subsection{QCD sum rules\label{sumrules}}
An alternative approach to obtain information 
on the hadronic spectrum within QCD 
and on the glueball properties in particular
is based on the ``QCD sum rules'' \cite{Shifman:1978bx} 
(see \cite{Shifman:1998rb} for a pedagogical
presentation and \cite{Colangelo:2000dp} for a recent review). 
It starts from a 
correlation function in Minkowski space involving gluonic or fermionic 
operators $O_i$ according to the considered $J^{PC}$ quantum numbers 
which is evaluated in the deeply space like region 
\begin{equation}
\Pi(Q^2)=i\int\ d^4x\ e^{iqx}\ \lan0| T\{O_1(x) O_2(0)\}|0 \ran
\quad (Q^2\equiv -q^2 \gg \Lambda_{QCD}^2). \label{correlationPi}
\end{equation}
In an application of the operator product expansion the short distance
contribution to the correlation function is calculated from QCD perturbation
theory whereas the long distance contribution is given in terms of 
``vacuum condensates''. These parameters, ultimately, have to be
determined from experiment, of special importance are the condensates of
lowest dimension, the quark and gluon
condensates 
$\lan\bar q q\ran$ and $\lan \frac{\alpha_s}{\pi} G^a_{\mu\nu}
G^a_{\mu\nu}\ran.$
One can estimate the low energy hadronic
spectrum by matching the correlator \eref{correlationPi}, as given by this
expansion,
with a dispersion relation of the type
\begin{equation}
\Pi(Q^2)=\frac{1}{\pi}\int_0^\infty ds \frac{\Im
\Pi(s)}{(s+Q^2)}; \label{disprel}
\end{equation}
in general, \eref{disprel} has to be modified 
by subtractions depending on the high energy behaviour of the absorptive
part $\rho(s)\equiv\Im \Pi(s)$ which adds terms with powers of $Q^2$ 
in~\eref{disprel}.
The absorptive part
can be represented as a sum over the low lying hadronic states and a
continuum contribution, in the simplest case of one resonance as
\begin{equation}
\rho(s)=\rho^{res}(s)+\theta(s-s_0)\ \Im \Pi^{QCD}(s),\quad 
\frac{1}{\pi}\rho^{res}=f_1^2\delta(s-m_{res}^2), \label{absorptive}
\end{equation}
where $s_0$ denotes the threshold of the QCD continuum.
For better sensitivity to low $Q^2$ of \eref{disprel} one
performs a ``Borel transformation'' which calculates the integral
\begin{equation}
 L^{(k)}(\tau)=\frac{1}{\pi} \int^{\infty}_0 dt t^k e^{-t\tau } \Im \Pi(t) 
\label{borel}
\end{equation}
with index $k=-1,0,1,\ldots$.
In this way a set of sum rules for index $k$ is obtained.

The successful application of this formalism to a
large variety of hadronic phenomena concerning 
conventional mesons and
baryons has given a good confidence to this
approach besides providing the necessary phenomenological parameters.
Consequently, this scheme has been applied to the analysis of glueballs as well.

It has been suggested that the valence gluons bound in a
glueball couple much stronger to the vacuum fields, than, for example, the
quarks in the $\rho$ meson where a description in terms of mean vacuum
fields is sufficient \cite{Novikov:1981xj}. Ultimately, this strong coupling to vacuum fields
leads to the higher mass scales for
glueballs. Particularly strong effects from vacuum fields occur in spin zero
systems of low energy, 
a problem which has been approached by the inclusion of instanton field  
configurations. Along this line of thought the radius
of the $0^{++}$ glueball is only 0.2 fm to be compared with 0.5 fm for the
$\rho$ meson \cite{Schafer:1994fd,Shifman:1998rb}.  

The analysis of sum rules
with lowest indices $k=-1,\ 0$ (referred to as ``subtracted''
\cite{Novikov:1981xj} and
``unsubtracted'' \cite{Narison:1984hu} sum rules) by Narison and Veneziano 
\cite{Narison:1988ts} leads to a consistent
solution including
two gluonic scalar states with masses (see \cite{Narison:2008nj} for an
update)
\begin{equation}
M_{gb1}\sim 0.9-1.1\ \GeV,\qquad M_{gb2}\sim\ 1.5-1.6\ \GeV
\label{mgb12}
\end{equation}   
Furthermore, for the state around 1 GeV, 
the application of certain low energy sum rules predict
a large width into $\pi\pi$ with \cite{Narison:1988ts,Narison:2008nj} 
\begin{equation}
\Gamma(M_{gb1\to
\pi^+\pi^-})\sim 0.7\ \GeV. \label{sigmawidth}  
\end{equation}
In an extension of this result a phenomenological 
scheme has been suggested for $f_0(500-1000)/\sigma$
\footnote{The higher
mass 1000 MeV refers to the ``Breit-Wigner'' or ``on-shell'' mass 
which are more appropriate for sum rules
(see section \ref{sigmamgam}).},  
$f_0(980)$ and a sequence of isoscalar mesons heavier than 1 GeV 
realizing a mixing of $q\bar q$ and gluonic states as in \eref{mgb12} and a 
radial gluonic excitation~\cite{Narison:2005wc}.

An approach emphasizing the role of instantons by Forkel \cite{Forkel:2003mk}
suggests a solution of sum rules with a single scalar state at a mass of
$M_{gb}(0^{++})\sim 1.25\pm 0.2$ GeV, in between the previous results \eref{mgb12}.
The same approach has also been applied to the pseudoscalar gluonium 
and gave the result $M_{gb}(0^{-+})=2.2\pm 0.2$ GeV, 
similar to the lattice predictions mentioned above.

More recently, a
sum rule study of the mixed scalar system of gluonium ($gg$) 
and quarkonium ($q\bar q$)
has been presented by Harnett et al. \cite{Harnett:2008cw} which also included
instanton contributions.
The analysis of the diagonal $q\bar q$ and $gg$ as well as the non-diagonal 
$qg$ sum rules yields a consistent solution  
when using two mass states in \eref{absorptive} 
at approximately 1 GeV and 1.4 GeV, similar to \eref{mgb12}. 
In this solution the couplings of the two hadronic states to 
$q\bar q$ and $gg$ ``constituents'' is comparable, corresponding to a near 
maximal
mixing, with a slight preference of the heavier state to couple to $gg$.
This would imply that there is only one glueball and one quark-antiquark state
in this scheme for the two hadronic states. The heavier mass state
could be related to $f_0(1500)$ 
and the lower mass state 
to either $f_0(500-1000)/\sigma$ or $f_0(980)$, both higher and lower mass 
states being mixed of
$gg$ and $q\bar q$.

\subsection{Conclusions from theory}
There is a general agreement in that the lightest gluonic state has
quantum numbers $J^{PC}=0^{++}$. One state is located around 1.4-1.7 GeV. 
A second
state is suggested in sum rules and in certain unquenched lattice
calculations near 1 GeV. In these analyses 
there is a strong, almost maximal mixing of 
the hadronic states near 1 and 1.6 GeV into the $q\bar q$ and $gg$
components. This agreement between both approaches is striking but not
universally accepted. In particular, the most recent lattice calculation
does not find this low mass state, but, given the computational limitations, 
its existence cannot be excluded either. A definitive calculation
requires a higher statistics, a continuum extrapolation and the inclusion of
fermionic operators. 

The next heavier states are expected with quantum
numbers $J^{PC}=0^{-+},2^{++}$ and with masses $\gtrsim 2$ GeV. States with
other quantum numbers including exotic, i.e. non-$q\bar q$ ones, are
expected for higher masses. Experimental analyses have been difficult 
so far in this mass region.
Therefore, the search for the scalar gluonic states looks particularly
promising despite the experimental and theoretical uncertainties.

\section{Experimental search strategies for glueballs}  

In the following we will mainly focus on the scalar glueball which is
expected to be the lightest one and which has been addressed in most research
studies. The isoscalar states with $J^{PC}=0^{++}$ listed as 
"established" by the 
PDG \cite{beringer2012pdg} below a mass of 2 GeV are
already listed in \eref{isoscalars}.
All these states have been considered as glueball candidates or mixed
glueball/quarkonium states in some of the models or theories. First we
discuss some conventional strategies for distinguishing glueballs.  
  
\subsection{Quark-gluon constituent structure of hadrons from their decays}

The 2-body strong decays of an unstable hadron reflect its constituent 
structure. Under $SU(3)_{fl}$ symmetry of the strong interactions 
a light quarkonic meson $M(q\bar q')$ decays 
by creation of quark anti-quark pairs from the vacuum with equal amplitudes
\begin{equation}
|0\ran\to |u\bar u\ran + |d\bar d\ran + |s\bar s\ran.
\label{qqprdecay} 
\end{equation}
into a superposition of meson states $M(q\bar q_i)M(q_i\bar q')$.
 A glueball can decay by creating $q\bar q$ pairs and form quarkonic mesons 
or by gluons and form gluonic mesons. An explicit model for such 
decays has been proposed in \cite{Amsler:1995td} and a similar approach 
in \cite{Anisovich:1996qj,Anisovich:2002ij} 
with small differences for glueballs. The role of OZI suppressed processes
has been studied by Zhao \cite{Zhao:2005im}.

The flavour symmetry can be broken in different ways. 
\begin{itemize}
\item {\it Strange quark suppression:}
because of the heavier strange quark mass 
the ratio 
\begin{equation}
\rho\equiv \lan s\bar
s|V|0\ran/\lan d\bar d|V|0\ran
\label{rhodef}
\end{equation}
of matrix elements for the creation of
$s\bar s$ versus $u\bar u$ or $d\bar d$ is suppressed, i.e. $\rho<1$. 
Empirically, $\rho\ge
0.8$ for established nonets, for
tensor mesons $\rho=0.96\pm 0.04$ is found in \cite{Amsler:1995td}
using form factors, in \cite{Anisovich:1996qj} $\rho=0.7-0.9$ 
without form factors.  
\item {\it Form factors:} they may provide momentum dependent effects,
representing, for example, inelastic channels at higher momenta or the
influence of angular momentum barriers. 
In \cite{Amsler:1995td} the choice is 
\begin{equation}
|F_{ij}(\vec q)|^2=
   \exp(-q^2/8\beta^2) \label{formfactor}
\end{equation}
which multiplies the phase space and threshold factor
\begin{equation}
S_p(q)=q^{2\ell+1} \label{threshold}
\end{equation}
for angular momentum $\ell$ and momentum $q=|\vec q|$ with $\beta=0.4$ GeV.
\item {\it Chiral effects:}  
For the decay of a scalar glueball $G$ the amplitude $M(G\to q\bar q)\propto m_q$
(quark mass $m_q$) according to a perturbative analysis using chiral symmetry 
which involves the
intermediate process $gg\to q\bar q$
\cite{Chanowitz:2005du}. In consequence, the decay branching ratio 
$r_{K\pi}=B(G\to
K^+K^-)/B(G\to \pi^+\pi^-)$ is enhanced over  $r_{K\pi}=1$ for flavour symmetry, 
whereas strange quark suppression would lead to $r_{K\pi}<1$. 
\end{itemize} 

\subsubsection{Quarkonium ($q\bar q$) decays.}
\setcounter{footnote}{1} 
In our comparison with glueball decays we 
consider here mainly isoscalar quarkonium states $|q\bar q\ran$ which 
are generally mixed from non-strange 
$|n\bar n\ran = \frac{1}{\sqrt{2}} |u\bar u + d\bar d\ran$ and 
strange quark $|s\bar s\ran$ components
\begin{equation}
|q\bar q\ran= \cos \alpha |n\bar n\ran - \sin\alpha |s\bar s\ran.
\label{mixing}
\end{equation}
Alternatively, the mixing can be described in terms of the $SU(3)_{fl}$ singlet and
octet components $|1\ran=\frac{1}{\sqrt{3}}|u\bar u+d\bar d+s\bar s\ran$
and $|8\ran=\frac{1}{\sqrt{6}}|u\bar u+ d\bar d -2 u\bar u\ran$.   
The mixing angle $\alpha$ in the quark-flavour basis is related to 
the nonet mixing angle $\theta$ in the singlet-octet basis by
\begin{equation}
\alpha=\theta + \arctan\sqrt{2}\simeq \theta + 54.7^\circ.
\label{nonetmix}
\end{equation}
For $\alpha=0\ (\pi/2)$ we obtain pure $n\bar n$ ($s\bar s$) states 
("ideal mixing") as approximately found for the vector mesons
$\rho,\omega$,$\phi$. The mixing of pseudoscalar mesons is written as
\numparts
\begin{eqnarray}
\quad|\eta\ran &= \cos \phi_{ps} |n\bar n\ran - \sin\phi_{ps} |s\bar s\ran,
\label{mixingeta}\\
\quad|\eta'\ran &= \sin \phi_{ps} |n\bar n\ran + \cos\phi_{ps} |s\bar s\ran.
\label{mixingetapr}
\end{eqnarray}
\endnumparts
According to a recent review \cite{DiDonato:2011kr}, the 
different analyses from radiative decays of vector and pseudoscalar mesons yield
values $\phi_{ps}\approx 42^\circ$ ($\theta_{ps}\simeq -13^\circ$), not far
away from
the pure flavour singlet-octet states ($\theta=0,-\pi/2$ or
$\phi_{ps}=54.7^\circ,\ -35.3^\circ$\footnote{Previously, a favoured value 
was quoted as $\phi_{ps}=37.4^\circ,\ \theta_{ps}=-17.3^\circ$
\cite{Amslerquark}.}). On the other
hand, the possibility of a gluonic component of $\eta'$ remains ambivalent.
Such a component can be added to $\eta'$, for example, by introducing an
additional mixing angle $\phi_{Gps}$ 
\beq  
|\eta'\ran = \cos\phi_{Gps}\, (\sin \phi_{ps} |n\bar n\ran + 
\cos\phi_{ps} |s\bar s\ran) +\sin\phi_{Gps}\,|gg\ran.
\label{mixingetaprg}
\eeq
For the last term with $Z_{\eta'}=\sin\phi_{Gps}$ one finds
typically  $|Z_{\eta'}|^2\lesssim 0.1$ but larger values up to
$|Z_{\eta'}|^2\lesssim 0.5$ are reported as well \cite{DiDonato:2011kr}.   
A more general approach with two mixing angles 
describing the mixing of meson states at the hadronic scale
and also the mixing of decay constants at short distances is discussed
in \cite{Feldmann:1998vh,Feldmann:1998sh}. Our analysis of $\chi_c$ decays
in the next subsection suggests that the conventional state mixing approach 
as above is appropriate in that case. 

\begin{figure}[t]
\begin{center}
\includegraphics*[angle=180,width=7.0cm,%
bbllx=5.0cm,bblly=11.0cm,bburx=16.5cm,bbury=20.0cm]
{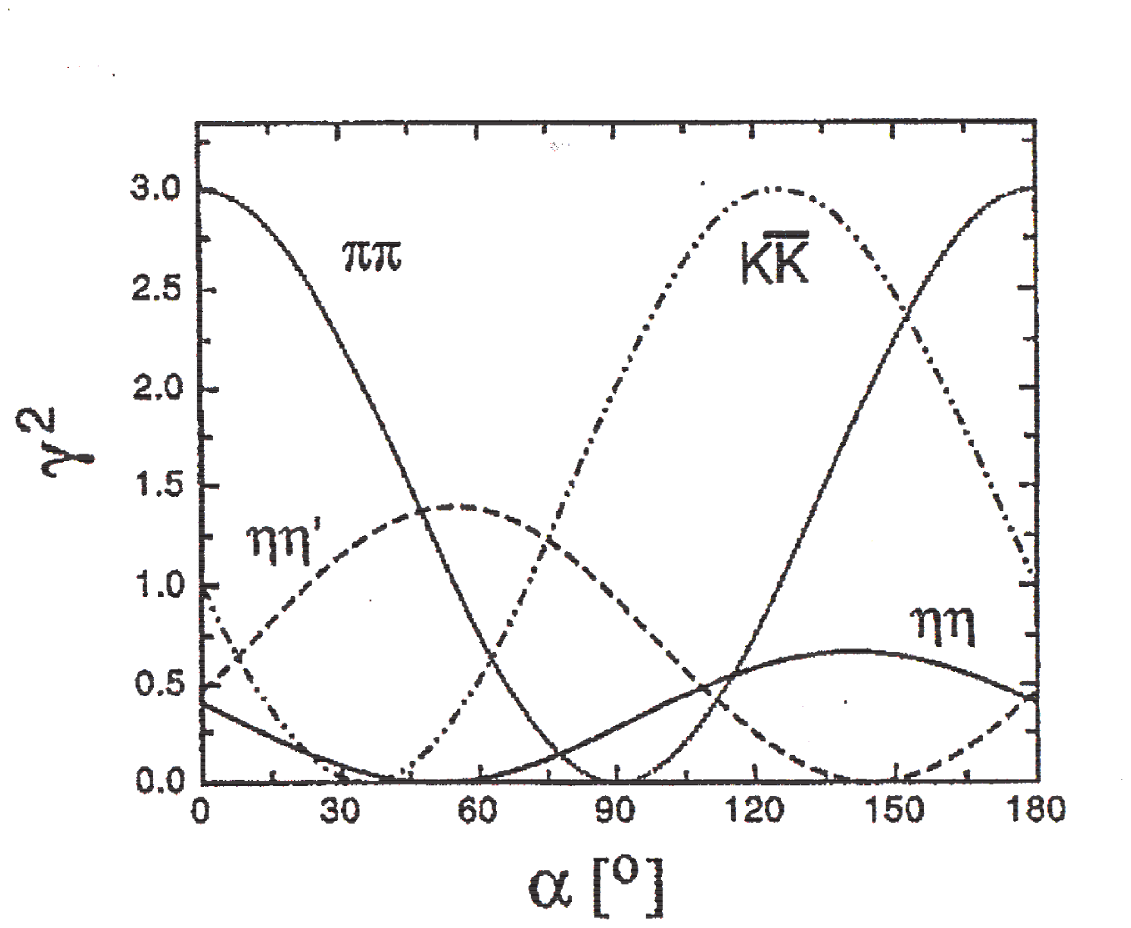}
\end{center}
\caption{\label{fig:mdecay} Couplings $\gamma^2_{ij}$ for isoscalar
$(q\bar q)$ meson decays as function of mixing angle $\alpha$ as in 
\eref{mixing} 
for $SU(3)_{fl}$ symmetry
($\rho=1$) from table \ref{tab:decay} with $r_G=0$ for $\phi_{ps}=37.4^\circ$
(figure from \cite{Amsler:1995td}).}
\end{figure}

According to the model \cite{Amsler:1995td} 
the partial width $\Gamma_{ij}$ of a quarkonium state into a 
pair of mesons $M_iM_j$
is computed from
\begin{equation}
\Gamma_{ij}=\gamma_{ij}^2 \times |F_{ij}(\vec q)|^2 \times S_p(\vec q)
\end{equation}
with \eref{formfactor}, \eref{threshold}; it 
depends on the above parameters $\alpha$ and $\rho$ through the couplings
\begin{equation}
\gamma_{ij}^2=c_{ij} |M_{ij}|^2 
\label{couplings}
\end{equation}
where $M_{ij}$ is the decay amplitude and  $c_{ij}$ is a weighting factor
for the different charge states.
For an isoscalar $(q\bar q$) decay these couplings $\gamma_{ij}^2$ 
are given in table \ref{tab:decay} for $r_G=0$ 
\cite{Amsler:1995td,Anisovich:1996qj};
the dependence
of $\gamma_{ij}^2$ on $\alpha$ is shown in figure~\ref{fig:mdecay}.

\begin{table}[b]
\caption{\label{tab:decay} Couplings $\gamma_{ij}^2$ 
for decay of an $I=0$ meson $f_0$ with mixed 
($q\bar q$)  and ($gg$) components 
\cite{Amsler:1995td,Anisovich:1996qj}; flavour mixing angles $\alpha$
as in \eref{mixing}, $\phi\equiv \phi_{ps}$ as in \eref{mixingeta}; 
glueball decay according to 
\eref{glueballdecaysa} \cite{Amsler:1995td}; 
$SU(3)_{fl}$ breaking parameters $\rho,R$ as defined in \eref{rhodef} and
\eref{Rstrange}; weights $c_{ij}$ included in
$\gamma_{ij}^2$; decay couplings $g_{ij}^2=g^2\gamma_{ij}^2/4$ 
for $(q\bar q)$ and
$r_G=\sqrt{3/(2+R^2)}G/g$ for $(gg)$ with $g,G$ from \eref{couplings_g} (based
on \cite{Anisovich:1996qj,Anisovich:2002ij}).}
\begin{indented}
\item[]\begin{tabular}{@{}ccc}
\br
decay channel & couplings $\gamma^2_{ij}$ for $|f_0\ran=\cos \phi_G |q\bar
q\ran + \sin \phi_G |gg\ran $ & 
 weights\\
\mr
$\pi\pi$      & $3 (\cos \alpha + r_G)^2$  & 3\\
$K\bar K$     & $(\rho \cos\alpha  - \sqrt{2} \sin \alpha +2r_G)^2$ & 4\\
$\eta \eta$   & $(\cos \alpha \cos^2 \phi -\sqrt{2}\rho \sin\alpha
                         \sin^2\phi +r_G(\cos^2\phi+R^2\sin^2\phi))^2$ & 1\\
$\eta \eta'$   & $2 \cos^2 \phi\sin^2 \phi(\cos\alpha  +\sqrt{2} \rho \sin\alpha
                      +r_G(1-R^2))^2$ & 2\\
$\eta' \eta'$   & $(\cos \alpha \sin^2 \phi -\sqrt{2}\rho \sin\alpha
                         \cos^2\phi+r_G (\sin^2\phi+R^2 \cos^2\phi))^2$ & 1\\
\br
\end{tabular}
\end{indented}
\end{table}

\subsubsection{Glueball decays.}  
\begin{figure}[t]
\begin{center}
\includegraphics*[angle=180,width=7.0cm]
{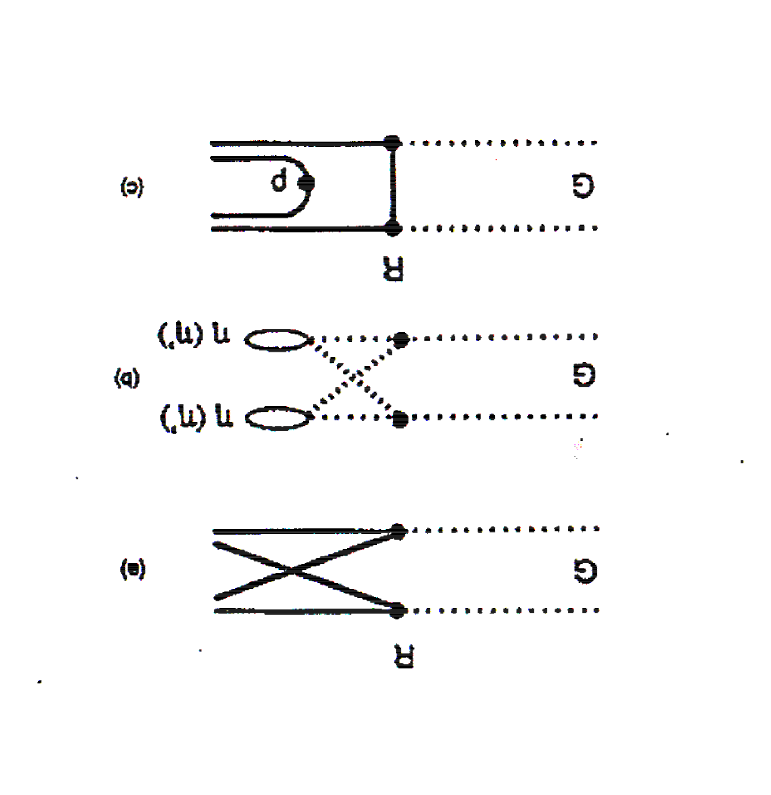}
\end{center}
\caption{\label{fig:gbdecay} Processes contributing to gluonium decay
(\ref{glueballdecaysa}-\ref{glueballdecaysc})
(figure from \cite{Amsler:1995td}).}
\end{figure}

Bound  
$gg$ states (``gluonia'') decay through a two
gluon intermediate state where several mechanisms can be envisaged
\cite{Amsler:1995td}
(see figure \ref{fig:gbdecay})
\numparts
\begin{eqnarray}
a) &\quad gb  \to gg  \to \bar q q \bar q' q' & \to M\ M'\label{glueballdecaysa}\\
b) &\quad gb \to gg \to g g g g          &\to \eta(\eta')\ \eta(\eta')
   \quad \text{or}\quad \to gb\ gb      \label{glueballdecaysb}\\
c) &\quad gb\to gg  \to \bar q q \to \bar q q \bar q' q' & \to M\
     M'\label{glueballdecaysc} 
\label{glueballdecays}
\end{eqnarray}
\endnumparts
In the decay rates of these processes 
the ratio of amplitudes 
\begin{equation}
R\equiv \lan s\bar s|V|g\ran/\lan d\bar d|V|g\ran
\label{Rstrange}
\end{equation} 
appears for the $q\bar q$ production from a gluon, 
where $R^2=1$ for $SU(3)_{fl}$ symmetry. For the decay of a mixed state
\beq
|f_0\ran=\cos \phi_G |q\bar q\ran + \sin \phi_G |gg\ran
\eeq
with $|q\bar q\ran$ mixed as in \eref{mixing} and $|gg\ran$ decaying as in
 (\ref{glueballdecaysa})  into pairs of pseudoscalars we write the decay
 couplings as \cite{Anisovich:2002ij}
\beq
g_{ij}= g\ d_{ij}^{\,q\bar q} +G\ d_{ij}^{\,gg};
\quad g=g_0 \cos \phi_G,\quad G=G_0 \sin \phi_G. \label{couplings_g}
\eeq
The couplings  $\gamma^2_{ij}$ for the decay of the mixed state with the
factor $g^2/4$ taken out and with $r_G=\sqrt{3/(2+R^2)}\,G/g$ 
are given
in table \ref{tab:decay}. For the glueball decay 
with flavour symmetry ($R=1$) one
finds from the term with $r_G\propto G$ the reduced partial widths 
$\gamma^2_{ij}$ 
\beq
 \gamma^2_{ij}(gb\to\pi\pi:K\bar K:\eta\eta:\eta\eta':\eta'\eta') 
\quad  = \quad3:4:1:0:1 \label{statweights}
\eeq
according to the statistical weights.

The two-gluon intermediate process 
also dominates the decays of scalar heavy quark, $c\bar c$, bound states with $C$ parity
$C=+1$
like $\chi_c$ in the perturbative analysis and one expects decay rates as in
\eref{statweights}. 
An update of this test for spin $J=0,2$ charmonia  
$\chi_{c0}, \chi_{c2}$~\cite{Amsler:1995td} 
is carried out in table 
\ref{tab:chidecay}.
We list the measured branching ratios $B$ into pairs of pseudoscalars
according to the PDG \cite{beringer2012pdg}, where recent contributions come 
from CLEO \cite{Adams:2008ab} and BES \cite{Bai:2001ha}.
From the model (but without
dynamical formfactor $F_{ij}(q^2)$, i.e. $\beta\to \infty$) 
one expects $B\propto c |M|^2 q^{2\ell+1}$. In table \ref{tab:chidecay}   
we also show
$B$ after division
by $cq^{2\ell+1}$, and after normalization
to that value for the $\pi\pi$ decay. These data are to be compared with 
the model predictions for $|M|^2=\gamma^2/c$  in table~\ref{tab:decay}
for ($gg$), i.e. for $g/G\to 0$, where recall $g$ and $G$ are given by
\eref{couplings_g}. 
These expectations are given in table \ref{tab:chidecay} as well for
different values of the strange quark parameter $R$ in
\eref{Rstrange}. 
We observe, that the expectations from $SU(3)_{fl}$ symmetry ($R=1$) 
are fulfilled for all decay processes of $\chi_{c0,2}$ within 10~-~20\% (only the upper bound for
$\chi_{c2}\to \eta'\eta'$ is a bit low). 
So there is no strong need for the additional gluonic processes
\eref{glueballdecaysb} affecting $\eta,\eta'$; nor is there any 
strong need
for dynamical form factors $F_{ij}(q^2)$; the weakness of angular momentum 
barrier factors (of Blatt-Weisskopf
type)  may be due to the small radius
of charmonium states of $O(1/m_c)$. 

\begin{table}[h]
\caption{\label{tab:chidecay} Branching ratios $B$ of $\chi_{c0}(3415)$ 
($0^{++}$) and $\chi_{c2}(3556)$
($2^{++}$) into pairs of pseudoscalars; also $B/(cq)$ or $B/(cq^5)$ resp. with charge 
weight $c$ and
momentum $q$ after normalization to the value for $\pi\pi$
(data as listed in \cite{beringer2012pdg}), together with 
the resp. theoretical
expectations $|M|^2=\gamma_{ij}^2/c$ for process 
$gg$ in table \ref{tab:decay} for different strange
quark parameters $R$
 (mixing angle $\phi_{ps}\approx 37.4^\circ$ for $\eta,\ \eta')$.}  
\begin{indented}
\item[]\begin{tabular}{@{}lcccccc}
\br
 &$\pi\pi$ & $K^+K^-$ & $K^0_sK^0_s$ & $\eta\eta$ &$ \eta\eta'$& 
   $ \eta'\eta'$ \\
weight $c$ &3&2&1&1&2&1 \\ 
\mr
$\chi_{c0}(0^{++})$&& &&&&\\
$B\times 10^3$ &$ 8.5\pm0.4$ &$6.06\pm0.35$ & $3.14\pm0.18 $&$ 3.03\pm0.21$ &
   $< 0.24$& $2.02\pm 0.22$ \\
$B/(cq)$& $1.\pm0.05$ &$ 1.13\pm0.06$ &$ 1.17\pm0.07$ & $1.14\pm0.08$ &
     $<0.048$ &  $0.87\pm 0.09$\\
\mr
 $\chi_{c2}(2^{++})$ && &&&&\\
$B\times 10^3$ &$ 2.43\pm0.13$ &$1.09\pm0.08$ & $0.58\pm0.05 $&$ 0.59\pm0.05$&
    $< 0.06$& $< 0.11$ \\
$B/(cq^5)$& $1.\pm0.06$ &$ 0.82\pm0.06$ &$ 0.88\pm0.08$ & $0.94\pm0.08$ &
     $<0.063$ &  $<0.32$\\
\mr
expect  && &&&&\\ 
R=1.0 & 1. & 1.& 1.& 1. & 0.& 1.\\
R=0.95 & 1. &  0.9 & 0.9 & 0.93 & 0.002 & 0.88 \\ 
R=0.90 & 1. &  0.81 & 0.81 & 0.86 & 0.008 & 0.77 \\ 
R=1.1  & 1. & 1.1 & 1.1 & 1.08 & 0.002 & 1.13\\
\br
\end{tabular}
\end{indented}
\end{table}

Looking at the substructures at the 10\% level 
we note that for the tensor state $\chi_{c2}$ the observed values for
$B/(cq^5)$ would prefer a slightly reduced $R\gtrsim 0.9$. In case of the
decays of the scalar $\chi_{c0}$ values of $B/(cq)$ for 
kaons are slightly enhanced by 10-20\%. This effect could
be a remote consequence of the chiral enhancement in scalar 
decays discussed above from the contribution of the intermediate 
process $gg\to q\bar q$ as considered for scalar glueballs 
in \eref{glueballdecaysc} (the argument is also
applicable for a decaying scalar $c\bar c$ state). The effect can be
simulated by choosing $R>1$ and we give in the last line of table
\ref{tab:chidecay} the predictions for $R=1.1$ which would fit the data 
for $\chi_{c0}$ generally better.

A modification of the $SU(3)_{fl}$ symmetry for decay rates (with $R=1$
in table \ref{tab:chidecay})
is expected in  schemes which include state mixing at hadronic scales and
mixing of decay
constants at short distances \cite{Feldmann:1998sh}; 
in that case the ratios of the $\chi_c$ decay rates
deviate considerably from unity, for example $B(\chi_{c0}\to
\eta\eta)/B(\chi_{c0}\to \pi^0\pi^0)=1.9$ and the same follows for the ratio
$(\eta'\eta')/(\pi^0\pi^0$). The data appear to be closer to the conventional 
meson state mixing approach as seen in table \ref{tab:chidecay}.

The remarkable success of the simple scheme based on $SU(3)_{fl}$
symmetry with phase space correction for decays into the same nonets
suggests
further tests with other nonets in the search 
for extra gluonic states, as will be discussed in section
\ref{symmetry_relations}.

\subsection{Enhanced and suppressed glueball production}
Not only the decay, also the production properties are characteristic for
the intrinsic structure of a hadronic state \cite{Robson:1977pm,Close:1987er}. 
\subsubsection{Gluon-rich processes.}
\label{gluon-rich}
There are processes which
provide a ``gluon rich'' environment with enhanced probability for glueball
production, see figure \ref{fig:gluonrich} for examples. 
These processes have been extensively explored
experimentally in the past \cite{Klempt:2007cp,Crede:2008vw}. We give
here a survey first over different processes and come back to specific problems later:
\begin{figure}[t]
\begin{center}
\includegraphics*[angle=90,width=13.0cm,bbllx=8.0cm,bblly=0.4cm,%
bburx=15.0cm,bbury=28.2cm]{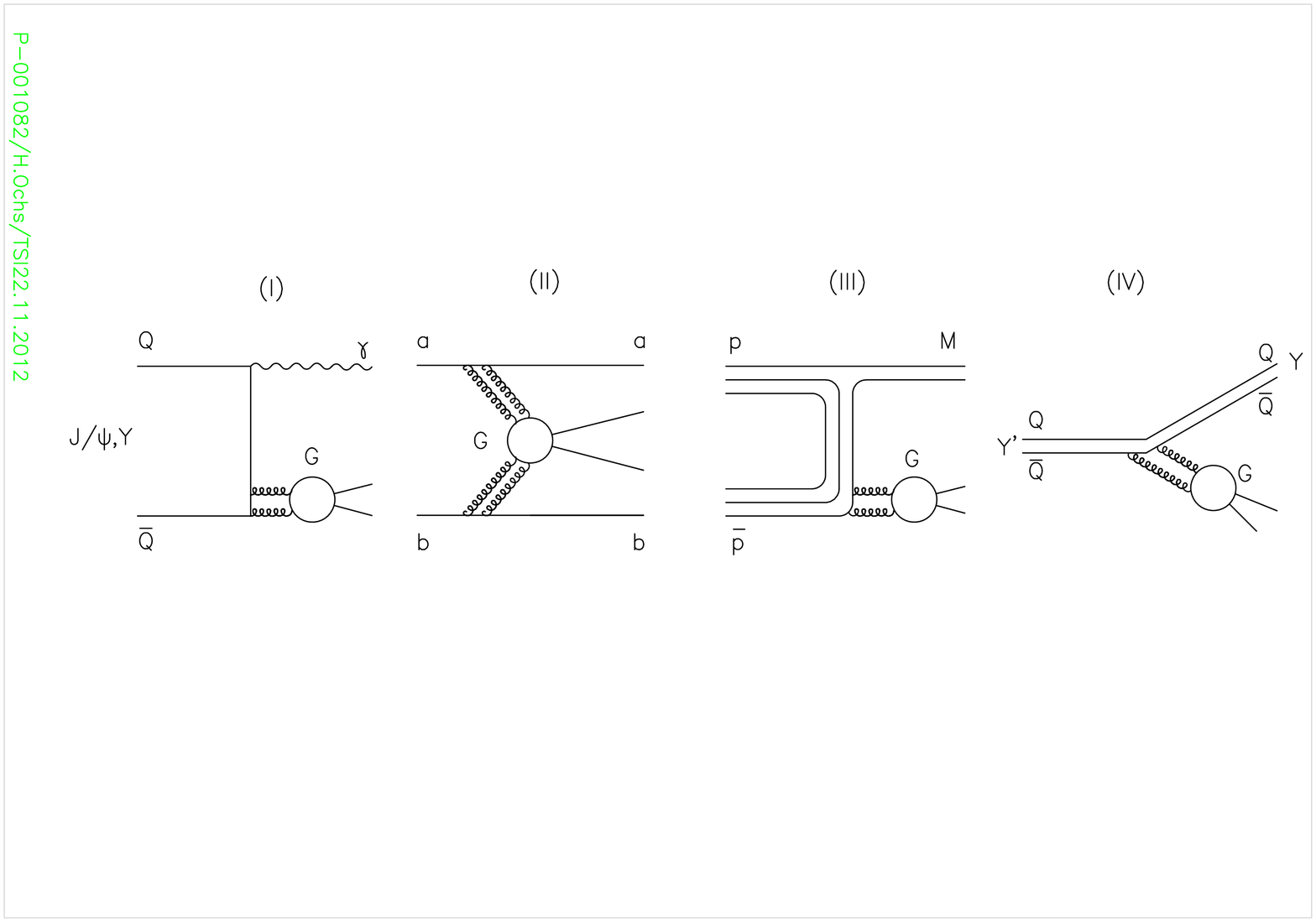}
\caption{\label{fig:gluonrich} Processes favouring glueball $G$ production,
where $J/\psi$ and $\Upsilon$ are respectively the lowest mass $c\bar c$ and
$b\bar b$ mesons with $J^{PC}=1^{--}$.}
\end{center}
\end{figure}
\begin{enumerate}
\item {\it Radiative $J/\psi$ or $\Upsilon$ decay:} In the perturbative approach the
($c\bar c$) or other heavy quarkonium $J^{PC}=1^{--}$ bound states decay 
predominantly 
through ($Q\bar Q)\to 3g$. Alternatively, the radiative decay 
($Q\bar Q)\to \gamma+ (2g)$
is possible with formation of an intermediate gluonium ($gg$) state
($(Q\bar Q)\to \gamma+ gb$).

\item {\it Central production of mesons:} In double diffractive high energy
processes the incoming hadrons scatter with small
momentum transfers and carry on the initial valence quarks. In Regge theory
this process 
is dominated by ``double Pomeron exchange''. If the Pomeron is 
viewed as a gluon dominated object then glueball production is enhanced in
this reaction ($p p \to p\ gb\ p$). The different contributing processes 
within QCD have been discussed in \cite{Khoze:2001xm}.

\item{\it $p\bar p$ annihilation:} The annihilation of quarks may proceed 
through intermediate gluons and the formation of glueballs ($p\bar p \to gb
+M$).

\item{\it Decay of excited heavy quarkonium $Y^{(n)}$ to ground state 
$Y$:}
In the example $Y^{(n)}\to Y +X$ the hadrons $X$ are 
emitted from intermediate gluons and therefore could be formed through an
intermediate glueball.

\item{\it Decay of heavy quark $b\to s g$:} This QCD process (through
``penguin'' diagram) may hadronize involving a glueball 
according to $B\to K gb$ \cite{Minkowski:2004xf}.

\item{\it Leading particle in gluon jet:} In analogy to the fragmentation
of the primary quark~$q$ of a $q$-jet into an energetic meson $M(q\bar q')$ 
which carries $q$ as valence quark, there may be the fragmentation of the
primary gluon of a gluon jet into an energetic meson $M(gg)$ which carries the initial gluon as
valence gluon $g\to gb+X$ (section~\ref{sectionleading}). 
\end{enumerate}
\subsubsection{Suppression of glueballs in $\gamma\gamma$ processes.}
Having neutral constituents a glueball couples
to photons only through loop processes and then it is suppressed 
in $\gamma\gamma$ reactions.

\subsection{Supernumerous states among $q\bar q$ nonets?}
Mesons with light quark constituents ($u,d,s$ quarks) are classified
in nonets of $3\times3$ states (octet+singlet). A well known example
is the  pseudoscalar nonet of lowest mass with $\pi,K,\eta$ near flavour octet 
and $\eta'$ near singlet.  
It is the aim of meson spectroscopy to establish the appropriate classification
of mesons. They should fit into nonets of $q\bar q$ states - possibly, 
there are also exotic states
like tetra-quark $qq\bar q\bar q$  or hybrid  $q\bar qg$ states. 
If there are glueballs in addition 
there should be supernumerous states which do not fit into a nonet
classification of the meson spectrum. 

\section{Spectroscopy of scalar mesons: $q\bar q$ nonets mixing with glueballs}

In the scalar sector different schemes have been
proposed for a spectroscopy with glueballs for the mesons below 2 GeV
including the isoscalars of \eref{isoscalars}.
\begin{table}[h]
\caption{\label{tab:route1} Spectroscopy with glueballs, route 1: glueball
near 1500 MeV.}
\begin{indented}
\item[]\begin{tabular}{cccl}
\br
 $I=0$ & $I=\frac{1}{2}$ &  $I=1$& \\
\br
             & $K^*(1950)$ && \\
\mr
{ {$f_0(1710)$}} &&& { 3 isoscalars:  $q\bar q$ nonet} \\
{ {$f_0(1500)$}} & $K^*(1430)$ & $a_0(1450)$ & 
   {+ glueball M$\sim$ 1.6 GeV}\\
{ { $f_0(1370)$?}} & & & \\ 
\mr
$f_0(980)$ & & $a_0(980)$& { light nonet}\\
           & $K^*(900)/\kappa$? & & {$q\bar q$\ 
    \ or $qq-\bar q \bar  q$}\\ 
$f_0(500)/\sigma$ & & &{ or\ $K\bar K$}\\
\br
\end{tabular}
\end{indented}
\end{table}

\subsection{Route 1: $(q\bar q)$ nonet - glueball mixing around 1500 MeV}
Lattice QCD in quenched approximation predicts the lightest glueball
in the mass range around 1600 MeV (see section \ref{lattice}). The discovery of the rather 
narrow $f_0(1500)$ 
in $p\bar p$ annihilation by the Crystal Barrel
Collaboration \cite{Anisovich:1994bi}  and by GAMS \cite{Alde:1987mp}
suggested at first the possibility of a gluonic state. A closer inspection
of decay mechanisms and branching ratios led Amsler and Close
\cite{Amsler:1995td} to propose a mixing scheme including 
$f_0(1370)$ and $f_0(1710)$ with a gluonic component as well \footnote{In the
scheme \cite{Amsler:1995td} the existence of the 
heavy state (now $f_0(1710)$) actually has
been predicted.}. In table \ref{tab:route1}
we list the meson states ordered according to mass and isospin. Between 1.0
and 1.9 GeV we find the states fitting into a $q\bar q$ 
nonet but with one supernumerous isoscalar which suggests the presence of a
glueball.

With more experimental results available and with further development of 
models various mixing schemes 
for the three isoscalars have been proposed (for a further discussion, 
see \cite{Crede:2008vw}). As an example, we present 
the result \cite{Close:1996yc} which includes data for $\gamma\gamma$
processes.
The physical mass states $f_0(m_i)$ are decomposed 
into the $q\bar q$ and $gg$ states or,
alternatively, into the $SU(3)$ eigenstates 
$|1\ran=(|u\bar u\ran + |d\bar d\ran + |s\bar s\ran)/\sqrt{3}$ and 
$|8\ran=(|u\bar u\ran + |d\bar d\ran - 2|s\bar s\ran)/\sqrt{6}$ and gluonium
$|gg\ran$
\begin{equation}
\begin{matrix} |f_0(1370)\ran \\ |f_0(1500)\ran \\ |f_0(1710)\ran \end{matrix}
=
\begin{pmatrix} 0.86 & 0.13 & -0.50\\
 0.43 & -0.61 & 0.61\\
 0.22 & 0.76 & 0.60 \end{pmatrix}
\ . \ 
\begin{matrix} 
 |n\bar n\ran\\ 
 |s\bar s \ran\\
 |gg\ran 
\end{matrix}
=  \begin{pmatrix}0.78 & 0.39 & -0.50\\
                 0.00 & 0.75 & 0.61 \\
                 0.62 & -0.49 & 0.60 \end{pmatrix}
\ . \ \begin{matrix} |1\ran\\ |8\ran\\ |gg\ran \\
 \end{matrix}
\label{mixingclose}
\end{equation}
One observes that in this scheme $f_0(1500)$ is a mixture only of octet
$q\bar q$ and glueball;
the glueball component is distributed over all three
$f_0$ states (last column in mixing matrix) with about equal amounts. In alternative schemes
either $f_0(1500)$ can be  dominantly gluonic \cite{Giacosa:2005zt} or 
$f_0(1710)$ while $f_0(1370)$ is near 
singlet \cite{Cheng:2006hu,Giacosa:2005zt}.

A possible problem with this mixing scheme are the doubts related 
to the very existence of $f_0(1370)$. This resonance is included in
global fits of parametric model amplitudes to a variety of channels
but it is not seen in any model independent
bin-by-bin phase shift analysis. The PDG under this entry does not provide
any established result on branching ratios or ratios thereof - contrary to
the well established nearby $f_0(1500)$.  
We come back to this problem in more detail in section \ref{f013701500}.

The light scalar mesons below 1 GeV can be grouped into another nonet
(see table \ref{tab:route1}) which includes the broad states 
$f_0(500)/\sigma$ and $K^*(800)/\kappa$, besides the narrow $f_0(980)$ and
$a_0(980)$. This meson nonet can been built from $q\bar q$ as usual, see,
for example  
\cite{Morgan:1974cm,Tornqvist:1995ay,vanBeveren:1998qe}. Alternatively, 
it can be given a substructure of diquarks ($qq-\bar
q\bar q$) according to the scheme by Jaffe \cite{Jaffe:1976ig} within 
the MIT bag model where the nonet is constructed as the direct product
of the anti-triplet 
of diquarks ($ud,\ us,\ ds$) and the triplet 
of anti-diquarks ($\bar u\bar d,\ \bar u\bar s,\ \bar d\bar s$). 
This construction explains naturally the ordering in mass
of the states where $a_0,f_0$ have two, $\kappa$ has one
and $\sigma$ has no  $s(\bar s)$ quark constituent
$$a^+=[su]\ [\bar s\bar d],\quad  f_0/a^0=([su]\ [\bar s \bar u] \pm [sd]\ [\bar s \bar
 d])/\sqrt{2},\quad   
  a^-=[sd] \ [\bar s\bar u]$$
\beq   \kappa^+=[su]\ [\bar u \bar d],\quad \kappa^0=[sd] \ [\bar u\bar d],
\quad \bar \kappa^0=[ud] \ [\bar s \bar d],\quad \kappa^-=[ud]\ [\bar s\bar
u] \label{tetraquark}\eeq
  $$\sigma=[ud]\ [\bar u \bar d]  $$  
Various applications of the 4-quark  model for the light
scalars have been considered for decay and production processes, in
particular with  photons \cite{Achasov:1987ts,Achasov:2009ee}. 
A picture of these scalar mesons as a mixture of tetra-quark states
(dominating in the light mesons) and heavy ($q\bar q$) states (dominating
the heavier mesons) has been proposed in \cite{Hooft:2008we} based on an instanton
induced effective Lagrangian theory.
The mixing between the mass eigenstates $f_0$ and $\sigma$ and the ideally
mixed states in \eref{tetraquark} is typically assumed small to be
consistent with the near mass degeneracy of $f_0$ and $a_0$. For example,
the corresponding mixing angle is $|\omega|<5^\circ$ in \cite{Hooft:2008we}.

The light scalar mesons can be established by the motion of phase shifts 
through $90^\circ$ in the appropriate two-body processes, 
except for the $\kappa$ with a phase in $K\pi$-scattering staying 
below $40^\circ$ (see section \ref{sectionkappa}).

\subsection{Route 2: light glueball around 1 GeV}
\label{route2}
\begin{table}[b]
\caption{\label{tab:route2} Spectroscopy with glueballs, route 2: light
glueball around 1 GeV.}
\begin{indented}
\item[]\begin{tabular}{cccl}
\br
 $I=0$ & $I=\frac{1}{2}$ &  $I=1$& \\
\br
             & $K^*(1950)$ && \\
\mr
$f_0(1710)$ &&& \\
\mr
$f_0(1500)$ & $K^*(1430)$ &  & { $q\bar q$ nonet: }\\
{ { ($f_0(1370)$?)}} & & &
{  $f_0(980)-f_0(1500)$ mixed } \\
$f_0(980)$ & & $a_0(980)$ \,\,\,\,&{ \hspace{1.8cm} like $\eta-\eta'$} \\
\mr
           & ($K^*(900)/\kappa$?) & & \\ 
{ { $f_0(500)/\sigma$}} & & 
          &\, { light glueball \ \ m$\sim$ 1 GeV}\\
\br
\end{tabular}
\end{indented}
\end{table}

The QCD results discussed above do not allow a definitive conclusion
about the mass and properties of the lightest scalar glueball. A light
glueball near 1 GeV is obtained from QCD sum rules; based on low energy
theorems also a large width is suggested which points to $f_0(500)$,
sometimes also referred to as $f_0(500-1000)$ according to the heavier
Breit-Wigner mass of $\sim 1000$ MeV
(see below and section 5.1.1). 
A light glueball near 1000 MeV is also
expected from certain lattice calculations \cite{Hart:2006ps} 
(see section 5.1.1), both states
$f_0(500-1000)$ or $f_0(980)$ are possible candidates. 

In view of the theoretical ambiguities at the end of the 90's,
a largely phenomenological approach to light scalars including a glueball 
has been pursued 
by Minkowski and Ochs \cite{Minkowski:1998mf} without reference to
the above specific QCD results. General properties, especially 
mixing patterns are studied basing
on an effective action for sigma variables \cite{Minkowski:1988rk}.
The lightest $q\bar q$ nonet
is constructed from the list of well established resonant
states, see table \ref{tab:route2}.
In particular, the existence of $f_0(1370)$ as an extra state
has not been accepted: the peaks appearing in the $\pi\pi$ mass spectrum
have
been interpreted as caused by a broad object centered around 1 GeV  
interfering destructively with the narrow resonances $f_0(980)$ and $f_0(1500)$
(``red dragon'' phenomenon). This
broad object has been related to $f_0(500)$: after separation of $f_0(980)$
the $\pi\pi$ elastic phase
shifts pass $90^\circ$ near 1000 MeV and can be represented locally by a
``Breit-Wigner'' resonance. 
A typical
fit for ``$f_0(1000)$'' yields (see also \cite{Au:1986vs})  
\begin{equation}
M_{BW}\approx  1000\ \MeV, \quad \Gamma_{BW}\approx \ 700\ \MeV. 
   \label{mass1GeV}
\end{equation}
The determination of the amplitude pole from the phase shifts is non-trivial
and we come back to this question
in section \ref{sigmamgam}. 

The lightest $q \bar q$ nonet in that
scheme includes the isoscalars $f_0(980)$ and $f_0(1500)$ with a mixing 
similar to $\eta'-\eta$. This has been motivated by the observed decay
rates $J/\psi\to \omega,\phi +X$ which  project out the strange and
non-strange quark components of $X$\footnote{It is assumed that the ``singly
disconnected diagram'', where the $q\bar q$ is produced in a single loop 
through one
intermediate gluonic exchange, dominates over the ``doubly disconnected
diagrams'' \cite{DiDonato:2011kr}.} and
are found comparable for
$X=\eta'$ and  $X=f_0(980)$. According to PDG \cite{beringer2012pdg}  
 (in units of $10^{-4}$):
\begin{eqnarray}
&B(J/\psi\to \omega\eta')=1.82\pm0.21; \quad 
    &B(J/\psi\to \phi\eta')=4.0\pm0.7 \\
&B(J/\psi\to \omega f_0(980))=1.4\pm0.5; \quad 
     &B(J/\psi\to \phi f_0(980))= 3.2\pm0.9 \label{Jf0omphi}
\end{eqnarray}
We write the flavour mixing of these scalar mesons as
\begin{eqnarray}
\quad|f_0(980)\ran &= \sin \phi_{sc} |n\bar n\ran + \cos\phi_{sc} |s\bar
s\ran, \label{mixingf0}\\
\quad|f_0(1500)\ran &= \cos \phi_{sc} |n\bar n\ran - 
    \sin\phi_{sc} |s\bar s\ran \label{mixingf0pr},
\end{eqnarray}
or inversely
\begin{eqnarray}
\quad|n\bar n\ran &=\sin \phi_{sc} |f_0(980)\ran + \cos \phi_{sc}
              |f_0(1500)\ran  \label{mixingnnbar}\\
\quad|s\bar s\ran &=\cos \phi_{sc} |f_0(980)\ran - \sin \phi_{sc}
              |f_0(1500)\ran.  \label{mixingssbar}
\end{eqnarray}
For estimates, the flavour
composition $|u\bar u, d\bar d, s\bar s\ran$ has been approximated by
\begin{equation}
\eta',f_0(980) \leftrightarrow |1,1,2\ran/\sqrt{6};\quad
\eta,f_0(1500) \leftrightarrow |1,1,-1\ran/\sqrt{3}.
\label{mixingscalars}
\end{equation}
The octet formed from $a_0(980)$, $K^*_0(1430)$ and $f_0'\equiv f_0(1500)$
is found to follow the Gell-Mann-Okubo mass relation (using
masses $984.7,\ 1412$ and $1507$ MeV) 
\begin{eqnarray}
m^2(f_0') &= m^2(a_0) &+ \tfrac{4}{3} (m^2(K_0^*)-m^2(a_0))\\
2.271     &=  0.970 &+ 1.365 = 2.335 \quad [\text{GeV}^2] \label{gmo}
\end{eqnarray}
within a few percent. 
 
The correspondence $\eta'\leftrightarrow f_0(980)$ and $\eta\leftrightarrow
f_0(1500)$  
is also characteristic of the ``Bonn quark model'' \cite{Klempt:1995ku,Koll:2000ke} 
 and the model of Nambu-Jona-Lasinio type \cite{Dmitrasinovic:1996fi} 
which include
instanton interactions with axial $U(1)$ 
symmetry-breaking \cite{'tHooft:1976fv};  these models 
explain the
reversed mass differences between the  octet and  singlet states 
in the scalar and pseudoscalar nonets. 

The extra state $f_0(500-1000)$ is then supernumerous and is taken as
glueball. This identification is supported at first view by the observation of this state
in all "gluon rich" processes (i)-(v), with the possible exception of
$J/\psi\to \gamma \pi^0\pi^0$ which is difficult to select from background
\cite{Ablikim:2006db}.\footnote{Preliminary data from BES-III suggest a
contribution from $f_0(500)$ centered at low masses around 500 MeV
\cite{Bes3}.}
Other results within this scheme 
are presented in \cite{Minkowski:1998mf} and on the appearance in 
$D,D_s$ and $B$ decays in \cite{Minkowski:2002nd} and \cite{Minkowski:2004xf}.
The $\kappa$ meson is not required in this scheme. Further discussion
follows in the subsequent sections on $f_0's$.

\subsection{Route 3: two supernumerous states, broad glueball
$f_0(1200-1600)$}
A global description of processes with mesonic final states 
$\pi\pi,K\bar K,\eta\eta,\eta\eta',4\pi$ with $J^{PC}=0^{++}$ 
in the mass range 280 -1900 MeV
in terms of a $K$ matrix
and also using a dispersion relation method 
has been presented
by Anisovich et al. \cite{Anisovich:2011zz} (earlier work in
 \cite{Anisovich:1996qj,Anisovich:2002ij}). Data from 
 $\bar p p,\bar p n$
annihilation into three mesons and meson pair production in $\pi p$
collisions have been included in a global fit. 
The appropriate mixing angle $\phi$ is
determined for each resonance from their decays 
(following table \ref{tab:decay}) where the glueball decay is
``flavour blind'' with a modification from strange quark suppression. The two isoscalar 
states in a
nonet are orthogonal in flavour space and the mixing angles of the
respective states should fulfill
\begin{equation}
\phi^{(I)}-\phi^{(II)}=\pm 90^\circ.
\label{nonetangle}
\end{equation} 
The results of the fits suggested the classification of the
isoscalar states as in table \ref{tab:route3}
with two $q\bar q$ nonets and one broad state 
 $ f_0(1200-1600)$ (width
$\Gamma=1100\pm140$) identified as glueball. Alternatively, one could exchange the
roles of $f_0(1300)$ and $f_0(1200-1600)$ which are both 
near flavour singlet. The $\sigma$ meson is another extra state and it is 
related to a confinement singularity of the $q\bar q$ potential. 
\begin{table}[h]
\caption{\label{tab:route3} Spectroscopy with glueballs route 3: 
two supernumerous states.}
\begin{indented}
\item[]\begin{tabular}{lll}
\br
[$f_0(980),f_0(1300)$] \qquad &[$f_0(1500),f_0(1750)$]& {$q\bar q$
     nonets}\\
$ f_0(1200-1600)$ \quad &$f_0(500)/\sigma$ & {glueball, extra state}\\ 
\br
\end{tabular}
\end{indented}
\end{table}

There is some similarity with route 2 in that the broad object is of gluonic
origin. The broad ``background'' under the narrow resonances in 
elastic $\pi\pi$
scattering is represented by two poles in route 3 and by one pole in route
2  corresponding to the appearance of only one loop 
of the amplitude in the complex plane after the narrow states are removed. 
Furthermore, both schemes work without $K^{*0}(900)/\kappa$.
On the other hand, 
the fits of route~3 include the state $f_0(1300)$ which is not included in
route~2.

\section{Properties of  $f_0(500)/\sigma$ - comparison with
$K^*_0(800)/\kappa$}
\subsection{Results on  $f_0(500)/\sigma$}
\subsubsection{Mass and width.}
\label{sigmamgam}
There are different definitions of the mass of an unstable particle. If
$P(s)\simeq 1/D(s)$ is its propagator a conventional
definition is the "on shell" mass with Re $D(s_0)=0$ in analogy to the
treatment of stable particles. A more fundamental definition is the 
``pole mass'' with $D(\bar s)=0$. As this definition is a basic property of
the $S$-matrix one expects in general better properties, for example, gauge
invariance in gauge theories; both quantities can be related 
(see, for example \cite{Kniehl:2008cj}). 

In the ``Breit Wigner'' approximation for masses 
around the resonance peak the scattering amplitude is written in terms
of resonance mass $m_0=\sqrt{s_0}$ and width $\Gamma$ as
\beq
T\sim \frac{1}{s_0-s-im_0\Gamma(s)}, \quad
\cot \delta=\frac{s_0-s}{m_0\Gamma(s)}
\label{BreitWigner}   
\eeq
and the ``on shell'' or ``Breit Wigner'' mass is defined by Re\,$T^{-1}=0$
or $s=s_0$, i.e. at this mass the amplitude becomes purely imaginary and the 
phase passes $90^\circ$; the width is obtained from 
$\frac{d\delta}{d\sqrt{s}}=\frac{2}{\Gamma}$ at $s=s_0$. In this way the Breit-Wigner mass and
width are experimentally observable quantities.
For  $f_0(500-1000)/\sigma$
one finds typically \eref{mass1GeV}. These Breit-Wigner 
quantities are also relevant in the applications of the QCD sum rules, where
they enter in the narrow width approximation to the dispersion relations as
in \eref{absorptive}.
The complex pole mass $\sqrt{\bar s}=\bar m-i\frac{\bar\Gamma}{2}$ is not directly 
accessible from experiment
but an analytic extrapolation of the amplitude
away from the region of experimental measurements is required.  

A description of the observed $\pi\pi$ 
phase shifts from threshold to $\sim 1400$
MeV has been achieved using a ``minimal meromorphic parameterization''
\cite{Minkowski:2008ri} which is defined for a resonance from its complex pole position and a
width $\Gamma (s)\propto q$. Three resonances have been included and a
background taking care of the Adler zero. Then the pole of the broad
resonance is close to the Breit Wigner result \eref{mass1GeV}.
 
Another procedure to determine the pole mass
is based on
the application of partial wave dispersion
relations (``Roy equations'' \cite{Roy:1971tc}) which follow from 
first principles of $S$
matrix theory such as analyticity, crossing symmetry and unitarity, 
together with 
experimental information on $\pi\pi$ scattering. 
The pole of the amplitude at complex energy 
$\sqrt{\bar s}=\bar m-i\ \bar\Gamma/2$ 
has been found in that way as \cite{Caprini:2005zr}
\beq
\bar m_\sigma=441^{+16}_{-8} \ \MeV,\quad \bar\Gamma_\sigma=544^{+18}_{-25}.
\label{sigmapol}
\eeq
Other applications of dispersion relations for an extended range of energy
($\sqrt{s} \lesssim 1400$ MeV) lead to similar results 
with $\bar m_\sigma=457^{+14}_{-13},\
\bar\Gamma_\sigma=558^{+22}_{-14}$ MeV \cite{GarciaMartin:2011jx} and
$\bar m_\sigma=442^{+5}_{-8},\ \bar\Gamma_\sigma=548^{+12}_{-10}$ MeV
\cite{Moussallam:2011zg} (For further discussion, see \cite{Amslerscalar}). 

The large difference
between the pole mass $\bar m_\sigma$ and the "Breit-Wigner mass" or "on shell
mass" $M^{BW}_\sigma$ where the phase shift
passes 90$^\circ$ can be understood in simple models such as the one by
Mennessier \cite{Mennessier:1982fk} which has been slightly modified for
recent applications \cite{Mennessier:2008kk,Mennessier:2010xg}. The analytical and unitary 
amplitudes for  
$\pi\pi\to \pi\pi,\ K\bar K$ are derived from a
$K$ matrix parameterization with $\sigma$ and $f_0(980)$ resonances 
where the real part of the amplitudes is derived from
dispersion relations. One finds for the pole and
``on-shell'' mass
\begin{equation}
\bar m_\sigma\simeq 422\ \MeV,\ \bar\Gamma_\sigma\simeq 580\ \MeV;\qquad 
    m_\sigma^{BW}\simeq 900\ \MeV,\
\Gamma_{\sigma\to\pi\pi}^{BW}\simeq 1000\ \MeV,  
\end{equation}
so $\bar m_\sigma$ and $\bar\Gamma_\sigma$ are found near \eref{sigmapol}.
The difference between pole and ``on shell'' mass is related in this model
to the analytic properties of the denominator $D(s)$ of the $\sigma$
propagator.\footnote{Earlier
phenomenological pole determinations 
by BES \cite{Ablikim:2004qna} using a variety of parameterizations 
gave the pole position
$\sqrt{\bar s}=(541\pm 39) -i\ (252\pm42)$ MeV.} 

\subsubsection{Phenomenological appearance of $f_0(500)/\sigma$.}
In $\pi\pi$ elastic scattering the $f_0(500)/\sigma$ contribution 
corresponds to a broad resonance with central mass near 1000 MeV where the
phase passes through  $90^\circ$. In decays
of mesons of low mass, such as $f_0(1500)\to \sigma\sigma$, 
only the low energy part of $\sigma$
below 1 GeV contributes because of phase space restrictions. In decays of
$B$ mesons with heavy $b$ quark the broad background appears
again which can be related to $f_0(500)/\sigma$, for example in $B\to K\pi\pi,\ B\to K \bar K K$  (Belle
\cite{Garmash:2004wa}) and $B^0_s\to J/\psi \pi\pi$ (LHCb \cite{LHCb:2012ae}). 
The Belle data have been parameterized in \cite{Minkowski:2004xf}
by a broad resonance with moving phase as in elastic scattering.
There are other decays, like $J/\psi\to \omega \pi\pi$
\cite{Augustin:1988ja,Ablikim:2004qna} or $D\to 3\pi$ \cite{Aitala:2000xu}  
where a low mass peak around 500 MeV appears. The appearance of this peak
in processes of relatively low energy and absence at higher energies 
may be related to a special dynamics not directly related to the $\sigma$
pole, for example to subthreshold $\rho$ exchange with primary 
$J/\psi\to \rho\pi$
and secondary $\rho\to
\omega\pi$ transition \cite{Wu:2006tb}. In 
decays like $\chi_c\to \sigma\sigma$ it is not obvious from the beginning, 
what mass spectrum the $\sigma$ decays will acquire. Some clue may come from
a comparison of $\chi_{c0}$ and $\chi_{c2}$ with the very different phase
space factors. Independently of the position of the mass peak, the final state
$\pi\pi$ phase should
follow the phase of elastic scattering if re-interactions are neglected. 
This was observed for $J/\psi\to \omega \pi\pi$
where the $S-D$ wave interference behaves smoothly over the peak region
around 500 MeV as reported in \cite{Ochs:2001vh}. 

\subsubsection{Couplings of $f_0(500)/\sigma$ to $\pi\pi$, $K\bar K$ and
$\eta\eta$.}
The intrinsic structure of $f_0(500)$  can be revealed by the decay branching
ratios. These can be 
extracted from a model amplitude
which fits data from both channels $\pi\pi$ and $K \bar K$. The decay ratio
of couplings $r_{\sigma \pi K}=\frac{|g_{\sigma K^+K^-}|}
{|g_{\sigma \pi^+\pi^-}|}$ has been
derived from an
extrapolation to the pole of an analytical amplitude or ``on shell'' 
along the real axis near $m_{BW}\sim1$ GeV.
A result of the first kind is obtained by Kaminski et al. 
\cite{Kaminski:2009qg} based on
a coupled channel analysis as
described in the previous section \ref{sigmamgam} using the analytical 
$K$-matrix model 
\cite{Mennessier:1982fk,Mennessier:2008kk}. They find 
$r_{\sigma K\pi}=0.52\pm 0.17$, i.e. a significant strange quark component in
the $\sigma$.
In table \ref{tab:sigma} we summarize the results on the ratios
$R_{\sigma K\pi}=\frac{g^2_{\sigma\to K\bar K}}{g^2_{\sigma\to \pi\pi}}$ 
 $=\frac{4}{3}r_{\sigma K\pi}^2$ and $R_{\sigma\eta\pi}$
 which are derived from fits to both the
 elastic and inelastic $\pi\pi$ scattering. In an approach based on dispersion
relations and a 3-coupled channel unitary model a similar result has been
 obtained by the same group $r_{\sigma  K\pi}=0.65\pm0.18$.  
\begin{table}[t]
\caption{\label{tab:sigma} Ratio of couplings 
$R_{\sigma K\pi}=g^2_{K\bar K}/g^2_{\pi\pi}$ and 
$R_{\sigma \eta\pi}=g^2_{\eta\eta}/g^2_{\pi\pi}$ for $f_0(500)/\sigma$.}
\begin{indented}
\item[]\begin{tabular}{lllll}
\br
processes       &$R_{\sigma K\pi}$ & $R_{\sigma \eta\pi}$& model & reference
mass\\
\mr
$\pi\pi\to \pi\pi/K\bar K$&$ 0.36^{+0.27}_{-0.20}$ & &K matrix \cite{Kaminski:2009qg}& 
  $\sigma$ pole / $m_{BW}$ \\
$\pi\pi\to \pi\pi/K\bar K/4\pi$& $0.56^{+0.36}_{-0.27}$ & &S matrix \cite{Kaminski:2009qg}&
 $\sigma$ pole\\
$\pi\pi\to \pi\pi/K\bar K/\eta\eta$& $0.6^{+0.1}_{-0.2}$ &$0.2\pm 0.04$ & 
   resonance model \cite{Bugg:2006sr}&
$m_{BW}\gtrsim 1$ GeV\\

\br
\end{tabular}
\end{indented}
\end{table}
A phenomenological description of 
$\pi\pi$, $K\bar K$ and $\eta\eta$ channels extending up to $1.6-2.0$ GeV 
based on extended inelastic Breit Wigner formulae for $f_0(500)/\sigma$ 
and inclusion of other resonances has been achieved by 
Bugg \cite{Bugg:2006sr}. Near threshold the process $\pi\pi\to K\bar K$
is dominated by $f_0(980)$ but a sizable $f_0(500)/\sigma$ component 
with $R_{\sigma K\pi}=0.6^{+0.1}_{-0.2}$ is necessary for a good fit.
We note however, that this quantity may be subject to additional systematic
uncertainties because of conflicting experimental phases just above the 
$K\bar K$ threshold: the fit by Bugg follows the phases by Etkin et al. 
\cite{Etkin:1981sg} which start rising near
$\varphi_{thr}\sim 170^\circ$ while the phases by 
Cohen et al. \cite{Cohen:1980cq}, 
preferred in  figure \ref{fig:s00kk} below, start decreasing from 
$\varphi_{thr}\sim 220^\circ$.\footnote{Bugg also obtained a
consistent result $R_{\sigma K\pi}=0.6\pm0.1$ 
from data on the radiative decay $\phi(1020) \to \gamma\pi^0\pi^0$
with the contribution of the virtual 
process $\pi\pi\to K\bar K$ within the $K\bar K$ loop model.}

The results for $R_{\sigma K\pi}$ in table \ref{tab:sigma} show consistency
between both methods. For the K-matrix model it has been found 
\cite{Kaminski:2009qg} that  $r_{\sigma K\pi}$ evaluated for the pole mass 
and for the on-shell (BW) mass coincide. The ratios in table \ref{tab:sigma}
are all derived from the same data and we estimate
\beq
R_{\sigma K\pi}=0.5\pm 0.2. \label{rsigmakpi}
\eeq
There is also a
small but significant decay into $\eta\eta$ with 
$R_{\sigma\eta\pi}=0.20\pm0.04$.

All these results indicate that $f_0(500)/\sigma$ is not a pure non-strange
object, but couples to a sizable $s\bar s$ (or gluonic) component. 
A pure flavour singlet 
or glueball would correspond to $r_{\sigma K\pi}=1$ or 
$R_{\sigma K\pi}=4/3$ which is above the results in table \ref{tab:sigma}. 
While some caution on these numbers is still appropriate because of possible
systematic errors there is also the possibility of a mixing with $f_0(980)$
as discussed in \cite{Narison:2005wc}. 
According to these model results,  a 4-quark model in its simplest form 
with a non-strange $\sigma$ \cite{Jaffe:1976ig} or with a
small mixing $r_{\sigma\pi K}<0.1$ as in \cite{Achasov:2012bh} is
not supported.

\subsubsection{Two-photon coupling of $f_0(500)/\sigma$.\label{twophoton}}
Following the general idea that glueballs couple weakly to photons one
may consider the two-photon decay of the tentative glueball as a crucial
measurement. The recent results from Belle  \cite{Uehara:2008ep,Mori:2006jj} 
on the reactions $\gamma\gamma\to
\pi^0\pi^0,\pi^+\pi^-$ have renewed the interest in this problem. The
two-photon width has been extracted by Pennington 
\cite{Pennington:2006dg} as $\Gamma(\sigma\to\gamma\gamma)=(4.1\pm0.3)\
\keV$. This result has been obtained from the application of partial wave dispersion
relations and using the position of the $\sigma$ pole \eref{sigmapol}
from \cite{Caprini:2005zr}. 
A more recent evaluation and the mean of various other
determinations \cite{Moussallam:2011zg} is a bit lower 
\begin{equation}
\Gamma(\sigma\to\gamma\gamma)=(2.08\pm0.20^{+0.07}_{-0.04})\ \keV.\ 
\label{sigmagg}
\end{equation}
A number of this order is in the range of expectations for
$q\bar q$ states but gluonic states require generally smaller decay widths,
for example $\Gamma(\sigma\to\gamma\gamma)\approx 0.2-0.3$ keV
\cite{Narison:1996fm}, an order of magnitude smaller than what is found
in \eref{sigmagg}. Therefore the glueball assignment has been questioned in
\cite{Pennington:2006dg}.

A possible solution of this problem has been suggested in \cite{Mennessier:2008kk}
within 
the analytic $K$ matrix model mentioned before
\cite{Mennessier:1982fk}.
Given a set of coupled multi-channel strong processes, the amplitudes for
the related electromagnetic processes are largely determined by unitarity
and constraints from dispersion relations, although with some ambiguities
from subtraction terms \cite{Omnes:1958hv}. Along this general
scheme, the calculation of the electromagnetic processes
$\gamma\gamma\to \pi\pi,K\bar K$, given the strong processes
$\pi\pi\to \pi\pi,K\bar K$ has been carried out in \cite{Mennessier:1982fk} 
based on a $K$ matrix model representing the amplitudes by a set of poles.
The subtraction ambiguities in electromagnetic processes correspond to the free ``direct
couplings'' of resonances to photons.

At low energies the process $\gamma\gamma\to \pi^+\pi^-$ is governed by 
one-pion exchange with the photons coupling to charged pions and this
process dominates by an order of magnitude over 
$\gamma\gamma\to \pi^0\pi^0$ where the initial photons couple to charged
pions and subsequently re-scatter:  $\gamma\gamma\to \pi^+\pi^-\to \pi^0\pi^0$. 
In the
resonant scattering $\gamma\gamma\to R\to \pi\pi$ there is the possibility
of $\pi\pi$ re-scattering and in addition of 
``direct coupling'' $\gamma\gamma\to R$. At higher energies the re-scattering
contribution decreases and the direct processes become dominant. They
correspond to interactions where the energetic photons resolve 
the constituents inside the resonances. An
example is the two-photon decay of $f_2(1270)$ and other tensor mesons 
which are well represented by the coupling of photons to the constituent
quarks \cite{Amslerquark}.

In \cite{Mennessier:2008kk} the ``direct coupling'' of $f_0(500)/\sigma$
has been determined from the simultaneous fits to the strong $\pi\pi$ and the
$\gamma\gamma$ processes below 700 MeV. Subsequently, the inelastic channels
including $K\bar K$ and a second resonance $f_0(980)$ has been included
\cite{Kaminski:2009qg,Mennessier:2010xg}. The following 
results at the resonance pole for $f_0(500)/\sigma$ are obtained
\begin{eqnarray}
\Gamma^\text{dir}_{\sigma}\to \gamma\gamma\simeq (0.13\ldots 0.19)\ \keV,\\
\Gamma^\text{resc}_{\sigma}\to \gamma\gamma\simeq (2.7\ldots 1.48)\ \keV,\quad
\Gamma^\text{tot}_{\sigma}\to \gamma\gamma\simeq (3.9\ldots 2.67)\ \keV
\end{eqnarray}
where the first number comes from \cite{Mennessier:2008kk}, the second one
from the average of the two numbers in \cite{Mennessier:2010xg}.
One can see that with this approach 
the $\gamma\gamma$ process is dominated by $\pi\pi$ re-scattering.
It is the direct coupling component which should be related to the
constituents of the resonance and this is an order of magnitude smaller than
the total width, quite compatible with gluonic constituents (like
$\Gamma\sim 0.2-0.3\ \keV$, as in \cite{Narison:1996fm}).
This result is based on a model consideration, but one may conclude that 
it is premature to reject the gluonic nature of $\sigma$ based on the
$2\gamma$ decay width.

\subsection{Comparing 
 elastic $\pi\pi$ and $K\pi$ scattering  - a problem with
the $\kappa$ resonance}
\label{sectionkappa}
In the nonet classification of the scalars below 1000 MeV (route 1) 
the broad $\kappa$ and
$\sigma$ particles appear. While $\sigma$ plays its role in all three
classification routes considered the $\kappa$ appears only in route 1
and would be supernumerous otherwise if a real resonance.
In this subsection we add a few comments on these broad objects in a
comparison.
 
At first sight there is a similarity in low energy $\pi\pi$ and $K\pi$
scattering and the strong 
low energy interaction suggests the existence of two
broad resonances  $\sigma$ and $\kappa$ with a width comparable to mass. 
The first  narrow resonance of low mass in $\pi\pi$ scattering appears with 
$f_0(980)$ and in $K\pi$ scattering with $K^*_0(1430)$. A closer look  
at the scattering phase shifts, however, reveals a considerable difference
between both scattering channels.

In the elastic scattering region we may parameterize the amplitude for the
production of a resonance $R$ above a broad background $B$, following
Michael \cite{Michael:1966pl}, 
as product of the respective $S$ matrices which satisfies unitarity
(also called ``Dalitz-Tuan method'')
\begin{equation}
S=S_R\ S_B,\quad \rm{or} \quad T= T_B+e^{2i\delta_B}T_R
\label{addresonances}
\end{equation}
where $S=1+2iT$ and $T=\sin \delta e^{i\delta}$. This approximation 
applies  below the inelastic thresholds of $\pi\pi\to K\bar K$ 
at $m_{\pi\pi}\sim 1000$ MeV and 
$K\pi \to K\eta'$ at $m_{K\pi}\sim 1450$ MeV
respectively (disregarding the negligible $K\eta$ production) and 
then the overall phase of $S=e^{2i\delta}$ is $\delta=\delta_R+\delta_B$.
According to \cite{Michael:1966pl} \eref{addresonances} is also applicable
for inelastic resonances and near elastic backgrounds.
If the contribution of the lowest narrow resonances $R$ to the phase
$\delta$ are
``removed'' the remaining background phase $\delta_B$ 
shows quite a different behaviour for the two channels 
(see figure \ref{fig:elastic}):
\begin{figure}[t]
\hspace{-0.5cm}\includegraphics[width=7.cm]{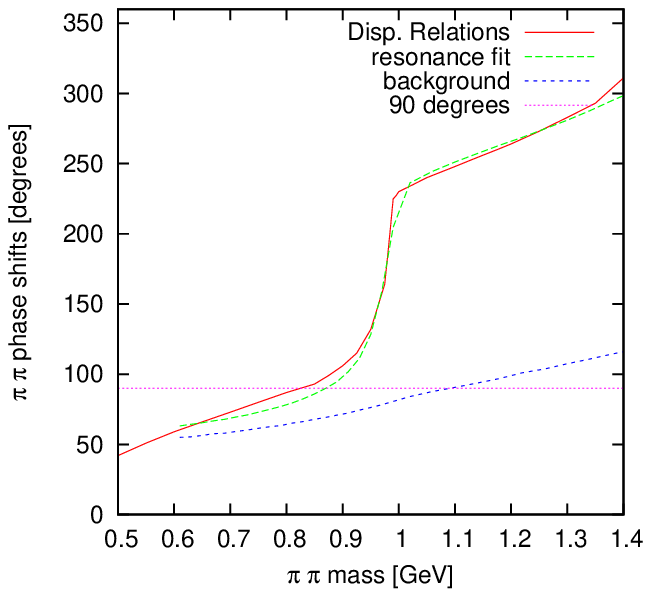}
\hspace{-2.cm}\includegraphics[width=7.cm]{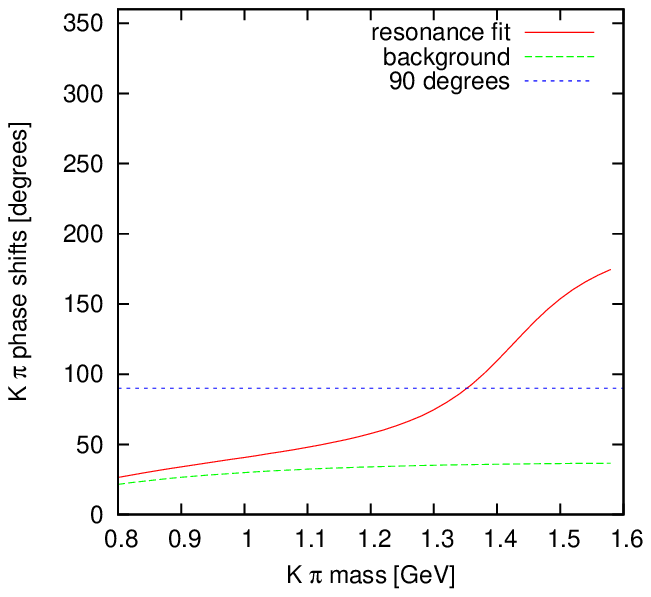}
\hspace{-1.cm}\includegraphics[width=3.9cm,bbllx=0.0cm,bblly=0.0cm,bburx=11.0cm,bbury=11.5cm]{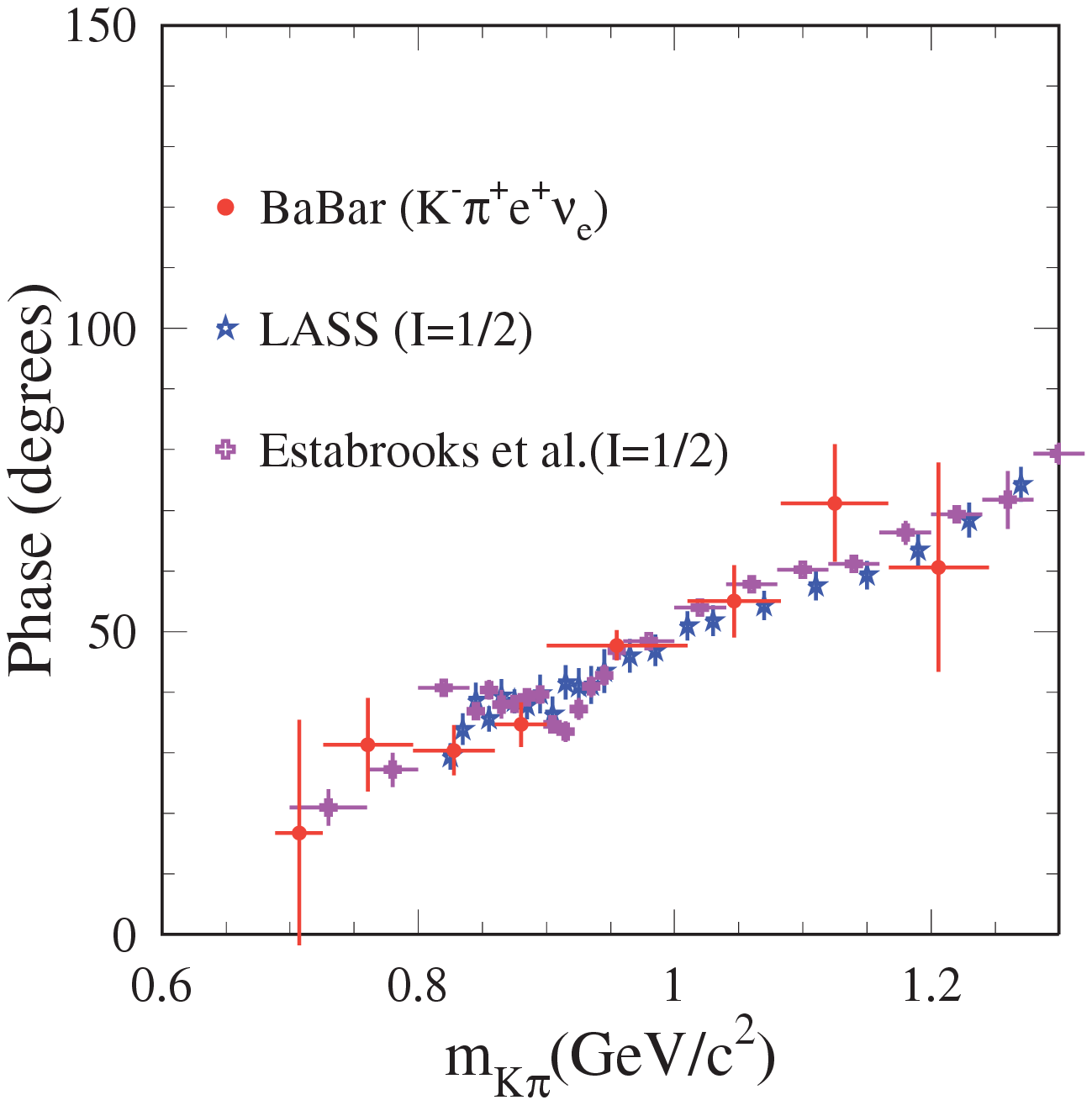}
\caption{\label{fig:elastic} 
Phase shifts of elastic scattering, Left: $\pi\pi$ phase shifts 
from Dispersion Relations
\cite{GarciaMartin:2011cn} and from resonance fit to CM-I data \cite{Hyams:1973zf}
($f_0(980)$ + background \cite{Ochs:2006rb});  Middle: $K\pi$ phase shifts
($K^*_0(1430)$ + background \cite{delAmoSanchez:2010fd,Dunwoodie}); Right: $K\pi$ phase shifts from 
BABAR
\cite{delAmoSanchez:2010fd} ($D^+\to K^-\pi^+e^+\nu_e)$, LASS
\cite{Aston:1987ir} ($K^\pm p\to K^\pm\pi^+n$) and 
Estabrooks et al. \cite{Estabrooks:1977xe}
($K^\pm p\to K^\pm \pi^- \Delta^{++}$), figure from
\cite{delAmoSanchez:2010fd}. The $K\pi$ 
background phase stays below $40^\circ$.}
\end{figure}
\begin{eqnarray}
m_{\pi\pi}&\approx 1000 \ \MeV: & \delta_B=90^\circ\\
m_{K\pi}&\approx 1400 \ \MeV: & \delta_B<40^\circ, 
\end{eqnarray}
i.e. in $\pi\pi$ scattering the ``background'' 
itself shows a resonant behaviour whereas  in $K\pi$ scattering the
background phase saturates already below 40$^\circ$ according to the present
experimental status of $K\pi$ scattering which presents itself as follows:

There is still a discussion on the appropriate description of low energy
$K\pi$ scattering.
Elastic $K\pi$ scattering amplitudes have been reconstructed from processes 
$K^\pm p\to K^\pm \pi^+n$ and $K^\pm p\to K^\pm \pi^-\Delta^{++}$ by 
isolating the respective
One-Pion-Exchange contribution \cite{Estabrooks:1977xe,Aston:1987ir}.
Following the LASS Collaboration \cite{Aston:1987ir}  
the  S-wave contribution with $I=\frac{1}{2}$ can be represented 
by the superposition as in \eref{addresonances} 
of the $K^*_0(1430)$ Breit-Wigner amplitude and a background which is 
parameterized by an effective range formula
($cms$ momentum $k$)
\begin{equation}
\cot(\delta_{\rm{BG}}) =\frac{1}{a k^2} + \frac{bk^2}{2};
\label{effrange}
\end{equation}
a recent reanalysis has been presented by the BABAR Collaboration
\cite{delAmoSanchez:2010fd,Dunwoodie} which 
gave the numbers $a=1.95\pm 0.09$ GeV$^{-1}$ and $ b=1.76\pm0.36$
GeV$^{-1}$ for the $I=\frac{1}{2}$ S-wave.

A model independent determination of the $K\pi$ phases in the elastic region 
is possible from the decays of $D\to K\pi e\nu_e$ and  
 $\tau\to K\pi \nu_\tau$ as in these decays, according to the Watson theorem
\cite{Watson:1954uc},  the $K\pi$ production phase agrees (modulo $\pi$)
with the elastic $K\pi$ scattering phases.

A measurement of all the decay distributions in 
$D\to K\pi e\nu_e$ by the BABAR Collaboration
\cite{delAmoSanchez:2010fd} has determined the $I=\frac{1}{2}$ $K\pi$ $S$-wave
phase shifts in the range $700-1400$ MeV. The results agree nicely 
with the earlier results on
$K\pi$ phases from the production off the proton target 
and confirm the description of the $S$-wave 
amplitude in terms of the $K^*_0(1430)$ and the effective range background
\eref{effrange} (see figure \ref{fig:elastic}, right panel). The resulting background phase approaches 38$^\circ$ 
at around 1400 MeV where it reaches a maximum.\footnote{The existence of 
an $S$-wave with a phase variation 
consistent with \eref{effrange} has been noted before \cite{Link:2005ge} 
in a limited mass range around $K^*(892)$ with lower statistics.}

An analysis of $\tau\to K\pi \nu_\tau$ decays has been presented by Belle
\cite{Epifanov:2007rf}.  The $K\pi$ mass spectrum requires contributions in
addition to $K^*(892)$ and a successful fit includes an $S$-wave with a
superposition of $K^*_0(800)/\kappa$ and $K^*_0(1430)$.  Also a
LASS-type fit with background \eref{effrange} was possible but 
found only marginally consistent with previous values for the 
parameters $a$ and
$b$. However, the measurement of further decay observables is required for an independent
determination of the $K\pi$ phase shifts and a test of the Watson theorem,
to compare with the results from $D\to K\pi e\nu_e$.
First results have been presented by BABAR with similar fits
\cite{Paramesvaran:2009ec}. At present we stay with the results presented in
figure \ref{fig:elastic}.

The situation with a low mass resonance $K^*_0(800)/\kappa$ is discussed
by the PDG \cite{Amslerscalar}. The existence of this state close to the $K\pi$
threshold and with a large width around 500 MeV is considered 
difficult to establish.
A positive result is obtained recently in an approach based on a 
dispersive representation of the $K\pi$ scattering 
amplitude ~\cite{DescotesGenon:2006uk}. After continuation into the complex 
energy plane a pole has been identified at
\beq
\bar m_\kappa = 658\pm 13\  \MeV,\quad \Gamma_\kappa = 557\pm 24\ \MeV, 
\eeq
just above $K\pi$ 
threshold at around 640~MeV. 

In summary, descriptions of the scalar $K\pi$ amplitude with a $\kappa$
pole built in are possible but the apparent slow phase motion of the 
corresponding amplitude which
saturates below 40$^\circ$ is untypical for a resonance. 
In this way a characteristic condition for
the existence of a resonance - the phase shifts are passing through 
90$^\circ$  - is not fulfilled. 
One could think of a repulsive background under the $\kappa$ with destructive
interference \cite{Ishida:1997wn}, but then the background phase ought to be larger than the
resonance phase ($|\delta_B|>|\delta_R|)$.
Alternatively, an enhanced but not
necessarily resonant low energy $K\pi$ interaction could reflect the
strong $t$-channel $\pi\pi\to K\bar K$ low energy ($\sigma$) forces.\footnote{This viewpoint has been 
envisaged in discussions with P. Minkowski.} 

\section{Intrinsic structure of $f_0(980)$}
\label{f0section}
\subsection{Mixing angle from various observables}
We discuss here several observables within the $q\bar
q$ constituent model and determine the
$q\bar q$ mixing angle defined as in \eref{mixingf0} following
earlier studies \cite{Minkowski:1998mf,Minkowski:2002nd}. Some of these observables have been
also considered in \cite{Anisovich:2000wb}; other determinations of the mixing 
angle, also in the context of a 4-quark picture, 
have been reported in \cite{Fleischer:2011au}, but no clear picture has emerged.
We consider the following ratios of branching ratios:

{\it (i) Decay into $\pi\pi$ and $K\bar K$.} This decay is often described
in terms of a Breit Wigner resonance formula, following Flatt\'e
\cite{Flatte:1976xu}, with amplitude 
\begin{equation}
 T_{ij} = \frac{g_ig_j\sqrt{\rho_i\rho_j}}
    {s_0-s-i(\rho_\pi g_{\pi}^2 + \rho_{K} g_{K}^2)}
\label{flatte}
\end{equation}
where $g_i$ denotes the couplings of the resonance to the  $\pi\pi$ and $K\bar K$
 channels and
$\rho_i=\frac{2k_i}{\sqrt{s}}$ the respective phase space with the momenta
 $k_i$ in the resonance rest frame.  The BES Collaboration has measured the
decays $J/\psi\to \phi(1020)\pi^+\pi^-/\phi(1020) K^+ K^-$ and determined
$g_K^2/g_\pi^2=4.21\pm0.33$ (errors added in quadrature) \cite{Ablikim:2004wn}.
 
Various results have been obtained from data on the process $\pi\pi\to K\bar
K$ together with $\pi\pi\to \pi\pi$ in an application of different coupled
channel approaches.
The data mainly used in these analyses 
because of their accuracy are from Cohen et al. on $K^+K^-$ 
\cite{Cohen:1980cq} and from Etkin et al. \cite{Etkin:1981sg} on $K^0_1K^0_1$ final
states. In table \ref{tab:g22/g12} we list some results on $g_K^2/g_\pi^2$ which included $K\bar
K$ data in the fits: 
\begin{table}[h]
\caption{\label{tab:g22/g12} Couplings $g_K^2/g_\pi^2$ in
\eref{flatte} for $f_0(980)$ from $\pi\pi\to K\bar K$ and from $J/\psi$
decay.}
\begin{indented}
\item[]\begin{tabular}{lllllll}
\br
  Martin &Estabrooks & Bugg & Kaminski & Mennessier & mean ($\pi\pi$) 
   & BES ($J/\psi$)\\
  $4.0\pm 0.6$ & 
    $3.2\pm 1.8$ & 2.3 & $3.9\pm 0.9$ & $9^{+11}_{-7}$&
    $ 3.3\pm0.6$ &$4.21\pm0.33$ \\
\br
\end{tabular}
\end{indented}
\end{table}
from Martin et al. 
\cite{Martin:1976vx} using several different models for resonances; from Estabrooks 
\cite{Estabrooks:1978de} based on a K matrix fit; from Bugg et al.
\cite{Bugg:1996ki} using a global parameterization based on a K matrix, see
also \cite{Anisovich:2000wb} for interpretation; from Kaminski et al.
\cite{Kaminski:1998ns} using a unitary model with separable interactions
(see also \cite{Kaminski:2009qg}) and the analytic K-matrix model by
Mennessier et al. \cite{Mennessier:2010xg}. As the ratio  $g_K^2/g_\pi^2$
depends essentially on the same $K\bar K$ data we average the 
weighted 
results as usual but assume a full correlation of the input data and obtain the mean in table
\ref{tab:g22/g12}. Finally, together with the independent BES measurement we find
\begin{equation}
f_0(980):\qquad g_K^2/g_\pi^2=4.0\pm 0.3;\quad g_K/g_\pi > 0.
\label{couplingsf0980}
\end{equation}

The same sign of the couplings $g_K$ and $g_\pi$ has been derived from the
fits \cite{Martin:1976vx,Estabrooks:1978de} 
to the $K^+K^-$ data by Cohen et al. In these data the relative sign of the
$\pi^+\pi^-\to K^+K^-$ and $\pi^+\pi^-\to\pi^+\pi^-$ amplitudes has been
fixed from the observation that near 1 GeV the $P$ wave in both channels is
governed by the tail of the $\rho$ and assuming these amplitudes to be in
phase.
This observation resolves the previously noted sign ambiguity (for example
in  \cite{Anisovich:2000wb,Mennessier:2010xg}).
The result \eref{couplingsf0980} can be expressed in terms of the mixing
angle $\phi_{sc}$ defined in \eref{mixingf0}. Using 
table \ref{tab:decay} with $\phi_{sc}=\alpha+\pi/2$ according to the
definition \eref{mixing} we find  
$ g_K^2/g_\pi^2\equiv \gamma^2_{K\bar
K}/\gamma^2_{\pi\pi}$ as 
\begin{equation}
 g_K^2/g_\pi^2 =(1+\sqrt{2}\cot \phi_{sc})^2/3
\end{equation}
in the symmetry limit $\rho=1$ and yields $\phi_{sc}=29.9^\circ\pm 1.3^\circ$ 
(for $g_K/g_\pi>0$, otherwise  
$\phi_{sc}=162.4^\circ\pm 0.5^\circ$ noting that
$g_{K^+K^-}/g_{\pi^+\pi^-}=(1+\sqrt{2}\cot \phi_{sc})/2$ from table
\ref{tab:decay}). 

{\it (ii) Decays $J/\psi\to f_0(980) \phi$ and $f_0(980) \omega$.}
In these decays the $s\bar s$ and $n\bar n$ quark components of
$f_0(980)$ are
projected out and according to \eref{mixingf0} their ratio can be related to
the quark mixing angle \cite{Minkowski:1998mf} assuming the dominance of the
``singly disconnected diagram'' \cite{DiDonato:2011kr}
\begin{equation}
R^{\phi/\omega}_{J/\psi}=\frac{B(J/\psi\to f_0(980) \phi)}
{B(J/\psi\to f_0(980)\omega)}: \quad  
      R_{J/\psi}^{\phi/\omega}=\frac{k_\phi}{k_\omega} \cot^2 \phi_{sc}.
\end{equation}
with $cms$ decay momenta $k_i$ of $\phi$ and $\omega$.
Taking these branching ratios from PDG as in \eref{Jf0omphi}
we find
$\cot^2\, \phi_{sc} = 2.46\pm1.12$
or $\phi_{sc}=\pm (33 \pm 7)^\circ$.
This result is similar to the corresponding ratio for $\eta'$, 
$R^{\eta'}_{J/\psi}=2.05\pm0.43$ using PDG values and this yields
$\phi_{ps}(\eta')=(35\pm 3)^\circ$ to be compared with the experimental 
mixing angle $\phi_{ps}= 42^\circ$ in \eref{mixingeta}.
This observation was the original motivation for the mixing ansatz for
scalar mesons in \cite{Minkowski:1998mf}.

{\it (iii) Two photon decays of $f_0(980)$ and $a_0(980)$.} 
These decays are sensitive to the charged constituents of these mesons. In an update of
\cite{Minkowski:1998mf} we consider the ratio $R_\gaga$ in a $q\bar q$
constituent model
\begin{equation}
R_{\gamma\gamma}^{f_0/a_0}=\frac{\Gamma_{\gaga}(f_0(980))}
     {\Gamma_\gaga (a_0(980))};\quad \Gamma_\gaga \propto (\sum_q c_qQ_q^2)^2
\label{Rgagadef} 
\end{equation}
with amplitudes $c_q$ for $q=(u,d,s)$, $(\sum c_q^2 =1)$, for the contributions from $q\bar q$ of charge $Q_q$ and
obtain with mixing \eref{mixingf0}
\begin{equation}
R_{\gamma\gamma}^{f_0/a_0}=\tfrac{25}{9} 
    (\sin \phi_{sc} +\tfrac{\sqrt{2}}{5} \cos \phi_{sc})^2.  \label{Rgaga}
\end{equation}
The two photon decays have been studied recently by the Belle experiment
for $f_0(980)$ \cite{Uehara:2008ep,Mori:2006jj} and $a_0(980)$
\cite{Uehara:2009cf}. We take here $\Gamma_\gaga^{f_0}=0.29^{+0.07}_{-0.06}
\ \keV$ for $f_0(980)$ from the PDG fit, 
in case of $a_0$ we use their branching ratio 
$\Gamma_{K\bar K}^{a_0}/\Gamma_{\eta\pi}^{a_0}=0.183\pm 0.024$
to obtain  $B(a_0(980)\to \pi\eta)=0.85\pm0.02$ and then
$\Gamma_\gaga^{a_0}=0.25^{+0.09}_{-0.05}\ \keV$;
finally, with \eref{Rgaga}
\begin{equation}
R_{\gaga}^{f_0/a_0}=1.16\pm 0.41,\qquad \phi_{sc1}=(23.7\pm8.1)^\circ ;\quad
\phi_{sc2}=(125.8\pm8.1)^\circ. \label{rgaga}
\end{equation}

{\it (iv) Charmed meson decays $D,D_s\to f_0(980)\pi$.}
The decays of $D^+$ and $D^+_s$ into
$f_0\pi^+$ and
$K^*_0(1430)\pi^+$ provide a further possibility to determine the
flavour structure of $f_0(980)$ and we follow here the investigation in
\cite{Minkowski:2002nd}. They consider the two ratios
\begin{equation}
R_{D1}=\frac{\Gamma(D_s^+\to f_0\pi^+)}{\Gamma(D^+\to f_0\pi^+)},\quad
R_{D2}=\frac{\Gamma(D_s^+\to f_0\pi^+)}{\Gamma(D^+\to K^*_0(1430)\pi^+)}
\label{ddecays}
\end{equation}
Assuming again the flavour composition \eref{mixingf0} for $f_0(980)$
and the $K_0^*(1430)$ to belong to the same nonet,
we write the amplitudes for the three processes involved as follows
\begin{eqnarray}
(a)\ & D_s^+\to  f_0\pi^+\hfill &  A=\cos\phi_{sc} V_{ud}V_{cs}^* a 
    \label{dsf0}\\
(b)\ & D^+\to  f_0\pi^+ \hfill &  B=\left[\frac{\sin\phi_{sc}}{\sqrt{2}} 
            V_{ud}V_{cd}^* (1+\epsilon) +\cos\phi_{sc}  V_{us}V_{cs}^* 
                \epsilon \right]\hspace{-0.1cm}a \label{df0}\\
(c)\ & D^+\to K^*_0(1430)\pi^+ \ &
            C= V_{ud}V_{cs}^*  (1+\epsilon)a .
  \label{damp}
\end{eqnarray}
Process ($a$) is given by the amplitude with Cabibbo-favoured decay 
 $c\to s W^+$, $W^+\to u\bar d \to \pi^+$ followed by the spectator
interaction $s\bar s\to f_0$ with strength $a$.
In the Cabibbo-suppressed process ($b$) there is the same direct
$\pi$-emission diagram as in ($a$) but 
one  has to add the ``colour suppressed'' amplitudes with decay
$c\to d\ W^+$, $W^+ \to u\bar d$, $d\bar d\to f_0$ and 
$c\to s\ W^+$, $W^+ \to u\bar s$, $s\bar s\to f_0$ both of relative size
$\epsilon$. Finally in ($c$) there is one direct and one colour suppressed
amplitude. 
In the approximation $V_{ud}=V_{cs}=\cos\vartheta_c$,
$V_{us}=-V_{cd}=\sin\vartheta_c$, with Cabibbo angle $\vartheta_c$ 
\begin{eqnarray}
R_{D1}& = & 2 \frac{q_a}{q_b} \cot^2\vartheta_c\cot^2\phi_{sc}
     \frac{1}{|1-(\sqrt{2}\cot\phi_{sc} -1)\epsilon |^2}\\
R_{D2}& = &  \frac{q_a}{q_c} \cos^2\phi_{sc}
     \frac{1}{|1+\epsilon |^2}
\label{r4r5}
\end{eqnarray}
where $q_i$ is the phase space for $S$ wave decays with pion momenta 
in the rest frame of processes $(a-c)$. 
Using the branching fractions established by the E791 Collaboration
\cite{Aitala:2000xu}
the following results have been found \cite{Minkowski:2002nd}
\begin{eqnarray}
\cot^2\phi_{sc} /|1-(\sqrt{2}\cot\phi_{sc} -1)\epsilon |^2 & = &
    1.26\, (1.0\pm0.4)  \label{Dmix1}\\
\cos^2 \phi_{sc} / |1+\epsilon |^2  & = & 0.52\, (1.0\pm0.3)
\label{Dmix2}
\end{eqnarray}
with the two solutions
\begin{eqnarray}
 \cot \phi_{sc1} & =  1.11^{+0.33}_{-0.20}, 
     & \phi_{sc1} = (42.0^{+5.8}_{-7.3})^\circ,\, \quad\  
     \epsilon_1  =  (2.85\pm5.35)\, 10^{-2}\\
\tan \phi_{sc2}  &= -0.34^{+0.46}_{-0.37}, \quad
     & \phi_{sc2} = 161.2^{+25.7}_{-16.5})^\circ,\quad
       \epsilon_2 =  0.31^{+0.004}_{-0.13}
\label{phieps}
\end{eqnarray}
It is satisfactory that the colour suppressed amplitudes are indeed found
small. 

\begin{table}[h]
\caption{\label{tab:summaryf0980} Quark structure of $f_0(980)$ from four
observables, mixing angle as in \eref{mixingf0}.}
\begin{indented}
\item[]\begin{tabular}{llllll}
\br
observable & exp. result & sol. $\phi_{sc1}$ & sol. $\phi_{sc2}$ & 
    $\eta'$-like as in \eref{mixingscalars}  \\
\mr
${g_K^2}/{g_\pi^2}$ &$4.0\pm 0.3$ &$(29.9\pm1.3)^\circ$ &
$[(162.4\pm0.5)^\circ]$ & 3\\
& $g_K/g_\pi>0$ &\quad $0^\circ < \phi_{sc}<90^\circ$ & &\\
$R^{\phi/\omega}_{J/\psi}$ & $2.29\pm1.04$ & 
          $(33\pm7)^\circ$ &  $(147\pm7)^\circ$ & 2\\
$R_\gaga^{f_0/a_0}$ 
& $1.16\pm 0.41$ &$(23.7\pm 8.1)^\circ$ &  $(125.8\pm8.1)^\circ$ &
$\frac{49}{27}=1.815$\\
$R_{D1}, R_{D2}$ & \eref{Dmix1},\eref{Dmix2} & 
    $ (42.0^{+5.8}_{-7.3})^\circ$ &$ 161.2^{+25.7}_{-16.5})^\circ $& \\
\br
\end{tabular}
\end{indented}
\end{table}
\subsection{Final result on mixing angle}
Our results are summarized in table \ref{tab:summaryf0980}.
There is a good agreement between the four determinations 
for mixing angle
solutions with $\phi_{sc1}<90^\circ$, contrary to $\phi_{sc2}>90^\circ$. 
This can be taken as an independent
confirmation of the relative sign of $g_K/g_\pi$ in \eref{couplingsf0980}.
Finally, we obtain as average $\phi_{sc}=(30.2\pm 1.2)^\circ$ which is
dominated by the observable $g_K^2/g_\pi^2$. In view of systematic
uncertainties, such as strange quark suppression effects, we estimate the
total error to be around 10\%, therefore finally
\begin{equation}
f_0(980):\qquad \phi_{sc}=(30\pm 3)^\circ.  \label{alphafinal}
\end{equation} 
The mixing angle for $f_0(980)$ is then near the mixing angle
\eref{mixingetapr}
for $\eta'$, i.e. $\phi_{sc}\approx \phi_{ps}=42^\circ$ as in 
route 2, or approximately 
with $\phi_{sc}=\arccos\sqrt{2/3}=35.3^\circ$ for
$|f_0(980)\ran \to (1,1,2)/\sqrt{6}$ as in \eref{mixingscalars}
(observables in table \ref{tab:summaryf0980} ``$\eta'$-like'') and so
$f_0(980)$ is close to a flavour singlet state; octet would have
$\phi_{sc}>90^\circ$.
One of the two solutions found in \cite{Anisovich:2000wb}, $23^\circ \leq
\phi_{sc} \leq 37^\circ$, is consistent with our result \eref{alphafinal}.

\subsection{Possible gluonic contribution}
While the data are nicely consistent with $f_0(980)$ being a 
pure quark state, we may
estimate a possible gluonic contribution as follows. The fit result 
\eref{alphafinal} which is largely dominated by the result on $g_K^2/g_\pi^2$
corresponds to a value $R_\gaga=1.53$ according to \eref{Rgaga}, so the actually measured value
in \eref{rgaga} is about 75\% of the nominal value. Taking into account a
gluonic component  in an extension of \eref{mixingf0} by
\begin{equation}
|f_0(980)\ran= \cos\beta\ (\sin \phi_{sc} |n\bar n\ran + \cos\phi_{sc} |s\bar
s\ran) + \sin\beta |gg\ran,
\end{equation} 
we can write a properly extended equation for $\Gamma_\gaga$ in
\eref{Rgagadef} where the additional gluonic term is of $O(\alpha_s^2)$.
Neglecting this term here we obtain with the above result 
$$R_\gaga\approx R_\gaga^{q\bar q}\cos^2\beta; \qquad 
\cos^2\beta\approx 0.75\pm 0.25,$$ 
i.e. the fraction of a possible 
gluonic component is about $\frac{1}{4}\pm \frac{1}{4}$. 
Note that the addition of a gluonic component would leave
$R_{J/\psi}^{f_0(980)}$ unchanged for fixed $q\bar q$ fractions but would
modify the other two observables considered.

Although with a considerable uncertainty,
this result suggests that more precise data on the above observables
could further strengthen the picture and provide better estimates
on the amount of glue
inside this meson. Recent measurements of decays $D_s\to 3\pi$ and
$D_s\to K\bar K \pi$
\cite{Aubert:2008ao,delAmoSanchez:2010yp,Mitchell:2009aa} 
have shown the potential of independent determinations of $g^2_K/g^2_\pi$. 
In  particular, BaBar has compared the
$\pi\pi$ and $K\bar K$ mass spectra just around 1 GeV 
and found good agreement in their 
shape \cite{delAmoSanchez:2010yp} up to about 1.1 GeV which should directly
determine this observable.

More precise results on $R_{J/\psi}^{f_0(980)}$ and $\Gamma_\gaga^{f_0(980)}$ 
could improve our knowledge of the flavour structure of $f_0(980)$ and are
accessible by ongoing or new experiments. A word of caution on the
interpretation of the $\gamma\gamma$ results is in order. The model
calculations suggested a 
considerable contribution from hadronic re-scattering in the $\sigma\to
\gamma\gamma$ decay (section \ref{twophoton}). On the other hand, 
the corresponding
calculations for $f_2(1270)$ have shown 
the dominance of the ``direct'' contribution as expected
\cite{Mennessier:2010xg}. Results on $f_0(980)$ in between the extremes
indicate 
re-scattering contributions but no strong difference between 
direct and PDG values for $f_0(980)\to \gamma\gamma$;
a strong model dependence on the contributing exchanges can be
observed \cite{Mennessier:2010xg}. Further studies, for example with
dispersive representations, could strengthen the conclusions.
In this presentation we assume the dominance of the direct contribution in \eref{Rgaga}
as for $f_2(1270)$.

\section{Experimental evidence 
for $f_0(1370)$ and $f_0(1500)$ - a reassessment}
\label{f013701500}
A crucial precondition of the mixing scenario in route 1 with a
supernumerous state in the mass region around 1600 MeV is the existence of
$f_0(1370)$. This section is devoted to a critical reassessment of the
evidence for this state 
in various processes and the comparison with $f_0(1500)$ nearby in
mass. In previous surveys supportive \cite{Bugg:2007ja} and sceptical views
\cite{Klempt:2007cp}  have been presented.

\subsection{Production in $p\bar p$ annihilation}
\begin{figure}[t]
\begin{center}
\includegraphics[width=5.5cm]{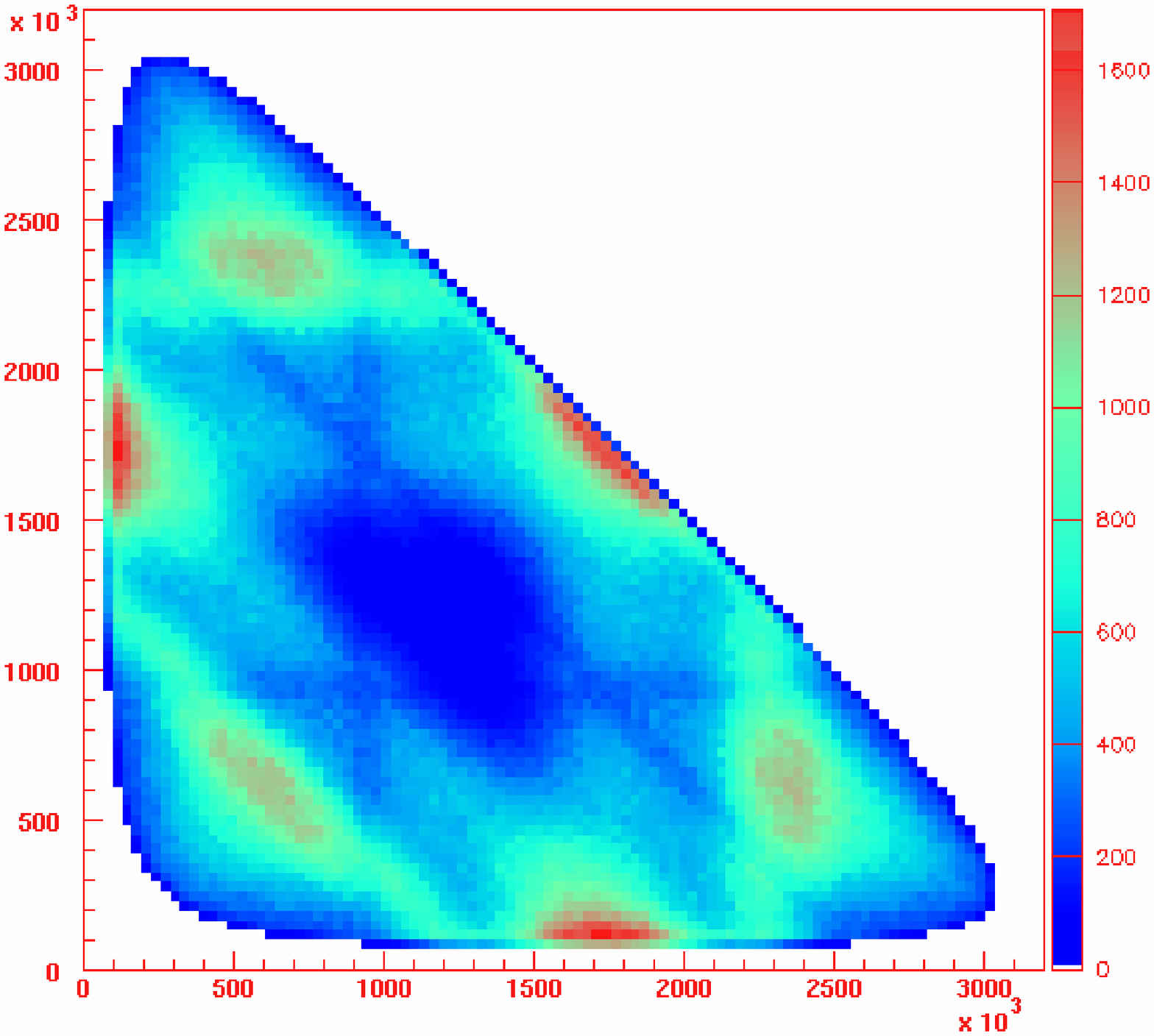}
\end{center}
\hbox{  }
\includegraphics[width=6.0cm]{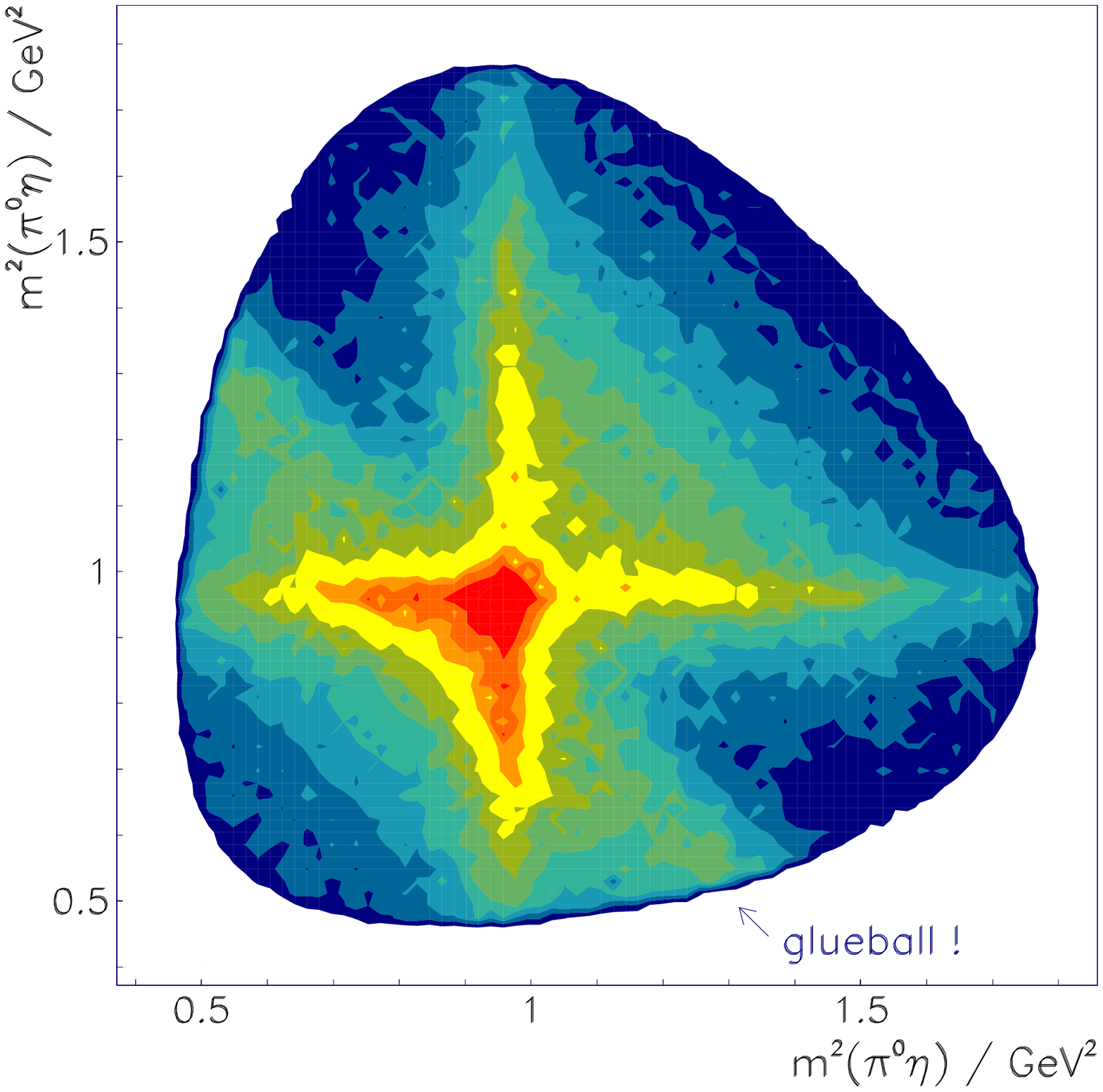}\quad
\includegraphics[width=8cm,bbllx=0.0cm,bblly=1.3cm,bburx=13.0cm,%
bbury=5.0cm]{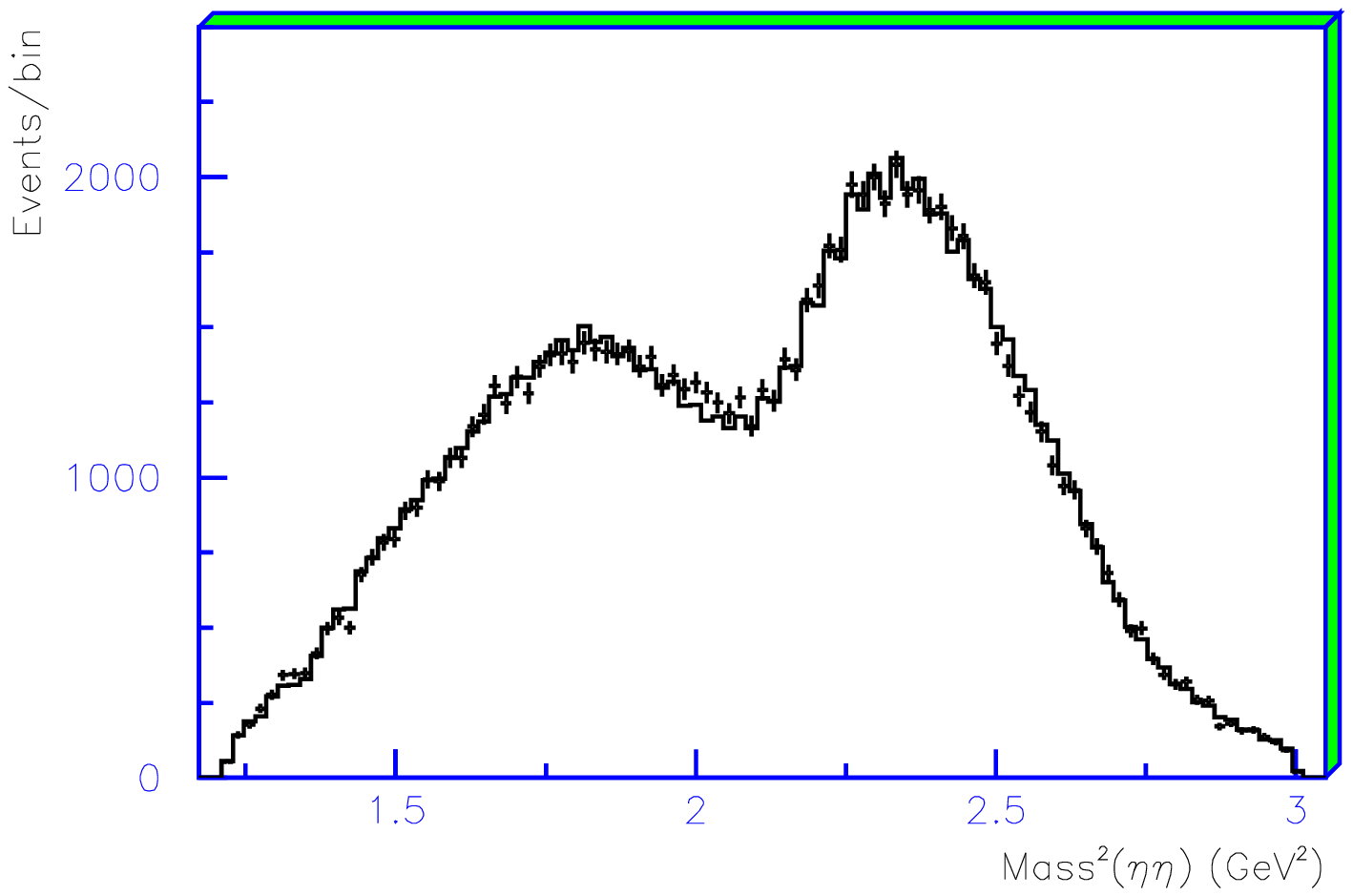}
\caption{\label{fig:dalitz3pi0} Upper panel: 
Dalitz plots for $p\bar p\to 3\pi^0$ annihilation at rest; Bottom: 
Dalitz plot for $p\bar p\to \pi^0\eta\eta$  together
with Mass$^2(\eta\eta)$ spectrum from Crystal Barrel Collaboration 
\cite{Amsler:1995bf} (figures from \cite{Klempt:2007cp}).}
\end{figure}

The state $f_0(1370)$ has been discovered in $p\bar p$ annihilation at rest 
by the
Crystal Barrel Collaboration (CBAR) and has been studied together with
$f_0(1500)$ in the reactions
\cite{Amsler:1995bf,Amsler:1995gf,Anisovich:1994bi}
\begin{equation}
a)\ p\bar p\to \pi^0\pi^0\pi^0,\quad b)\ p\bar p\to \pi^0\eta\eta,\quad
c)\ p\bar p\to \pi^0\pi^0\eta. \label{pbarpreactions}
\end{equation}
The analysis is based on the Dalitz plots shown
in figure \ref{fig:dalitz3pi0}. 
The $3\pi^0$ Dalitz plot has sixfold symmetry and is based on 
$\sim$ 700000 events. The coloured version of the top plot shows events in
bands near the corners at
$m^2(\pi^0\pi^0)\approx 2.25\ (=1.5^2)$ GeV$^2$
which corresponds to the isotropic decay of the $S$ wave resonance
$f_0(1500)$ (at fixed (vertical) mass squared $s_{23}$ the (horizontal)
$s_{12}$ depends linearly
on the decay angle $\cos \theta_{23}$). 
At the mass near $m^2=1.370^2=1.9$ GeV$^2$ there is no band of 
comparable strength, the two bumps at the edges of the plot correspond to
$f_2(1270)$ with its peaked angular distribution according to spin 2. 
The evidence for the second
resonance $f_0(1370)$ is based on the results of a multi-resonance $K$ matrix fit.

A more direct evidence is suggested from reaction (\ref{pbarpreactions}b).
In the $\pi^0\eta\eta$ Dalitz plot (Bottom left of 
figure~\ref{fig:dalitz3pi0}) one observes the horizontal and vertical
bands from $a_0(980)$, in the diagonal along the arrow ``glueball'' there is
again a band with flat density related to $f_0(1500)\to \eta\eta$. 
In parallel, a second band appears to the right (at lower $\eta\eta$ mass), but
the band does not continue to the edges of the plot as the $f_0(1500)$ band
does. The projection in $m^2(\eta\eta)$ (Bottom right of
figure~\ref{fig:dalitz3pi0}) shows 
peaks at higher mass with contributions from $f_0(1500)$ 
and at the lower mass near $m\sim \sqrt{1.7}=1.3$ GeV, and this peak has
been related to
$f_0(1370)$. The overlap with the $a_0(980)$ bands makes it difficult to
directly judge the relevance of these peaks and one relies on the fitting
procedure. 

\begin{figure}[t]
\includegraphics*[width=15.0cm]{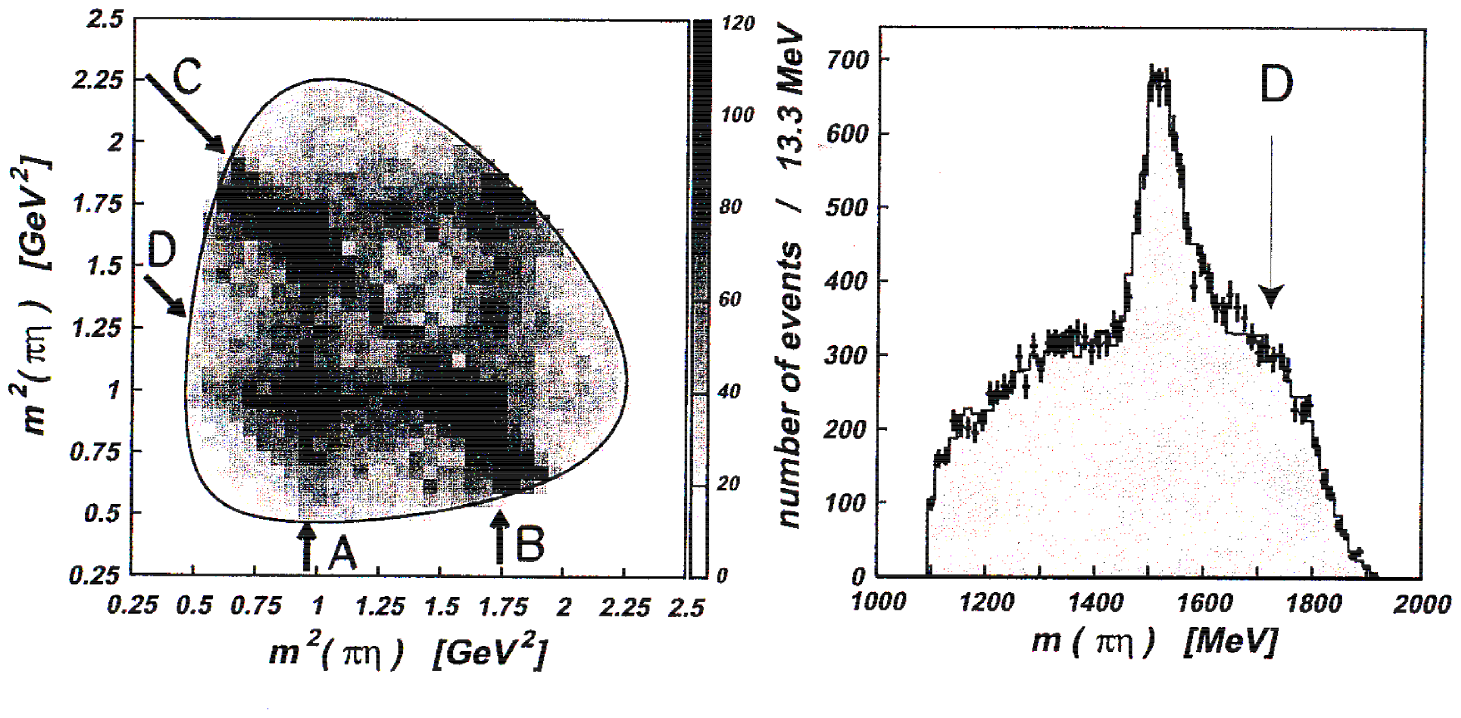} 
\begin{center}
\includegraphics*[width=10.cm]{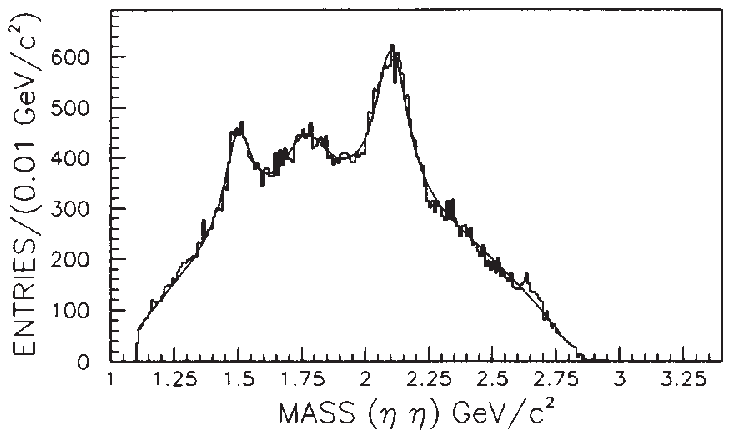}
\end{center}
\caption{\label{fig:dalitzpietet1} Upper part left: Dalitz plot for 
$p\bar p\to \pi^0\eta\eta$ at $\sqrt{s}=2.0$ GeV (Crystal Barrel);
right: mass$^2(\eta\eta)$ spectrum (figures from \cite{Amsler:2002qq}). 
Bottom:
$\eta\eta$ mass spectrum in $p\bar p \to \pi^0\eta\eta$ at $\sqrt{s}=3.0$
GeV, figure from \cite{Armstrong:1993fh} (Experiment E-760).}
\end{figure}

A clarification can be expected from a change of primary energy which would
shift the diagonal bands at fixed $\eta\eta$ mass away from the crossing $\eta\pi^0$
bands. Such a measurement has been performed with 900 MeV antiprotons
($\sqrt{s_{p\bar p}}=2.05$ GeV) again by CBAR experiment 
\cite{Amsler:2002qq}. The $\eta\eta\pi^0$ Dalitz plot with two-fold symmetry 
is shown in
figure \ref{fig:dalitzpietet1}. Indeed, the diagonal band (C) which
corresponds to $f_0(1500)$ is now shifted away from the crossing $a_0(980)$
bands (A). The new bands (B) are related to $a_2(1320)$. A second diagonal 
band, between $f_0(1500)$ and the crossing point of the $a_2(1320)$ bands 
would be expected for $f_0(1370)$, but there is no indication of any second
band.
Also the peak in the $\eta\eta$ mass spectrum has changed into a shoulder
easily related to the resonances in the $\eta \pi^0$ channel.
It is difficult to invent a mechanism which would make disappear a strong
$f_0(1370)$ signal visible at  $\sqrt{s_{p\bar p}}=2m_p$ by increasing
the cms energy by only 0.17 GeV.

At yet higher energy the process $p\bar p\to \pi^0\eta\eta$ 
has been measured by Experiment E760
at Fermilab \cite{Armstrong:1993fh}. At $\sqrt{s_{p\bar p}}=3$ GeV  the mass spectrum 
in figure \ref{fig:dalitzpietet1} (lower part) again shows a clear peak
at 1500 MeV suggestive of $f_0(1500)$ (see also discussion in 
 \cite{Klempt:2007cp}), but the shoulder at low mass has now
disappeared entirely and there is no indication of $f_0(1370)$. More recently, 
Experiment E835 has
analyzed data on the same reaction at $\sqrt{s_{p\bar p}}=3.4$ GeV and
determined the spins of the observed resonances \cite{Uman:2006xb}. 
As a result they identified
their lowest mass peak with $f_0(1500)$ and their second peak as $f_0(1710)$.

At all energies in these reactions $f_0(1500)$ shows up clearly 
as band in the Dalitz plot
and sometimes as peak in mass spectra. On the other hand, no such ``direct''
evidence is visible for $f_0(1370)$. The effect 
appears only as a result of a global resonance fit 
at the lowest primary energy where there is a large overlap with other
resonance phenomena. 


\subsection{Search in phase shift analysis of $\pi\pi$ scattering}
A method which does not rely on parametric global fits is the determination
of the complex scattering amplitude in steps of increasing energy 
directly from the data. Any resonance
decaying into the $\pi\pi$ final state has to appear also in the elastic
$\pi\pi$ scattering. The scattering amplitude
$T_\ell(s)=xm_0\Gamma(s)/(s_0-s-im_0\Gamma(s))$ with a Breit-Wigner 
resonance pole at $s_0-im_0\Gamma$ 
moves along a circle for increasing (real) mass $m=\sqrt{s}$ around $(0,
ix/2)$ with radius $x/2$
in the complex plane (``Argand
diagram''). This circular behaviour is smoothly
modified by background effects 
and it is a characteristic signature of resonant behaviour to look for.
The aim of phase shift analysis is the determination of 
the complex scattering amplitude from the scattering data. 

\begin{figure}[h]
\begin{center}
\includegraphics*[angle=90,width=3.0cm,bbllx=9.5cm,bblly=11.5cm,%
bburx=15.5cm,bbury=19.5cm]{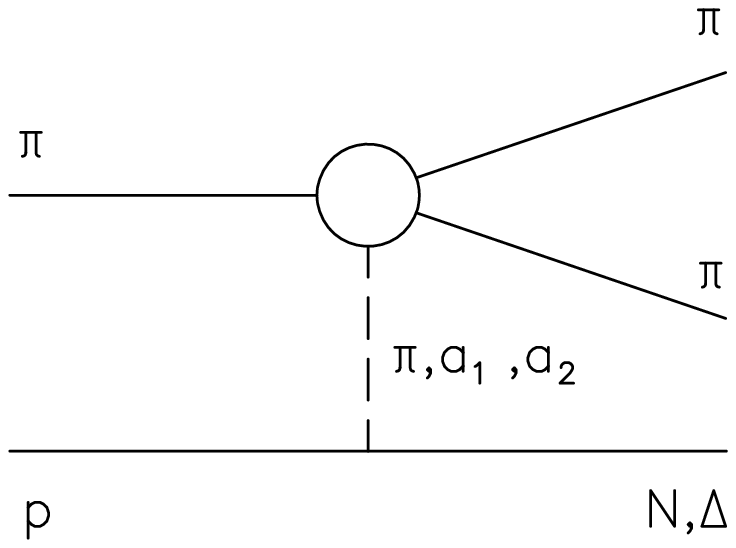}
\caption{\label{fig:pionexchange} One-Pion-Exchange: $\pi\pi$ scattering contributes to $\pi
p\to \pi\pi N (\Delta)$ reaction.}
\end{center}
\end{figure}

\subsubsection{Methods to reconstruct $\pi\pi$ scattering amplitudes.}
In this subsection
we will recall the main assumptions underlying the reconstruction 
of the $\pi\pi$ partial wave amplitudes
$T_\ell$ from measurements of the reaction 
\begin{equation}
\pi^- p \to \pi^- \pi^+ n\quad 
\label{pipin}
\end{equation}
(and others) using the 
One-Pion-Exchange model (OPE) \cite{Chew:1958wd} (see figure
\ref{fig:pionexchange}). 
The incoming pion scatters  off a pion from the pion cloud surrounding the
target proton at small momentum transfers
$t=(p_{in}^p-p_{out}^N)^2$. The theoretical description of the
production process is based on Regge theory and the first aim is to isolate
the one-pion exchange contribution from other exchange contributions 
(for a review, see \cite{MartinBR1976}). The helicity amplitudes for process
\eref{pipin} in the ``t-channel'' (Gottfried Jackson) frame (quantization
axis along exchanged pion) obtain contributions
\begin{equation}
f^\ell_{\lambda\lambda'\mu} = N\frac{\sqrt{2\ell+1}}{\sqrt{q}}\ T_\ell
\ \delta_{\lambda,-\lambda'}\ (m_{\pi\pi}\frac{\sqrt{-t}}{m_\pi^2-t}
\delta _{\mu0}  +
\frac{c_A}{2}\sqrt{\ell(\ell+1)}\ \delta_{n,0})+\ldots
\label{OPE}
\end{equation}
with the $\pi\pi\to\pi\pi$ partial wave amplitude
$T_\ell=\sin\delta_\ell e^{i\delta_\ell}$ as a factor; $N$ is a known
normalization, $m_{\pi\pi}$ is the $\pi\pi$ mass and
$q=\sqrt{m_{\pi\pi}^2/4-m_{\pi}^2}$ . The first term with
the pole at $t=m_\pi^2$ corresponds to the OPE Born-term with the pion pair produced
with helicity $\mu=0$. The nucleon helicity
flip amplitudes with $\lambda=-\lambda'$ dominate over non-flip amplitudes 
at high energies and vanish like $\sqrt{-t}$. Simple models suggest corrections from absorption at small
impact parameters of the collision for net helicity flip
$n=|\lambda-\lambda'+\mu|=0$, such as the second term in \eref{OPE}
\cite{Ochs:1972mc,Estabrooks:1972tw,Ochs:1973th,Estabrooks:1974vu,Grayer:1974cr}.
Furthermore, contributions from $a_2$ exchange are found important at larger $t$
while $a_1$-exchange had been neglected at the time. The determination of
the $\pi\pi$ amplitudes $T_\ell$ can be based on \eref{OPE} (see, for
example \cite{Estabrooks:1974vu}).

Alternatively, one may use $t$-integrated angular distributions
$W(\cos\theta,\phi)$, represented
by the spherical harmonic moments $\langle Y_L^M\rangle$ as function of 
$m_{\pi\pi}$ which are expressed by the density matrix $\rho$
and helicity amplitudes \cite{Grayer:1974cr}
\begin{eqnarray}
\frac{d\sigma}{dm_{\pi\pi}d\Omega}&=& N \sum_J \sum_{M=-J}^{J} 
\langle Y_J^M \rangle \Re Y_J^M(\theta,\phi) \label{YLMdef}\\
\langle Y_J^M \rangle &= & \sum_{\ell,\ell',\mu,\mu'} c
^{J\ell\ell'}_{M\mu\mu'}\Re \rho^{\ell\ell'}_{\mu\mu'}\\
N\rho^{\ell\ell'}_{\mu\mu'}&=&\frac{1}{2}\sum_{\lambda\lambda'}
f^{\ell}_{\lambda\lambda'\mu}f^{\ell' *}_{\lambda\lambda'\mu'}
\end{eqnarray}
It is useful to introduce amplitudes and the density matrix for asymptotic
natural $(+)$ and unnatural exchange $(-)$ which do not interfere
in the observables considered here  using the parity relations 
$f^\ell_{\lambda\lambda'\mu}=(-)^{\lambda+\lambda'+\mu}
f^\ell_{-\lambda-\lambda'-\mu}$ 
\begin{eqnarray}
g^{\ell (\pm)}_{\lambda\lambda'\mu}
  &=&\frac{1}{2}(f^\ell_{\lambda\lambda'\mu}\mp (-)^m
   f^\ell_{\lambda\lambda'-\mu})\\
\rho^{\ell\ell'(\pm)}_{\mu\mu'}&=&\frac{1}{2}(\rho^{\ell\ell'}_{\mu\mu'}
  \mp (-)^{m'}\rho^{\ell\ell'(\pm)}_{\mu-\mu'})
\end{eqnarray}
In general, there are more density matrix elements than measured moments. 
Motivated by
the models for OPE with absorption and amplitudes as in \eref{OPE} some
simplified assumptions for the structure of density matrix elements have
been proposed \cite{Ochs:1972mc}:

(i) ``Spin coherence:'' There are only nucleon flip amplitudes. In case
of experiments with unpolarized target the same consequences on observables
follow already from the
proportionality between flip and non-flip amplitudes 
\begin{equation}
g^{\ell (\pm)}_{++\mu}=\alpha^{(\pm)}g^{\ell (\pm)}_{+-\mu}.
\label{spincoherence}
\end{equation}
This relation also
implies for the density matrices Rank $(\rho^{(\pm)})\leq 1$
while Rank $(\rho^{(\pm)})\leq 2$ would be allowed for in general. Then one
can also write
\begin{equation}
\rho^{\ell\ell'(\pm)}_{\mu\mu'}=\tilde g^{\ell (\pm)}_{\mu}
\tilde g^{\ell (\pm)*}_{\mu'} \label{densitymatrix}
\end{equation}
with $\tilde g^{\ell (\pm)}_{\mu}=\sqrt{1+|\alpha^{(\pm)}|^2}g^{\ell
 (\pm)}_{+-\mu}$.
At this point the sum over nucleon helicities has disappeared and the
relation between the observable moments and the $\pi\pi$ production amplitudes 
is very similar to on shell $\pi\pi$ scattering and one uses for the
amplitudes $ \tilde g^{\ell (\pm)}_{\mu}$ the spectroscopic notation
$L_\mu^{(\pm)}=S^{(-)}_0,P^{(-)}_0,P^{(\pm)}_1\ldots$ 

(ii) ``Phase coherence:'' A second simplification suggested 
by absorbed OPE \eref{OPE} is the appearance of the same phase $\delta_\ell$ 
for all helicity amplitudes with the same $\pi\pi$ spin~$\ell$. 

These two assumptions are motivated only for small $t$; they could be violated
by the presence of other exchanges like $a_1$ or $a_2$ exchanges. Both
assumptions together actually lead to constraints between the observable
moments $\langle Y_L^M \rangle$ which are rather well satisfied
experimentally at 17 GeV $\pi p$ collisions
\cite{Hyams:1973zf}. Also the reduction in rank of the density matrix 
has been confirmed experimentally near the $\rho$ meson mass peak 
\cite{Grayer:1973rn}.

Alternatively, one can relax the assumption on phase coherence and introduce 
extra phases between the different helicity amplitudes
\cite{Estabrooks:1972tw}. Measurements of relative phases, such as
$\Delta^\ell=\delta_1^{\ell (-)}-\delta_0^{\ell (-)}$, 
can check the quality of the phase coherence 
approximation and the dominance of a single production mechanism for spin
$\ell$.
Further discussions of the two-pseudoscalar system and the ambiguity
problems are given in \cite{Chung:1997qd}.

As an example we give the relations for identical particles with even
angular momentum \cite{Hyams:1973zf,Chung:1997qd}.
\begin{eqnarray}
\sqrt{4\pi} \lan Y_0^0\ran & = |S_0^-|^2+|D_0^-|^2 + |D_1^-|^2 + |D_1^+|^2
   \nonumber\\
\sqrt{4\pi} \lan Y_2^0\ran & = \tfrac{\sqrt{5}}{7} (2 |D_0^-|^2 + |D_1^-|^2
     + |D_1^+|^2) + 2|S_0^-||D_0^-|\cos (\phi_{S_0^-} - \phi_{D_0^-})
  \nonumber\\
\sqrt{4\pi} \lan Y_2^1\ran & = \tfrac{\sqrt{10}}{7} 
|D_1^-||D_0^-|\cos (\phi_{D_1^-} - \phi_{D_0^-})
+ \sqrt{2}|S_0^-||D_1^-|\cos(\phi_{S_0^-} - \phi_{D_1^-})
  \nonumber\\
\sqrt{4\pi} \lan Y_4^0\ran & = \tfrac{6}{7} |D_0^-|^2 -
  \tfrac{4}{7}( |D_1^-|^2 + |D_1^+|^2) \nonumber\\
\sqrt{4\pi} \lan Y_4^1\ran & =\tfrac{2\sqrt{15}}{7}|D_0^-||D_1^-|\cos
 (\phi_{D_0^-} - \phi_{D_1^-}) \label{ylmequations}
\end{eqnarray}
We stress again, that this kind of equations without reference to 
nucleon helicities rely on the assumption of ``spin
coherence'' for unpolarized target experiments,

The validity of the results obtained from production
experiments based on the OPE model has been demonstrated 
recently by the nice agreement 
between $K\pi$ phase
shifts obtained from $K p \to K\pi n$ and from decays $D\to K\pi e
\nu_e$ using the Watson theorem (see subsection \ref{sectionkappa} and
figure \ref{fig:elastic}).

\subsubsection{Results on $\pi^+\pi^- \to \pi^+\pi^-$.}
A measurement of $\pi^-p \to \pi^-\pi^+n$ with unpolarized target 
has been carried out by the CERN-Munich 
Collaboration \cite{Grayer:1974cr}. They measured the moments $\langle
Y_L^M\rangle$ in the range $0.6 < m_{\pi\pi}<1.8$ GeV from which the partial
waves with $\pi\pi$ spin $\ell\leq 3$ have been determined. The overall phase
of the amplitude cannot be determined directly but it is inferred from the
Breit-Wigner phase of the leading resonances $\rho(770)$, $f_2(1270)$ and 
$\rho_3(1690)$. In general, the moments with $L\leq L_{max}$ determine the partial wave 
amplitudes only up to discrete ambiguities. Writing the amplitude 
 $T(s,z)=f(s)\Pi_{i=1}^{L_{max}/2}(z-z_i)$ as function of $z=\cos \theta$
one notes that the measurement of $|T(s,z)|^2$ at fixed $s=m_{\pi\pi}^2$ 
does not determine the sign of the imaginary part of the zeros $z_i$
(``Barrelet zero'' \cite{Barrelet:1971pw}) and so one obtains $2^{L_{max}/2}$
different solutions. Some can be removed by unitarity constraints. 
Subsequently, an experiment with polarized target has been carried out
by the CERN-Krakow-Munich Collaboration \cite{Becker:1978kt,Becker:1978ks}.
The
following results have been presented on $\pi\pi$ scattering 
amplitudes $T_\ell$ above
1 GeV which are of interest for our discussion of $f_0(1370)$.

(i) The first analysis by Hyams et al. \cite{Hyams:1973zf} determined the moments averaged over
$|t|<0.15$ GeV (CM-I). These moments have been fitted by 
amplitudes $g^{\ell (\pm)}_\mu$ $(\mu=0,1)$ as in
\eref{YLMdef}-\eref{densitymatrix} parameterizing the $\pi\pi$ amplitudes as
function of $m_{\pi\pi}$ by a $K$ matrix. In addition,
local
deviations from this solution have been studied. The $S$ wave featured
the narrow $f_0(980)$ interfering with a broad background passing $90^\circ$
near $m_{\pi\pi}\approx 1$ GeV. 
A classification in terms of amplitude zeros has been given but no study of
alternative solutions has been attempted.

(ii) The first full study of ambiguities from CM-I data 
has been carried out by Estabrooks and Martin
\cite{Estabrooks:1974qd,Estabrooks:1975cy}. 
They obtained four different phase shift solutions A-D classified according
to the Barrelet zeros. A $K$ matrix fit to the S waves of the four solutions
 and taking into account $K\bar K$ data as well, 
yielded three scalar resonances $f_0(800),\ f_0(1005)$ and $f_0(1540)$
with width around 1000, 8 and 200 MeV resp. \cite{Estabrooks:1978de}.
A subsequent analysis in which the phase of the amplitude has been
determined using the analyticity constraints of dispersion relations 
produced modified solutions $\alpha,\beta\ (\beta')$ starting from A,B
\cite{Martin:1977ff}. Because of earlier $\pi^0\pi^0$ data the solution C
as starting solution has been rejected at the time but has been favoured later.
 
(iii) A second analysis 
by the CERN-Munich group \cite{Hyams:1975mc} (CM-II)
included in the fits to the primary data a complete error correlation matrix 
using larger mass bins 
and obtained in
that way $\pi\pi$ amplitudes more strongly constrained 
than those of \cite{Hyams:1973zf}. Also
in these fits some $t$ dependence is parameterized. The ambiguities are
analyzed and again 4 solutions are identified. The smaller error bars of 
zeros Im $z_i$ do not allow the crossing of Im $z_1$ at 1.5 GeV as in the
previous analysis (i) in \cite{Hyams:1973zf}; therefore these earlier results
should not be trusted above $\sim 1.4$ GeV.

(iv) A further experiment with transversely 
polarized target has been performed by 
the CERN-Krakow-Munich Collaboration \cite{Becker:1978kt,Becker:1978ks}
with the aim to check the assumptions underlying the previous experiments.
Large polarization effects have been found which prove the existence of
sizable nucleon non-flip amplitudes of $a_1$-exchange type, so far neglected. 
However, the more general
property of ``spin coherence'' is still approximately valid, i.e. amplitudes
with nucleon spin flip and non-flip are proportional for different 
di-meson spin orientations $\ell,\mu$ to a good approximation in the
average over $t$ as in \eref{spincoherence}, so that the previous analyses 
can be maintained (see also discussion
in \cite{Minkowski:1998mf}). In \cite{Kaminski:2006yv} it has been shown
that the $\pi\pi$ $S$ wave phases in the region 1000-1400 MeV agree quite
well from the polarized target experiment \cite{Kaminski:1996da},
from analysis (i) \cite{Hyams:1973zf} and solutions $(---)$ or $(-+-)$ from
\cite{Hyams:1975mc}, corresponding to solutions A,C in \cite{Estabrooks:1975cy}. 
We therefore consider only these two solutions in the following.

\subsubsection{Results on $\pi^+\pi^- \to \pi^0\pi^0$.}
\begin{figure}[t]
\begin{center}
\includegraphics*[width=7cm,bbllx=0cm,bblly=0.5cm,bburx=15.0cm,
bbury=15.0cm]{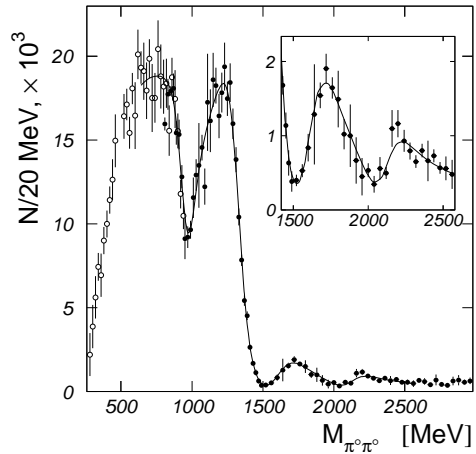}
\end{center} 
\caption{\label{fig:S00mass}  Dependence of $S$ wave
intensity on $\pi^0\pi^0$-mass in $\pi^- p \to \pi^0\pi^0 n$.
 The first two dips reflect the destructive interference 
of a broad
background with the narrow $f_0(980)$ and $f_0(1500)$, 
figure from \cite{Alde:1998mc} (GAMS Collaboration).} 
\end{figure}

Partial wave amplitudes of the charge exchange reaction have been obtained from
the production process $\pi^-p\to \pi^0\pi^0n$ at incident momentum
18.3 GeV by experiment E853 at the Brookhaven National Lab
\cite{Gunter:2000am} and at 100 GeV by the GAMS Collaboration at CERN
\cite{Alde:1998mc}. Both experiments measured the moments integrated over
small $|t|$ and determined the partial wave amplitudes using methods as
discussed above. The E853 experiment found for $m_{\pi\pi}\lesssim 1.5\ \GeV$ 
one solution with phase coherence ($\phi_{D_0}-\phi_{D_-}\approx 0$) and another 
one with large relative phase; above that mass the errors are rather large. 
Also at larger $-t> 0.4\ \GeV^2$ the phase coherent solutions disappeared. 
GAMS found solutions with phase coherence in
the full mass range studied up to 3 GeV and imposed this condition
subsequently in the phase shift analysis. The mass spectrum of the $S$ wave is shown in
figure \ref{fig:S00mass}. 

\subsubsection{Selection of a unique isoscalar $\pi\pi$ $S$ wave.}
\begin{figure*}[t]
\includegraphics*[angle=-90,width=5cm,bbllx=3cm,bblly=1.5cm,bburx=19.5cm,
bbury=19.5cm]{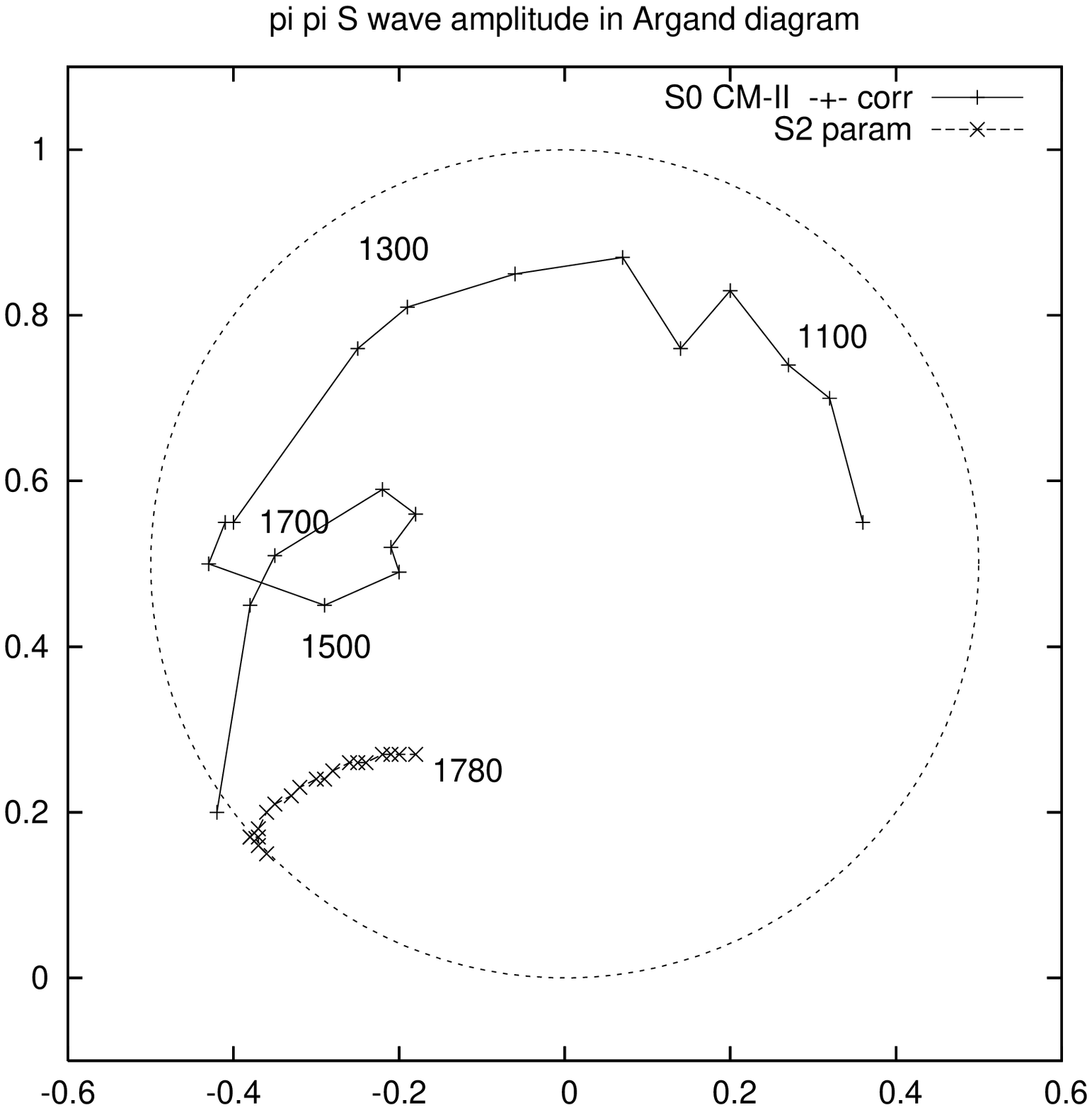}
\includegraphics*[angle=-90,width=5cm,bbllx=3cm,bblly=1.5cm,bburx=19.5cm,%
bbury=19.5cm]{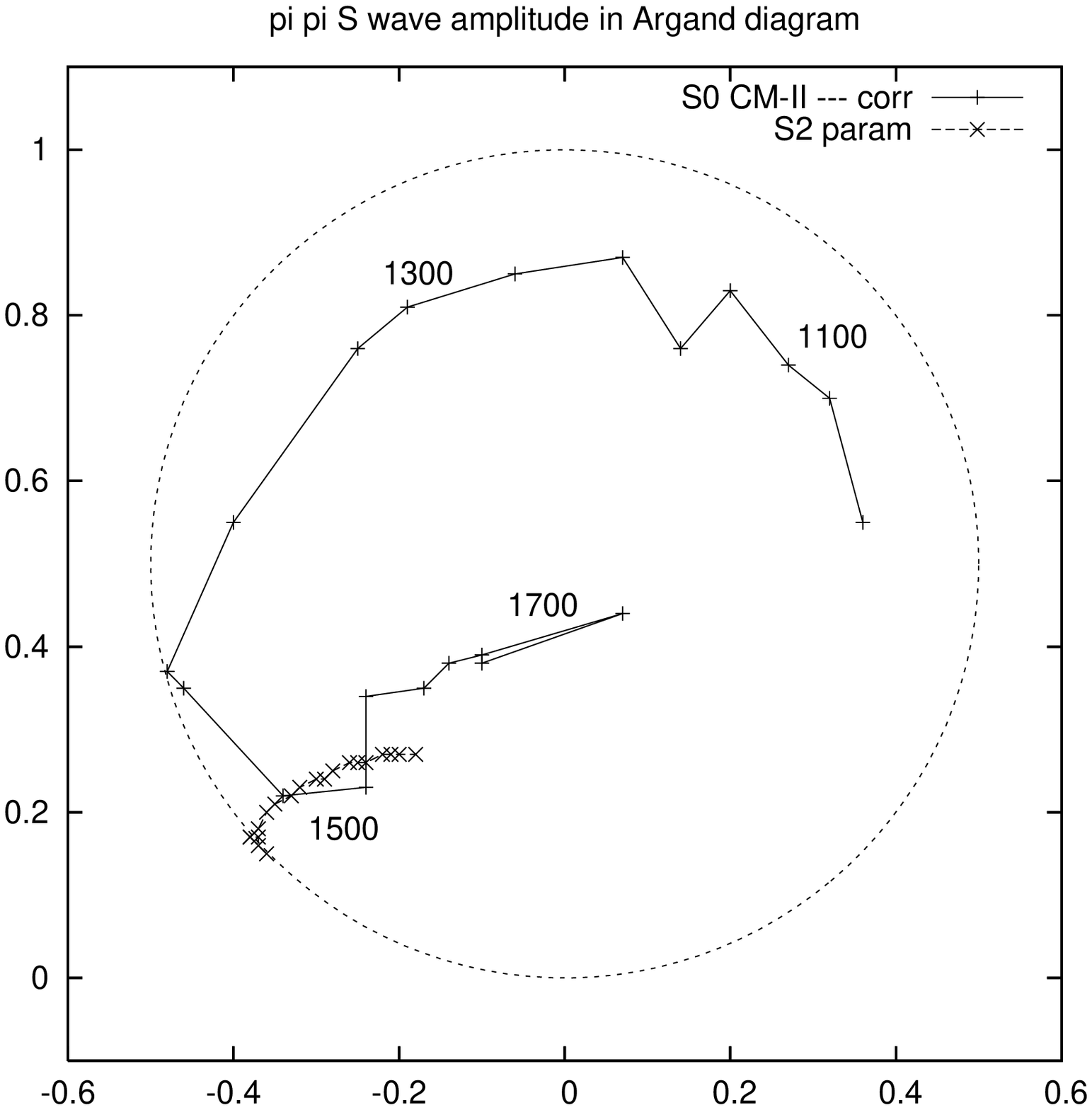}
\includegraphics*[angle=-90,width=5cm,bbllx=3cm,bblly=1.5cm,bburx=19.5cm,%
bbury=19.5cm]{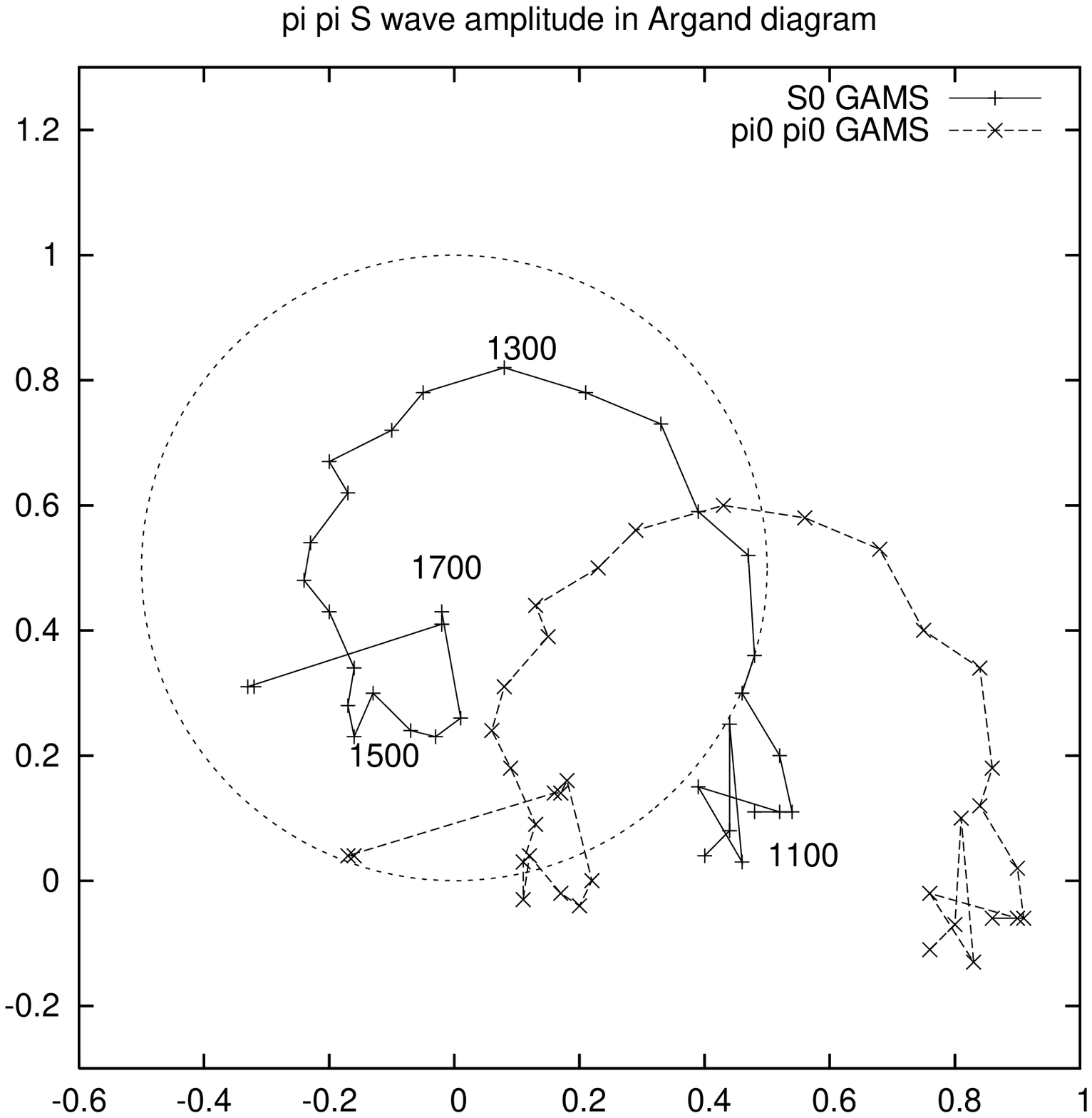} \\
\caption{\label{fig:swave00} Argand diagrams for $\pi\pi$ $S_I$ waves: 
$S_0$ solutions  $(-+-)$ and
$(---)$ from $\pi^+\pi^-\to \pi^+\pi^-$ (CM-II) but using 
new inelastic $S_2$ as shown; right panel: $S_{00}$ amplitude for
 $\pi^+\pi^-\to \pi^0\pi^0$ (GAMS) and component $S_0$ thereof
(figures from \cite{Ochs:2006rb}).}
\end{figure*}

\setcounter{footnote}{1}
A further selection between the two phase shift 
solutions  $(---)$ and $(-+-)$ or A and C, respectively, left over 
from the $\pi^+\pi^-$ analysis
can be obtained by comparing to $\pi^+\pi^-\to \pi^0\pi^0$ with its different ambiguity
structure. Comparing CERN-Munich and GAMS data which have good accuracy
up to $m_{\pi\pi}\leq 1.8$ GeV a unique solution has been established
\cite{Ochs:2006rb,Ochs:2010cv}.\footnote{These analyses are preliminary
concerning the parametric S matrix fits.} A new (inelastic)
$I=2$ amplitude $S_2$ for the mass range $1000 \leq m_{\pi\pi} \leq 1800$ MeV 
has been constructed from data \cite{Durusoy:1973aj,Cohen:1973yx} and 
has been used to obtain updated $I=0$ amplitudes
$S_0$ (see Argand diagrams left 
in figure \ref{fig:swave00} for new $S_0$ and $S_2$). The diagram rightmost 
in figure \ref{fig:swave00} shows 
the $S$ wave amplitude $S_{00}$ for the process $\pi^+\pi^-\to \pi^0\pi^0$
which has been reconstructed from the magnitude $|S_{00}|^2$, in the given
normalization, the phase
$|\phi_S-\phi_{D_0}|$ and the Breit Wigner phase and normalization 
of $f_2(1270)$ in the
$D_0$ wave. 
The $I=0$ amplitude $S_0$
is obtained from $S_{00}=S_0-S_2$
and is shown in the same diagram (normalization $|S_I-\frac{i}{2}|\leq
\frac{1}{2}$).

The amplitude $S_0$ obtained from $\pinullpinull$ data should agree with 
$S_0$ from $\pipluspiminus$. This is only true approximately for the $(-+-)$
solution: there is a circular motion from 1000 to 1740 MeV with a
smaller circle superimposed above 1400 MeV. Solution $(---)$ would turn
into the center of the circle instead and so it has been rejected. There are
discrepancies between the remaining solutions $S_0$ which have been related 
to systematic uncertainties in the overall phase, the $I=2$ amplitude and
influence from higher partial waves. These uncertainties over large mass
scales should not affect the nature of the local resonance phenomena like
$f_0(1370)$ and $f_0(1500)$.

\begin{figure*}[t]
\includegraphics*[angle=0,width=5cm,bbllx=3cm,bblly=1.5cm,bburx=10.0cm,
bbury=8.2cm]{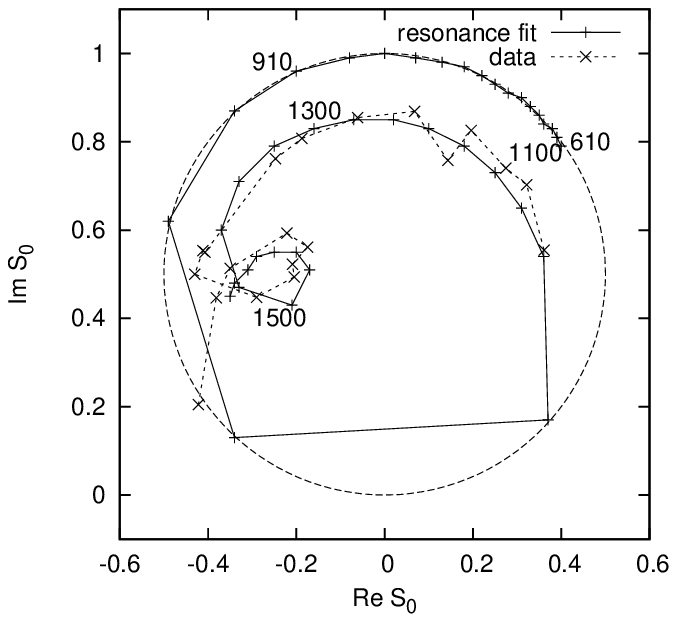} 
\includegraphics*[angle=0,width=5.0cm,bbllx=3cm,bblly=1.5cm,bburx=9.2cm,%
bbury=7.5cm]{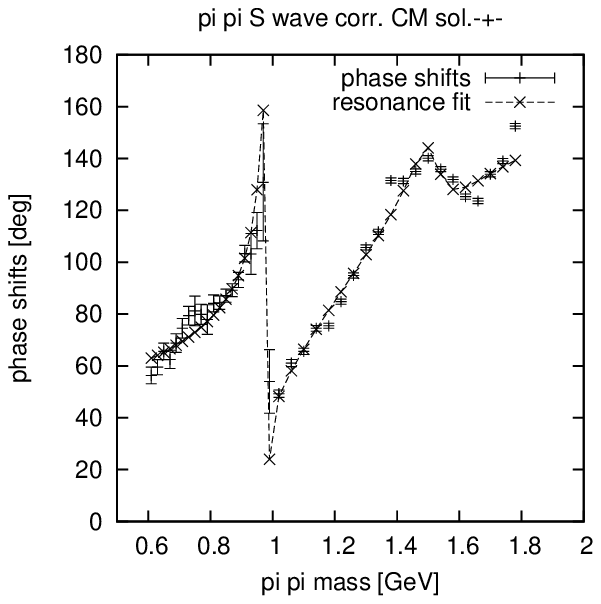} 
\includegraphics*[angle=0,width=5.0cm,bbllx=3cm,bblly=1.5cm,bburx=9.5cm,%
bbury=7.9cm]{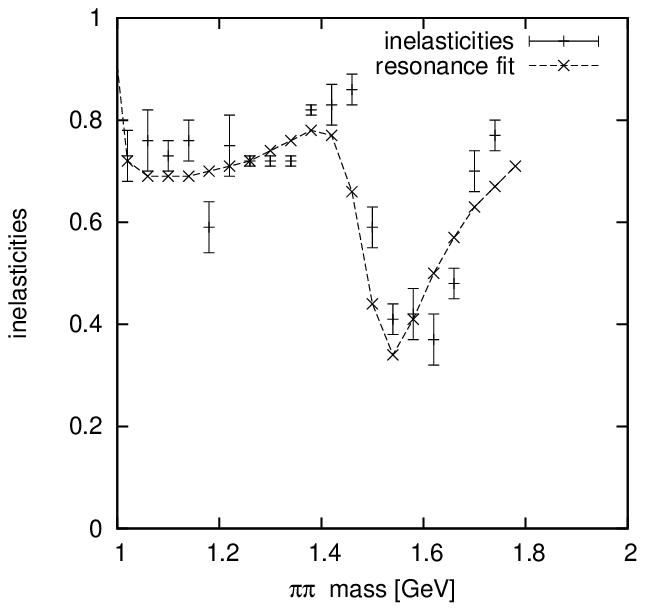} \\
\caption{\label{fig:resonances} Resonance fit \eref{smatrix} 
in comparison with data (CM-I/II):
Argand diagram for corrected $S_0$ wave,
phase shift $\delta_0^0$ and inelasticity $\eta^0_0$ (figures from
\cite{Ochs:2006rb}).}
\end{figure*}

The amount of agreement is nevertheless impressive, in particular concerning
the existence of $f_0(1500)$ which shows up clearly as a small circle
in both analyses. Mass and width are roughly compatible with the PDG values
\cite{beringer2012pdg} (m=1505 MeV, $\Gamma$=109 MeV). The partial width
$x_{\pi\pi}=\Gamma_{\pi\pi}/\Gamma_{tot}$ in the resonant amplitude
$T_0=x_{\pi\pi} m_0\Gamma/(s_0-s-im_0\Gamma)$ can be 
estimated from the depth 
of the $f_0(1500)$ circle in figure \ref{fig:swave00} (CM-II), see
also inelasticity $\eta_0^0$ in figure \ref{fig:resonances} ($\Delta
x=2\Delta \eta_0^0$): 
\begin{equation}
f_0(1500): \qquad  x_{\pi\pi}= 0.25\pm0.05 \ (\text{CM-II}),
    \qquad   x_{\pi\pi}=0.349 \pm 0.023\ (\text{PDG}). \label{xpipi1500}
\end{equation}
Here the first result (CM-II) is determined from Im $T_0$ of the 
resonant elastic partial wave amplitude and the second one (PDG) 
from all inelastic channel cross sections; 
both should agree because of the optical theorem and they roughly do within
30 \%.

In order to represent the selected solution $S_0(-+-)$ 
in terms of a minimal number of 
resonances an ansatz for a unitary 3-resonance $S$ matrix for 
the 3 channels ($\pi\pi,K\bar K,4\pi$) has been fitted to the CM-II data.
In a generalization of \eref{addresonances} 
the $S$ matrix is written 
as product of three unitary matrices for resonances
$S_R=1+2iT_R$
\begin{eqnarray}
S&=&S_{f_0(980)}S_{f_0(1500)}S_{\rm broad}\label{Smatrixfull}\\
T_R&=& [m_0^2-\mpipi^2-i(\rho_1 g_1^2 +\rho_2 g_2^2+ \rho_3 g_3^2)]^{-1}
   \nonumber\\
   & &   \phantom{aaa}\times\rho^{\frac{1}{2}T} (g_ig_j)\rho^\frac{1}{2}
    \label{smatrix}
\end{eqnarray}
where $\rho_i=2k_i/\sqrt{s}$ and $(g_ig_j)$ is the matrix with real constant couplings.
This $S$ matrix behaves locally like 
a superposition of a
narrow inelastic resonance over a largely elastic background as in
\eref{addresonances}, but it does not fulfill the symmetry requirements for
the $S$ matrix exactly, therefore further tests are necessary.
Also, the formulae \eref{Smatrixfull},\eref{smatrix}  do not represent 
the correct behaviour near $\pi\pi$ threshold and cannot be
used for the determination of the pole of $S_{\rm broad}$.  
 
These formulae have been fitted in \cite{Ochs:2006rb} 
to the data from CM-II for $\mpipi>1000$ MeV and the
phases from CM-I in $600<\mpipi<1000$ MeV for definiteness. 
As can be seen
in figure \ref{fig:resonances} the three resonances give a good overall
description of the data. The broad state corresponds to a Breit-Wigner
resonance with mass parameter
$m_0=1100$ MeV and a total width of similar size which we relate to the high
mass tail of $f_0(500)$.

\subsubsection{Search for $f_0(1370)$ in $\pi\pi$
scattering.\label{respipi}}
The fit in figure \ref{fig:resonances} reproduces the main features
related to the rapid phase variations for the two narrow resonances 
at 1000 and 1500 MeV above a
broad background which can also be described as resonant state. In
particular, there is no indication of an additional state around 1370 MeV
which should become noticeable as an extra circle and the related variations 
in phase and inelasticity comparable to $f_0(1500)$. 
If we estimate from  figure
\ref{fig:resonances} that an extra state
with more than 1/3 of elasticity $x_{\pi\pi}$ of $f_0(1500)$ should have been 
noticed,
we arrive at the limit
\begin{equation}
f_0(1370):\qquad  x_{\pi\pi} < 0.1 \quad (\text{CL}=95\%) \quad
 \text{(CM-II data)} \label{xpipi1370}  
\end{equation}
The Argand diagram of solution C in \cite{Estabrooks:1974qd}
shows a similar behaviour. The K matrix fits to those data require
$f_0(1500)$ but no additional resonance near 1300 MeV
\cite{Estabrooks:1978de}.

It should be noted that in figure \ref{fig:resonances} the data points 
represent in each mass interval, together with the results from the other
partial waves an (almost) perfect fit to the original $\lan Y_L^M\ran$ data at this mass.
There is no correlation between data points at different masses.
An additional resonance should therefore become visible in such energy
independent analyses and the results in \eref{smatrix},\eref{xpipi1370}
don't depend on the parameterization chosen for the fit.

The present evidence for $f_0(1370)$ comes almost 
exclusively from fitting energy dependent
parameterizations to the angular moments or Dalitz plots. For example, Bugg
\cite{Bugg:2007ja} has fitted the CM-I moments (and other data) 
to a superposition of 
resonances with a
sizable $f_0(1370)$ contribution in $\pi\pi$ scattering. 
The $\pi\pi$ $S$ wave
amplitude shows an extra
circle between 1000 and 1500 MeV. From the size of this circle 
in comparison to the elastic
circle below 1000 MeV we conclude that the elastic width of $f_0(1370)$ 
in this fit is about $x_{\pi\pi}\sim 0.35$ 
(figure 16 in \cite{Bugg:2007ja}). Such an effect can hardly
be consistent with the $\eta-\delta$ ``data'' in figure \ref{fig:resonances}
which in turn fit the original moments in all mass bins separately. 
Accordingly, although the overall fit to the moments
(figures 14,15 in \cite{Bugg:2007ja}) looks    
reasonable, there are significant 
local deviations between fit and data 
in all moments, also in the region around 1300 MeV.

In the K matrix analysis by Anisovich and Sarantsev
\cite{Anisovich:2002ij} $f_0(1370)$ appears with partial width 
$x_{\pi\pi}\sim 0.19 - 0.26$, depending on the solution (Table 6 in
\cite{Anisovich:2002ij}), and it is
larger or equal to the $\pi\pi$ width of $f_0(1500)$ with
$x_{\pi\pi}\sim 0.19 - 0.23$. Decay widths into $\pi\pi$ of both states with 
equal strength are hard to reconcile with the
energy-independent result \eref{xpipi1370} from CM-II and the amplitude
structure in figure \ref{fig:resonances}.

The GAMS Collaboration has performed a partial wave decomposition
of their $\pi^0\pi^0$ data \cite{Alde:1998mc} which has been used 
to reconstruct the $S$ wave amplitudes (our figures \ref{fig:swave00} and
\ref{fig:resonances} from \cite{Ochs:2006rb}). 
The Collaboration also
determined the contributions of individual resonances 
to the $S$ wave mass spectrum 
$|S|^2$ (figure \ref{fig:S00mass}).  
A model amplitude is fitted to the spectrum which represents 
a superposition of a real
Gaussian-type background and Breit-Wigner resonances with arbitrary
production phases. With the following production cross sections 
$\sigma \times BR$ a good description has been obtained (see curves in figure
\ref{fig:S00mass})
\begin{equation}
 f_0(980)\ofsp:\ 5.4 \pm 1.2\ \nb,\,
 f_0(1300)\ofsp:\ 70 \pm 15\ \nb, \,
 f_0(1500)\ofsp:\ 12\pm 3\ \nb,\,
 f_0(2010)\ofsp:\ 3 \pm 1\ \nb
\label{gamsresonances}
\end{equation}
The fit requires the by far largest cross section for 
$f_0(1300)$ which is not at all visible in the amplitude 
analysis of our figure
\ref{fig:swave00}. This conflict arises using a 
fitting procedure which tries to derive 
a two-dimensional (complex) amplitude
with several interfering components from a 
one-dimensional mass spectrum which ends up with ambiguities. One can
anticipate that the model amplitude for the $S$ wave 
would not fit the relative phases
to the $D$ wave represented by the $f_2(1270)$ resonant amplitude
which enter the amplitude analysis. It should also be noted that
the second dip near 1500 MeV 
in the $S$ wave mass spectrum is generated not only by $f_0(1500)$
but also by the large background from the isotensor $S_2$ (see figure 
\ref{fig:swave00}).

\subsubsection{Inelastic scattering $\pi\pi\to K\bar K, \ \eta\eta$.}
\label{inelasticpipi}
Data on $\pi\pi\to K\bar K$ have been analyzed as well.
The experiments with highest statistics are : 

(i) Argonne experiment on
$\pi^-p\to K^-K^+n$ and $\pi^+n\to K^- K^+ p$ at 6 GeV for 
$m_{K\bar K}<1.55$ GeV by
Cohen et al. \cite{Cohen:1980cq}. Their complete analysis yields 8 different
solutions; they are all eliminated but one. They find one S wave resonance
with $m_0=1425\pm 15$ MeV and width $\Gamma=160\pm 30$ MeV. From a
determination of the zeros of Det $(K^{-1})$ they exclude the existence of
more than one
resonance below 1500 MeV. A peak in the mass spectrum near 1300 MeV
associated with a smooth phase behaviour is
explained by the interference between the background from $f_0(980)$ and the
resonance at 1425 MeV (figure 33 of \cite{Cohen:1980cq}). 
Their Argand diagram suggests a negative
relative coupling between background and resonance, as also concluded in
\cite{Minkowski:1998mf} for the Brookhaven experiment.   

(ii) Brookhaven experiment on $\pi^-p \to
K^0_s K^0_s n$ at 23 GeV and for $m_{K\bar K}<2.4$ GeV 
by Etkin et al. \cite{Etkin:1981sg}. Their best fit includes a resonance at
$m_0=1463 \pm 90$ MeV and width $\Gamma=118\pm ^{+138}_{-16}$ in perfect
agreement with $f_0(1500)$ from PDG. Their global fit with $f_0(980)$ and $f_0(1500)$
(but without $f_0(500)$) 
represents the $S$ wave phase $\delta_0^0$ well, but not the 
peak near 1300 MeV.

\begin{figure*}[t]
\includegraphics*[angle=0,width=14cm,bbllx=1.5cm,bblly=1.3cm,bburx=14.5cm,
bbury=7.0cm]{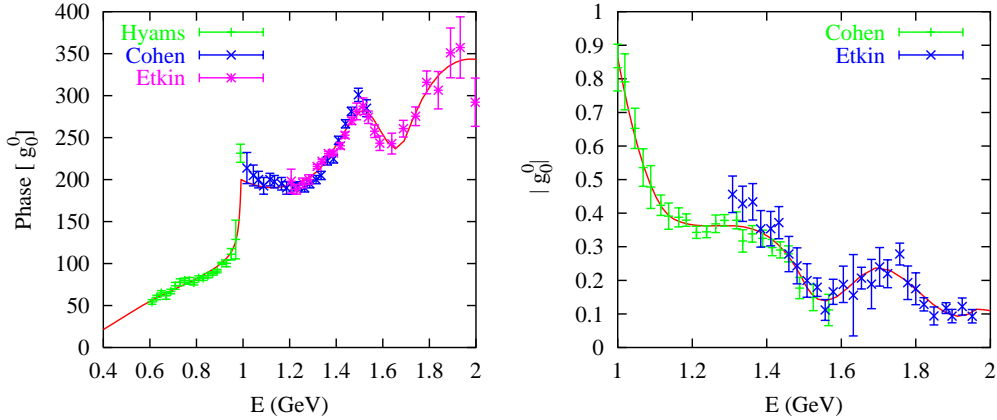} 
\caption{\label{fig:s00kk} Phase and magnitude of $I=0$ $S$ wave amplitude for
$\pi\pi\to K\bar K$ vs. $E_{\pi\pi}$ (data 
\cite{Hyams:1973zf,Cohen:1980cq,Etkin:1981sg}); curves for $E_{\pi\pi}>1$ GeV 
are 
polynomial fits (figure from \cite{Buettiker:2003pp}).}
\end{figure*}

These results are summarized by B\"uttiker et al. \cite{Buettiker:2003pp}
in  figure \ref{fig:s00kk}. The phase shifts at the $K \bar K$ threshold 
join smoothly with those from elastic $\pi\pi$ scattering which 
coincide below threshold because of unitarity 
(see, for example \cite{Johannesson:1974ma}). 
Near the $K \bar K$ threshold the
presence of a $P$ wave - the tail of the $\rho$ meson - in 
 $\pi^+\pi^-\to K^+K^-$ (Cohen et al.) allows the accurate determination of
 the $S$ wave. As there are conflicting results in this region 
a preference is given
for low energies $E<1.3$ GeV to the data by Cohen et al. 
The $\pi\pi$ phase shifts obtained in a dispersion relation approach
\cite{GarciaMartin:2011cn} also prefer the upper branch of phases which
approach $225^\circ$ at $K\bar K$ threshold.

Subsequent descriptions in terms of resonances 
including the broad object $f_0(500)$ yield a 
good qualitative description of $\pi\pi$ and $K\bar K$ data without
an additional resonant state near 1300 MeV: a K matrix fit
\cite{Estabrooks:1978de} 
with three poles and a fit with the ansatz \eref{smatrix} \cite{Ochs:2010cv}.
On the other hand, a contribution from $f_0(1370)$ is suggested in
\cite{Bugg:2006sr} with an effect hardly visible in the phase motion.
Indeed, the phase behaves very smoothly around 1370 MeV in the data (Left
panel of figure \ref{fig:s00kk}), while there is a strong effect from
$f_0(1500)$ similarly to the elastic $\pi\pi$ channel, and, of course, from
$f_0(980)$.  

(iii) Finally, the inelastic process $\pi^+\pi^-\to \eta\eta$ has been measured by
Binon et al. (GAMS Collaboration)\cite{Binon:1983ny} (more recent results in
\cite{Binon:2004yd}). Also in this case, a significant effect
is seen for $f_0(1500)$ with mass and width consistent with PDG. The state 
interferes with the broad background, opposite in
sign to $K\bar K$ \cite{Minkowski:1998mf}. It forms  a circle in the Dalitz
plot also seen in \cite{Bugg:2006sr}. A clearly visible signal from
$f_0(1370)$ can hardly be identified.

\begin{figure}[t]
\includegraphics[angle=0,width=6.2cm]{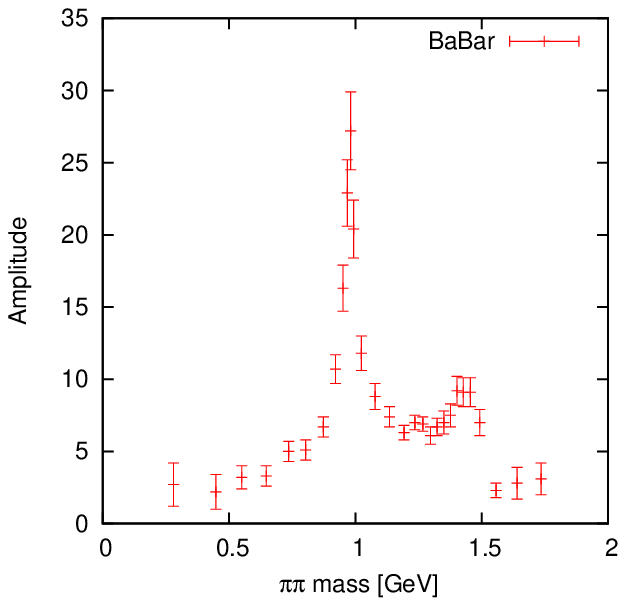} \hspace{-1.7cm}
\includegraphics[angle=0,width=6.2cm]{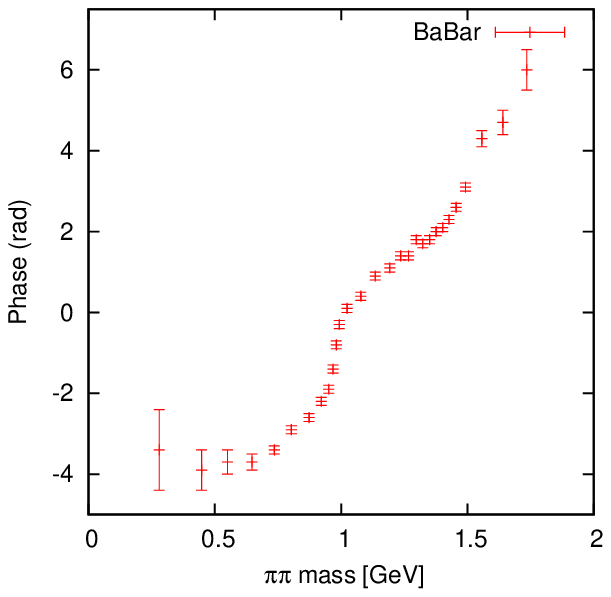} \hspace{-1.7cm}
\includegraphics[angle=0,width=6.2cm]{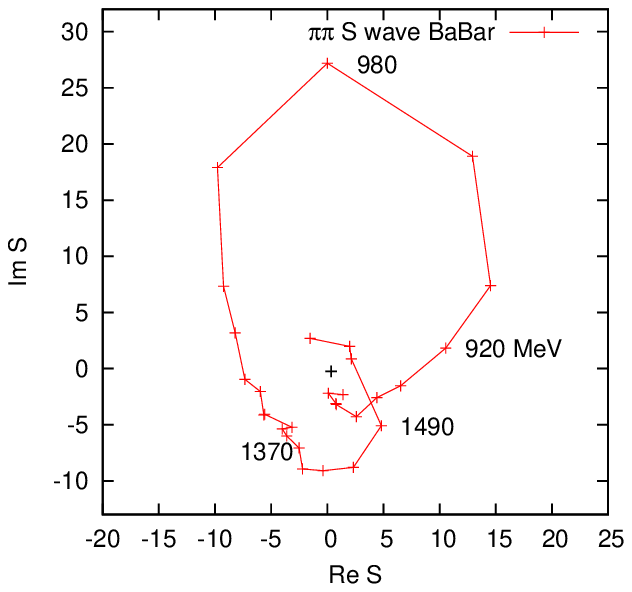}
\caption{\noindent $\pi\pi$ $S$ wave amplitude and phase extracted from decays
$D_s^+\to \pi^+\pi^-\pi^+$ (BaBar Collaboration \cite{Aubert:2008ao});
right panel: Argand diagram for $\pi\pi$ amplitude
(the phase is normalized to $\pi/2$ at the $f_0(980)$ peak). 
The points are not equidistant in mass. 
\label{fig:dsdec}}
\end{figure}   

\subsection{Decays of $D$ and $B$ into $f_0$ mesons}
In the weak decays of heavy quark mesons some well defined $q\bar q$ 
states evolve from the intermediate weak and strong interaction processes 
and they finally can form isoscalar mesons. Then, these processes
may help to clarify their intrinsic structure.
\subsubsection{Decay of $D_s$ mesons.\label{Dsmesons}}
The dominant subprocess for $f_0$ production 
in the decay $D_s^+\to \pi^+\pi^-\pi^+$ is identified as 
\beq
D_s^+\to \pi^+ + s\bar s;\quad s\bar s\to f_0\to \pi^+\pi^-.
\eeq
Earlier studies of this process by E791 \cite{Aitala:2005yh} and FOCUS
\cite{Link:2003gb} Collaborations have shown a strong signal from $f_0(980)$
and some resonant structure associated to $f_0(1370)$ or $f_0(1500)$. 
In a dedicated analysis by Klempt, Matveev and Sarantsev
\cite{Klempt:2008ux} of E791 data no positive evidence 
for $f_0(1370)$
besides $f_0(1500)$ was found, although such a contribution has not been
excluded.    

In a more recent high statistics investigation of the decay $D_s\to 3\pi$ by
the BaBar Collaboration \cite{Aubert:2008ao} 
the known resonances $\rho,f_2$ have been represented by
Breit-Wigner forms while the $S$ wave amplitude $A_S$ 
is parameterized in a model independent way by an
interpolation between 30 points in the complex plane
\footnote{This method has been used first by the Fermilab E791 Collaboration
\cite{Aitala:2005yh}.} 
$$A_S (m_{\pi\pi})= {\rm Interp}(c_k(m_{\pi\pi}) e^{i\phi_k(m_{\pi\pi})})|_{k=1\ldots
30}.$$ 
The resulting amplitudes  and phases are shown in 
figure \ref{fig:dsdec}. One observes clearly the peak from $f_0(980)$
and another peak below 1500 MeV. The phases show two regions of strong
variation with mass, near 1000 MeV and near 1500 MeV (middle panel). 
Near 1370 MeV the phase motion
per mass interval is near minimum, without resonant signal. 

The Argand diagram (right panel) shows a large circle peaking at 980 MeV 
with a small circle near 1500 MeV superimposed curving outward.
This behaviour corresponds to a negative relative coupling of 
$f_0(1500)$ with respect to the background from $f_0(980)$ (and some 
$f_0(500)$) according to formula \eref{addresonances} with
$T\simeq T_B-T_R e^{2i\delta_B}$.
  
This is to be contrasted to elastic $\pi\pi$ scattering
in figure \ref{fig:resonances} where the small circle at 1500 MeV is
curving inward corresponding to positive coupling. 
The same situation with different sign of coupling has been discussed
before in $\pi\pi\to K\bar K$  and  $\pi\pi\to \eta \eta$
\cite{Minkowski:1998mf} and can be interpreted in terms of the near flavour octet 
composition of
$f_0(1500)$ as in route 2 according to \eref{mixingssbar}. 
There is no
evidence for an extra circle in the Argand plot near 1370 MeV. 
However, as the initial
coupling to $s\bar s$ is not known one cannot derive a limit on the
branching fraction of $f_0(1370)\to \pi\pi$. The shift of the $f_0(1500)$
peak towards smaller masses is due to the interference with the background
amplitude from $f_0(980)/f_0(500)$.
The negative relative coupling of $f_0(980)$ and $f_0(1500)$ to $s\bar s$ 
is also reflected by the clear depletion in the
Dalitz plot of $D_s\to 3\pi$ at the $f_0(980)-f_0(1500)$ band crossing 
\cite{Aubert:2008ao}, and this has been seen already in the data
from the E791 Collaboration \cite{Aitala:2000xt}, see also the discussion
of phases in \cite{Minkowski:2002nd}.

\begin{figure}[t]
\includegraphics[angle=0,width=9cm]{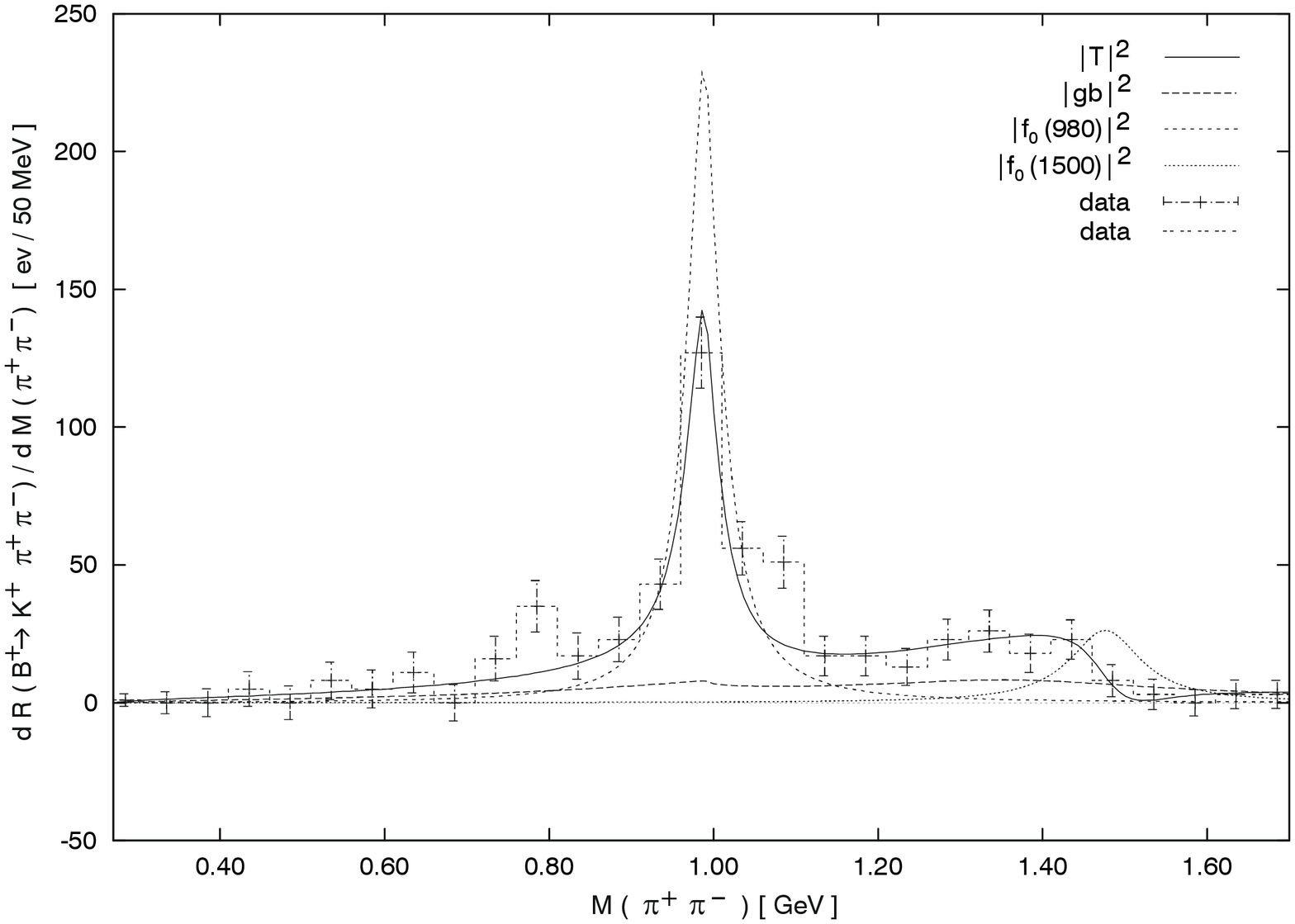} \hspace{-1cm}
\includegraphics[angle=0,width=9cm]{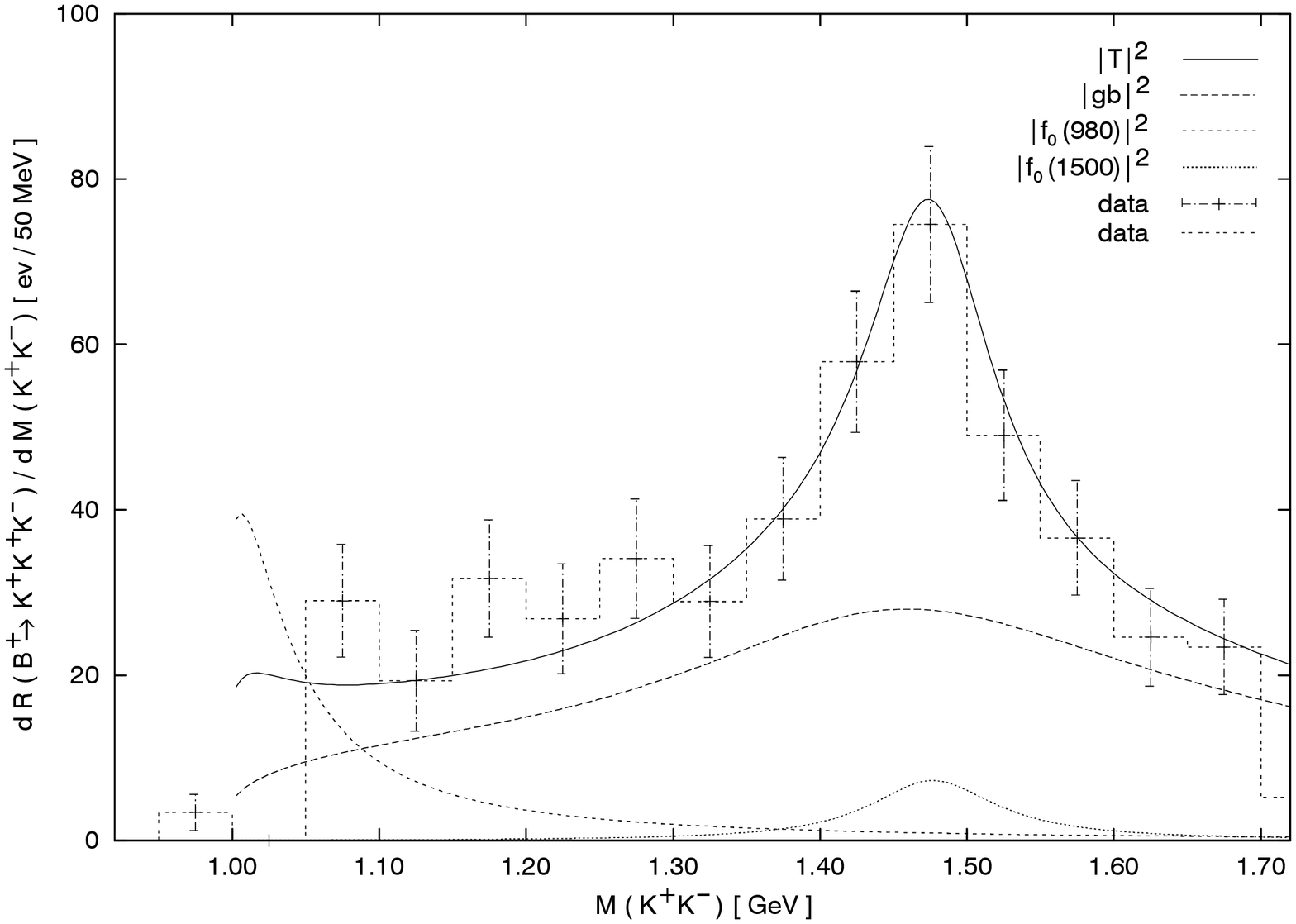}
%
\caption{\noindent $\pi^+\pi^-$ and $K^+K^-$  mass spectra in $B$-decays 
(Belle \cite{Garmash:2004wa})
 in comparison with a model amplitude $|T|^2$ 
of the coherent superposition of $f_0(980),\ f_0(1500)$ and a glueball ($gb$)
 forming the broad background. 
Also shown are the individual resonance terms
$|T_R|^2$. The background ($gb$) and $f_0(1500)$ interfere
destructively in $\pi\pi$ and constructively in $K^+K^-$ final states
(figure from \cite{Minkowski:2004xf}).
\label{fig:bdec}}
\end{figure}   

\subsubsection{Decay $B^+\to K^+ f_0 $, $f_0\to \pi\pi, \ K\bar K$.}
The $\pi\pi$ and $K\bar K$ mass spectra from these decays 
as measured by Belle \cite{Garmash:2004wa} are shown in
figure \ref{fig:bdec}. There is a
clear signal of $f_0(1500)$ in the $K^+K^-$ spectrum, but not in
$\pi^+\pi^-$, despite the 4 times larger decay probability in favour of the 
$\pi\pi$ channel. Therefore, a new state $f_X(1500)$ has been suggested
\cite{Garmash:2004wa} which looks like $f_0(1500)$ 
except for the branching ratios.\footnote{
In the analysis of $B\to K\bar K K$ by BaBaR \cite{Aubert:2006nu} the mass peak near
1500 MeV has been fitted by $X(1550)$ but with a large width of $257\pm
33$ MeV, twice as large as of $f_0(1500)$. We note however, that the fit
- contrary to the analysis \cite{Minkowski:2004xf} - 
corresponds to an approximately destructive 
interference of this  resonance with background. The data are not so well
reproduced in the resonance region and another
solution with smaller width may exist.}

This effect can also be explained \cite{Minkowski:2004xf} 
by taking into account the different quark
composition of $f_0(1500)$ near flavour octet as in \eref{mixingscalars} 
and 
$f_0(500)$ (or $f_0(980)$) taken as flavour singlet (or glueball) 
along route 2 
which leads to an  opposite sign of interference in the strange and
non-strange components: one can obtain a constructive interference
of both components in $K^+K^-$ and then 
one finds a destructive one in $\pi^+\pi^-$.
In this model represented by the curves 
in figure \ref{fig:bdec} 
the broad background from $f_0(500)$  and the narrow
resonances are superimposed as in \eref{addresonances} following Michael 
\cite{Michael:1966pl} 
\begin{eqnarray}
T_{\pi\pi}&\sim T_{gb}+c_1 T_{f_0(980)}e^{2i\delta_{gb}} + & 
         c_2T_{f_0(1500)}e^{2i\delta_{gb}}, \nonumber \\
T_{K\bar K}&\sim T_{gb}+c_3 T_{f_0(1500)}e^{2i\delta_{gb}}+ & 
    c_4 T_{f_0(1500)}e^{2i\delta_{gb}}.
\label{DalitzTuan}
\end{eqnarray}
Here $T_i$ denotes Breit-Wigner amplitudes and $\delta_{gb}$ 
the $S$ wave phase shift corresponding to $f_0(500)$ which
varies between $90^\circ$ and  $130^\circ$ in the considered mass range. 
The phase difference of the $K\bar
K$ component, i.e. of $c_2$ and $c_4$ in \eref{DalitzTuan} near $\pi$, 
explains the different structures around
1500 MeV: the strong peak at 1500 MeV in $K^+K^-$ and the small enhancement
at 1400 MeV followed by a drop towards the resonance position at 1500 MeV in
the $\pi\pi$ spectrum. The interpretation in terms of subprocesses has some
complexity by the contribution of different diagrams 
\cite{Minkowski:2004xf} and we do not discuss this further here. 
\begin{figure}[h]
\begin{center}
\includegraphics*[angle=90,width=12.0cm,bbllx=7.0cm,bblly=0.0cm,%
bburx=13.0cm,bbury=29.5cm]{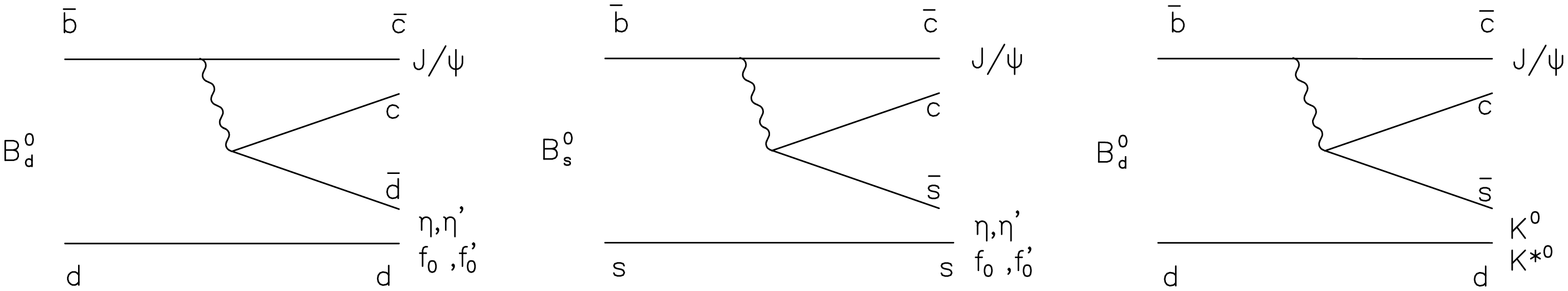}
\caption{\label{fig:BtoJ} Leading diagrams for decays $B_{d,s}^0\to J/\psi\to \eta,\eta';\
f_0,f_0'$ and $B_{d}^0\to J/\psi+K^0(K^{*0})$.}
\end{center}
\end{figure}
\begin{figure}[t]
\begin{center} 
\vspace{0.3cm}
\includegraphics[angle=0,width=9cm,bbllx=0,bblly=540,%
bburx=530,bbury=844,clip=]{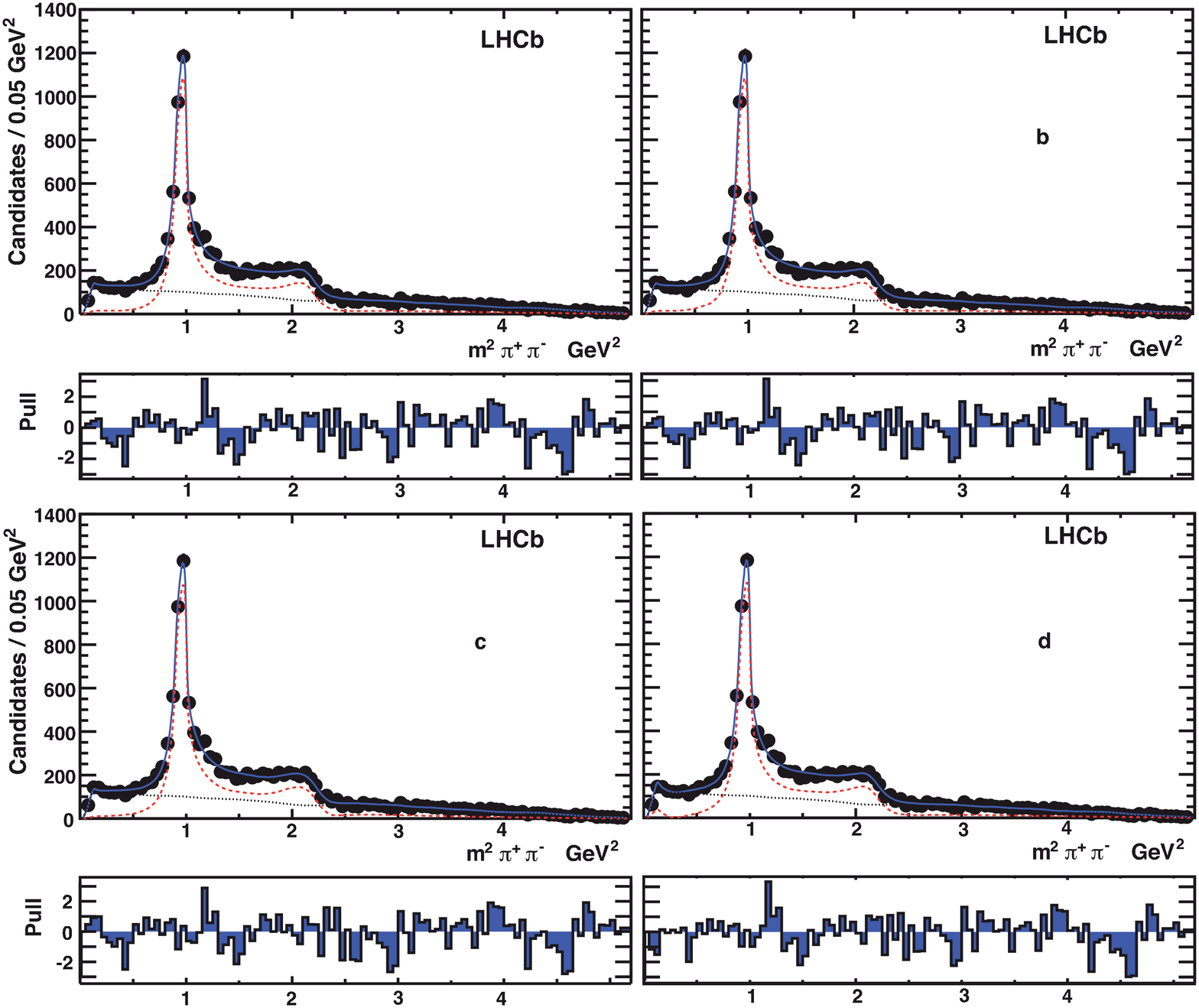}
\end{center}   
\caption{Mass squared spectrum $m^2(\pi^+\pi^-)$ in $\bar B^0_s\to J/\psi
\pi^+\pi^-$ decays; the dotted
and dashed lines indicate the background and signal contributions
respectively, figure from \cite{LHCb:2012ae} (LHCb Collaboration).
\label{fig:bdecJpsi}}
\end{figure}

\subsubsection{Decay $\bar B^0_s\to J/\psi f_0,\ f_0 \to \pi\pi$.}
\label{B0stopipi}
This
process has been measured recently by Belle \cite{Li:2011pg} at the $Y(5S)$
resonance in $e^+e^-$ annihilation and with higher
statistics by the LHCb Collaboration in pp
collisions at 7 TeV \cite{LHCb:2012ae}.
It proceeds dominantly 
through the intermediate steps 
\begin{equation}
B_s\to J/\psi + s\bar s;\quad s\bar s \to f_0\to \pi\pi. 
\end{equation}
The dominance of the $s\bar s$ intermediate state has been explored in
particular in connection with the $\eta-\eta'$ mixing 
\cite{Datta:2001ir,Fleischer:2011au,Liu:2012ib,Knegjens:2012vc}
(see figure \ref{fig:BtoJ}, middle panel).

The $\pi^+\pi^-$ mass spectrum has been
described by both Collaborations in a model fit with contributions from 
$f_0(980)$ and
$f_0(1370)$  above a broad background in the $\pi\pi$ $S$ wave. The LHCb
Collaboration also included $f_2(1270)$ in the minimal fit (at a 10\% level) 
and investigated
the inclusion of small contributions from other
states which led to only minor improvements.
The mass spectrum shown in figure \ref{fig:bdecJpsi} shows the narrow peak
from $f_0(980)$ and an enhancement followed by a drop near the position of the
$f_0(1500)$ resonance ($m_{\pi\pi}^2=2.25 \GeV^2$). 
In their preferred solution they find
mass and width of the second resonance called ``$f_0(1370)$'' as
$m_{f_0(1370)}=1475.1\pm 6.3$ MeV, and $\Gamma_{f_0(1370)}=113\pm11$ MeV.
Comparing these numbers with the PDG listings we find the identification of
this state with $f_0(1500)$ to be more appropriate: the width agrees well
with  $\Gamma_{f_0(1500)}=109\pm7$ MeV while it is much smaller than the
PDG estimate  $\Gamma_{f_0(1370)}=300-500$ MeV. The mass is smaller by 2\% 
in comparison with
the PDG value $m_{f_0(1500)}=1505\pm 6$ MeV, but the mass may be affected by
additional systematic effects from the interference with other
contributions.
Also, the 
``$f_0(1370)$'' mass is found at the upper edge of the large range suggested by PDG:
$m_{f_0(1370)}=1200-1500$ MeV. Therefore we identify the heavier resonance
in $\pi\pi$ as $f_0(1500)$ and not as $f_0(1370)$ in the $\bar B^0_s$ decay. 
An additional experimental check is possible by comparison with the 
$K^+K^-$ channel using the respective known branching ratios; the proper
inclusion of interference effects is important according to the lesson from
$B\to K\bar K K$ above. Indeed, a
signal at 1500 MeV is observed \cite{Aaij:2011ac} but there is an overlap
with $f_2'(1525)$. We also note the similarity of the mass spectra in
figures \ref{fig:dsdec} and \ref{fig:bdecJpsi} with peak below 1500 MeV. 

\subsubsection{Mixing of $f_0(1500)-f_0(980)$ from $B_s$ and $B_d$ decays.}
The branching ratios for $B_s\to J/\psi f_0$ also allow tests concerning the
intrinsic structure of these states. In route 2 (see table \ref{tab:route2})
they belong to the same multiplet and the coupling
of $s\bar s$ into $f_0(980)$ and $f_0(1500)$ according to
\eref{mixingscalars} is approximately like $2/\sqrt{6}$ and $-1/\sqrt{3}$.
This expectation is compared with the measurement of branching ratios
in table \ref{tab:bdecffpr}.

\begin{table}[h]
\caption{\label{tab:bdecffpr} Fit fractions of $\bar B^0_s\to
J/\psi f_0(980)$ and $J/\psi f_0(1500)$ for two solutions found by LHCb 
\cite{LHCb:2012ae}, also rescaled for unseen decays (using
B($f_0(1500)\to\pi\pi)=0.349\pm0.023$ \cite{beringer2012pdg}
and B($f_0(980)\to
\pi\pi)=0.75\pm0.12$ \cite{Ablikim:2005kp}; predictions (route 2) without
phase space corrections; ($f_0'/f_0\simeq ((-\sin\phi_{sc})/\cos\phi_{sc})^2$).}
\begin{indented}
\item[]\begin{tabular}{rllll}
\br
preferred solution & $\pi^+\pi^-$ fraction \%& corrected \% & route 2 &
Phase\\
\mr
$\bar B^0_s\to J/\psi f_0(980)$ & $107.1 \pm 3.5$ & $214.2 \pm 34.9$ & 
   $\simeq (\sqrt{2/3})^2$ & 0.\\
$\to  J/\psi f_0(1500)$ & $32.6 \pm 4.1 $ & $141.6\pm 20.1 $ & 
   $\simeq (-\sqrt{1/3})^2$ & ($241.5\pm 6.3) ^\circ$\\
 NonResonant       &$ 12.84\pm2.32 $ & $19.26\pm  3.48$  & & ($217.0 \pm
 3.7)^\circ$\\
$f_0(1500)/f_0(980)$ &    & $0.66\pm 0.14$ & 0.5 $\times$ PS & \\
\mr
Alternate solution & &&&\\
\mr
$\bar B^0_s\to J/\psi f_0(980)$ & $100.8 \pm 2.9$ & $201.6 \pm 32.5$ &
   $\simeq (\sqrt{2/3})^2$ & 0.\\
$\to  J/\psi f_0(1500)$ & $7.0 \pm 0.9 $ & $30.4\pm 4.4 $ &
   $\simeq (-\sqrt{1/3})^2$ & ($181.7\pm 8.4) ^\circ$\\
 NonResonant       &$ 13.8\pm2.3 $ & $20.70\pm  3.48$  & & ($232.2 \pm
 3.7)^\circ$\\
$f_0(1500)/f_0(980)$ &    & $0.15\pm 0.03$ & 0.5  $\times$ PS & \\

\br
\end{tabular}
\end{indented}
\end{table}
The Belle Collaboration finds the branching ratios 
$B(\bar B^0_s\to J/\psi f_0(980))=1.16^{+0.43}_{-0.31}\times 10^{-4}$ and
$B(\bar B^0_s\to J/\psi f_0(1370))=0.34^{+0.14}_{-0.15}\times 10^{-4}$ if we add the
errors in quadrature. After corrections for unseen decay modes we find 
(taking the second resonance as $f_0'\equiv f_0(1500)$)
\begin{equation}
r_{f_0'/f_0}^s=\frac{p^3 B( \bar B^0_s\to J/\psi f_0(1500))}{(p')^3 B(\bar B^0_s\to J/\psi f_0(980))},\quad 
r_{f_0'/f_0}^s=
\begin{cases}
1.18\pm 0.25) & \text{LHCb preferred solution},\\
0.15\pm 0.05) & \text{LHCb alternate solution},\\
1.12\pm 0.61  & \text{Belle}.
\end{cases}
\label{rf0}
\end{equation}
with $p(p')$ the momenta of $f_0(980)$ ($f_0(1500)$) in the $B_s$ system.
Combining the LHCb preferred solution with Belle yields
\beq
r_{f_0'/f_0}^s=1.17\pm0.23. \label{rf00pr}
\eeq
It is interesting to compare this result with the corresponding ratio for
decays into $\eta$ and $\eta'$ which should agree approximately for route 2. 
This ratio has been measured recently by
\cite{Belle:2012aa} as $R_\eta=B(B_s\to J/\psi\eta')/B(B_s\to
J/\psi\eta)=0.73\pm0.14\ (stat.)\pm0.02\ (syst.)$ which leads to the
phase space corrected ratio
\beq
r_{\eta/\eta'}^s=
\frac{p_{\eta'}^3\ B(B_s\to J/\psi\eta)}{p_\eta^3\ B(B_s\to J/\psi\eta')}
   = 1.10\pm 0.22.
\eeq
This number is in a remarkable agreement indeed with the result 
in \eref{rf00pr} supporting the similar flavour mixing of 
$f_0(1500)-f_0(980)$ 
and $\eta - \eta'$ with inverse mass ordering 
as in \eref{mixingscalars} for route 2. 

These ratios can be compared also directly with the expectations 
according to the quark-gluon structure of these mesons. If we allow for a gluonic
component with mixing as in \eref{mixingetaprg} we find (see also
\cite{Knegjens:2012vc})
\beq
r_{\eta/\eta'}^s=\frac{\tan^2\phi_{ps}}{\cos^2\phi_G^{\eta'}} \quad \to \quad
\cos\phi_G^{\eta'}=0.86\pm0.09.
\label{cosphig}
\eeq
where $\phi_{ps}=42^\circ$ has been used. This is easily compatible with
having no gluonic component in $\eta'$, alternatively, one can admit a contribution
with $\phi_G^{\eta'}\simeq (30.7^{+8.8}_{-12.5})^\circ$.

Accordingly, introducing a gluonic component for  $f_0'\equiv f_0(1500)$ as
\beq
|f_0'\ran=\cos \phi_G|(\cos \phi_{sc}|n\bar n\ran - \sin\phi_{sc}|s\bar
s\ran) + \sin \phi_G|gg\ran
\eeq
we find for route 2 
\beq
r_{f_0'/f_0}^s=
\cos^2\phi_G^{f_0'}\tan^2\phi_{sc} \quad  \to \quad 
\cos\phi_G^{f_0'}=1.87\pm0.32.
\eeq
where we have used our result $\phi_{sc}=30^\circ$ in \eref{alphafinal}
and $r_{f_0'/f_0}^s=1.17$ in \eref{rf00pr}. 
We find that
this result for $\cos\phi_G^{f_0'}$ exceeds the physical region by
$2.7\sigma$. This discrepancy may indicate a deficiency of the model with
gluons mixed only into $f_0(1500)$ (if not of the $f_0-f_0'$ 
mixing model altogether),
but looking at the success of the relation $r_{\eta/\eta'}^s\approx
r_{f_0'/f_0}^s$
one could also think of an underestimation of errors in the branching
fractions of $B_s$ decays, of the correction factors for unseen decays or
our mixing angle $\phi_{sc}$. An interesting test of the resonance analysis 
with $f_0(1500)$ would be the
corresponding analysis of the decays
$B_s\to J/\psi K\bar K$ where different interference phenomena appear, as we
could witness already in the $B\to K+X$ decays (see figure~\ref{fig:bdec}). 

The production
phase of $f_0(1500)$ measured with respect to $f_0(980)$ in the $s\bar s$
component selected here is expected in route 2 as 
$\Delta \phi^s=180^\circ$ which is near the observed results in table
\ref{tab:bdecffpr}. For illustration, consider our approximate representation
\eref{mixingscalars}: in the $n\bar n - s\bar s$ coordinate system 
$f_0(980)=(1,\sqrt{2})/\sqrt{3}$ and $f_0(1500)=(\sqrt{2},-1)/\sqrt{3}$,
so the relative phase for the $s\bar s$ component is $\Delta\phi^s=\pi$
and for the $n\bar n$ (or $d\bar d$) component it is $\Delta\phi^d=0$.
The phase in the LHCb preferred solution is found as
\beq
\Delta \phi^s=(241.0\pm 6.3)^\circ
\eeq 
with an additional shift by $\sim 60^\circ$ 
which could be related to a moving background phase as in the model of the type
\eref{DalitzTuan}
(the $f_0(500)$ component in figure
\ref{fig:resonances} moves from about $80^\circ$ to $140^\circ$).

The flavour
mixing $f_0(980)-f_0(1500)$ can be measured also from $B_d$ decays, 
a measurement which may become feasible in the near future.
In close analogy to the Cabibbo-suppressed decays $B_d\to J/\psi \eta^{(')}$ \cite{Datta:2001ir} 
one can study the corresponding decays into $f_0/f_0'$. The dominant
subprocesses in this case are (figure \ref{fig:BtoJ}, left panel)
\beq
B_d\to J/\psi+ d\bar d,\quad d\bar d\to f_0^{(')},\ \eta^{(')}.
\eeq
One obtains for
$B_d$ decays 
in the standard mixing scheme
\beq
r_{\eta/\eta'}^d=
\frac{p_{\eta'}^3\ B(B_d\to J/\psi\eta)}{p_\eta^3\ B(B_d\to
J/\psi\eta')},\quad
r_{\eta/\eta'}^d=\cot^2\phi_{ps}  
\eeq
In the same way, assuming an $f_0(980)-f_0(1500)$ mixing as in route 2 
\beq
r_{f_0'/f_0}^d=\frac{p^3 B(  B^0_d\to J/\psi f_0(1500))}{(p')^3 B(
B^0_d\to J/\psi f_0(980))},\quad
r_{f_0'/f_0}^d=\cot^2\phi_{sc}      
\eeq
with the same mixing angle $\phi_{sc}$. For these mixing schemes
one obtains the constraints
\beq
r^d_{\eta/\eta'}\times r^s_{\eta/\eta'}=1,\qquad r^d_{f_0'/f_0}\times r^s_{f_0'/f_0}=1.
\eeq
Also in route 2 one expects $\phi_{ps}\approx \phi_{sc}$ or
\beq
r_{f_0'/f_0}^d \approx r_{\eta/\eta'}^d.
\eeq
For the production phase of $f_0(1500)$ relative to $f_0(980)$ we obtain
in $B_d\to J/\psi\pi\pi$ decays $\Delta \phi^d=0$, or, 
above a background with moving phase as
in $\pi\pi$ scattering 
\beq
\Delta \phi^d\approx 60^\circ.
\eeq

\subsubsection{Mixing angles of isoscalars $f_0$ from $B_d$ and $B_s$
decays.}
\label{Bddecay}
The determination of the individual flavour 
mixing angles $\phi$ of isoscalars like
$f_0(980)$ and $f_0(1500)$, but also of $f_0(500)$, i.e. the ratio of
$n\bar n$ and $s\bar s$ amplitudes of these states, 
is possible from the ratio of $B_s$ and $B_d$ decays.
As in case of $\eta,\eta'$ \cite{Datta:2001ir}, one can obtain the mixing
angle $\phi_{sc}$ for $f_0(980)$ or the mixing angle $\phi_{sc}'$
for $f_0(1500)$ from the ratios
\begin{eqnarray}
r_{f_0}&=\frac{p^3_s \Gamma( B_d\to J/\psi f_0(980))}
{p^3_d \Gamma( B_s\to J/\psi f_0(980))}=\frac{1}{2} (F_{CKM})^2 \cdot
\tan^2\phi_{sc};\\
r_{f_0'}&=\frac{p^3_s \Gamma( B_d\to J/\psi f_0(1500))}
{p^3_d \Gamma( B_s\to J/\psi f_0(1500))}=\frac{1}{2}(F_{CKM})^2 \cdot
\cot^2\phi_{sc}', \label{f0prdirect}
\end{eqnarray}  
where $F_{CKM}=V_{cd}/V_{cs}$ is calculated for the dominant tree diagram
and penguin diagram contributions and with the CKM matrix elements from PDG
one finds $|F_{CKM}|\simeq 0.225$. 
The mixing angle $\phi_{sc}$ for $f_0(980)$ should agree with our
determination in \eref{alphafinal}, the angle  $\phi_{sc}'$ for $f_0(1500)$
may be different if the two states are in different multiplets.

Under the assumption that $f_0(1500)$ and $K^{*0}(1430)$ 
belong to the same multiplet, the mixing angle for $f_0(1500)$ 
can be determined also from the ratio
\beq
r_{f_0'}^K=\frac{p^3_d\ \Gamma( B_s\to J/\psi f_0(1500))}
{p^3_s \Gamma( B_d\to J/\psi K^{*0}(1430))}=\sin^2\phi_{sc}'
\eeq
in analogy to a corresponding ratio of $\Gamma(B_s\to J/\psi \eta)/\Gamma(B_d\to
J/\psi K^0)$ (see figure  \ref{fig:BtoJ}). 
It is interesting to note, that these determinations of mixing angle between
the different flavour components are independent of a possible presence of a
gluonic component.

In the same way one could also study the flavour content of
$f_0(500)/\sigma$.

We have no direct predictions from the other classification routes discussed.
Some aspects of the tetra-quark model for $f_0(980)$ are elaborated in
\cite{Fleischer:2011au}.

\subsection{Decay of $J/\psi\to \phi(1020) f_0,\ f_0\to \pi\pi,\ K\bar K$} 
In the $\pi\pi$ final state of this $J/\psi$ decay the BES Collaboration
\cite{Ablikim:2004wn} observed a peak around 1335 MeV followed by a drop
near 1500 MeV. In their model fits they include, besides $f_2(1270)$, 
the scalar $f_0(1500)$ and require $f_0(1370)$ as well 
with mass $m=1350\pm 5 $ MeV and width  $\Gamma= 265 \pm 40$ MeV. 

The $J/\psi$ decay is expected to proceed dominantly (according to the
``singly connected diagram'') through a gluonic intermediate
state and, triggered by $\phi(1020)$, the $\pi\pi$ final state evolves 
from an intermediate $s\bar s$ configuration. Therefore, we expect a 
similar final state interaction as in the decay $B_s\to J/\psi \pi\pi$
(see discussion in section \ref{B0stopipi} 
and figure \eref{fig:bdecJpsi}) and in $\pi\pi\to K\bar K$
(see discussion in section \ref{inelasticpipi} and figure \ref{fig:s00kk}).
Indeed, the $\pi\pi$ spectra in these processes with an
enhancement near 1400 MeV and a drop near 1500 MeV look all rather similar.
There, evidence was only given for a single resonance interfering with some
smooth background amplitude and the resonance was compatible with $f_0(1500)$.
It is therefore important for a definite conclusion about an extra state
 ``$f_0(1370)$'' to analyze
the observed peak in $J/\psi\to \phi+\pi\pi$ decays 
with higher statistics in a bin-by-bin phase shift analysis.

In the $K^+K^-$ channel the BES data \cite{Ablikim:2004wn} do not show any enhancement around 1350
MeV and the branching fraction ratio
\beq
R_{K\bar K/\pi\pi}=\frac{B(f_0(1370)\to K\bar K)}{B(f_0(1370)\to \pi\pi)}=0.08 \pm 0.08
\label{rf01370kkpipi}
\eeq
has been determined. Other measurements of this ratio 
listed by PDG \cite{beringer2012pdg} 
are spread over a range up to $R_{K\bar K/\pi\pi}\lesssim 1.0$, therefore 
\beq
B(f_0(1370)\to K\bar K)\lesssim 0.1.
\eeq
The measurement
$\Gamma(\eta\eta)/\Gamma(4\pi) = 4.7\pm2.0\times 10^{-3}$
\cite{Barberis:2000cd,beringer2012pdg} implies this number
to be an upper limit for $B(f_0(1370)\to \eta\eta)$.
So, as an estimate
\beq
B(f_0(1370)\to 2\ {\rm pseudoscalars})\lesssim 20\%
\eeq  
and almost all decays should go into higher multiplicities.

\subsection{Central production in hadron hadron collisions}
\begin{figure}[h]
\begin{center}
\includegraphics*[angle=90,width=4.0cm,bbllx=9.5cm,bblly=10.5cm,%
bburx=14.5cm,bbury=18.5cm]{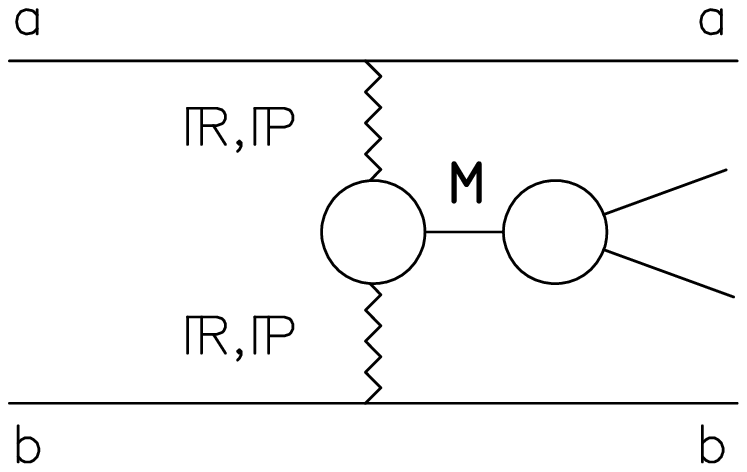}
\caption{
Central production of meson M in $a+b$ collisions with two rapidity gaps 
and exchange of Reggeons and/or Pomerons.
\label{fig:central}}
\end{center}
\end{figure}

Central production processes are selected by requiring two gaps in rapidity
$y=\tfrac{1}{2}\ln\tfrac{E+p_\parallel}{E-p_\parallel}$, empty of particles, 
adjacent to the
direction of the incoming hadrons, and small momentum transfers 
$t=(p_i-p_f)^2 $ between 
the incoming and outgoing forward hadrons $a$ and $b$
(see figure \ref{fig:central}).  For such processes a description 
in terms of Regge theory applies.
The exchange of Reggeons $\reg$ or Pomerons $\pom$ determines the dependence of
the respective cross sections on the primary energy $\sqrt{s}$.
At high energy $\sqrt{s}$ the Double Pomeron Exchange ($\pom\pom$)  
process dominates
with an approximately constant cross section whereas subleading
contributions fall like $\sigma(\pom\reg)\sim 1/\sqrt{s}$ and
$\sigma(\reg\reg)\sim 1/s$. 
While the Reggeon is considered a $q\bar q$ object, the Pomeron is thought
rather to be dominantly gluonic, more discussion follows below in section
\ref{pomeronpartons}.
Therefore the strength of central meson production 
can give a hint towards its possible gluonic component
\cite{Robson:1977pm,Close:1987er}.
In particular, a search for scalar mesons in many different channels has
been performed. 

\subsubsection{The $\pi\pi$ system.}
\begin{figure}[t]
\begin{center}
\includegraphics*[width=8.0cm,bbllx=0.0cm,bblly=0.0cm,bburx=19.0cm,%
bbury=25.0cm]{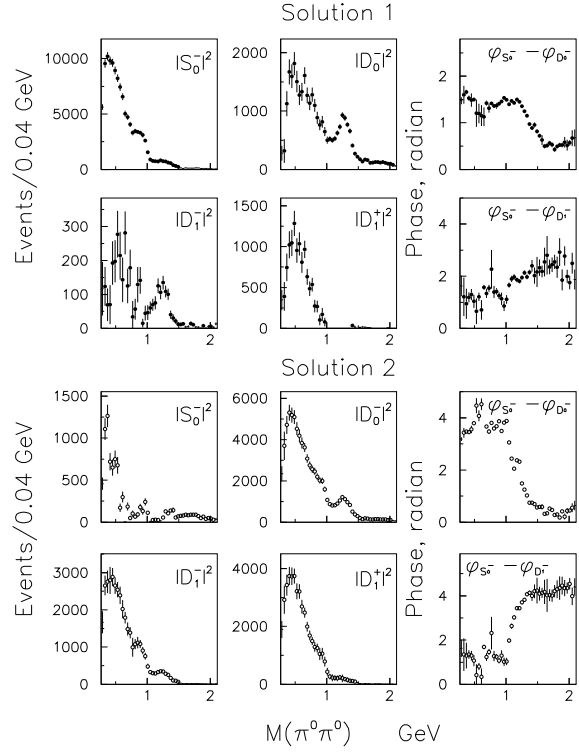}
\caption{
Partial wave amplitudes for the centrally produced $\pi^0\pi^0$
system in $pp\to
p(\pi^0\pi^0)p$ at $\sqrt{s}=450$ GeV, figure from \cite{Barberis:1999ap}(WA102
Collaboration).\label{fig:centralpi0pi0}}
\end{center}
\end{figure}

has been studied extensively at the ISR
(Axial Field Spectrometer) \cite{Akesson:1985rn},
at the SPS collider by WA102 \cite{Barberis:1999ap} and by GAMS
\cite{Alde:1997ri},
and now at the LHC by ALICE \cite{Schicker:2011ug}.
In figure \ref{fig:centralpi0pi0} we show as an example an amplitude
analysis of the $\pi^0\pi^0$ system including $S$ and $D$ waves which has
been carried out using experimental 
moments $\lan Y_L^M \ran$ and relations \eref{ylmequations}. 
Two distinct solutions appear. Solution 2 with a $D$ wave at
threshold much stronger than $S$ wave is considered unphysical.

The following features of the physical solution 1 in figure
\eref{fig:centralpi0pi0} are noteworthy:\\
(i) {\it Low mass peaks in $S$ and $D$ waves.} In the $S$ wave intensity 
there is a strong peak at low mass
$m_{\pi\pi}\sim 400 $ MeV but without associated phase variation.
In addition, there is also a peak in the
$D_0^-$ wave at 500 MeV much below the first known resonance $f_2(1270)$,
very different to elastic $\pi\pi$ scattering, for example.
As pointed out in \cite{Minkowski:2002nf} a very similar phenomenon  
has actually been observed in $\gamma\gamma\to \pi^+\pi^-$ (not so in
$\pi^0\pi^0$), where 
the low mass $S$ and $D$ wave peaks are
related to the dominant one pion exchange process 
at low energy with direct coupling of photons to charged pions. 
The pion exchange pole near the physical region of scattering at $t=m_\pi^2$ 
causes the enhanced forward-backward scattering generating the large 
$D$ wave at
first. This analogy suggests the appearance of one pion exchange in the
subprocess 
$$\pom\pom\to \pi\pi$$ 
as well, which adds to the direct resonance formation as in figure
\ref{fig:central}. A Regge calculation of this exclusive process has been
presented recently \cite{Lebiedowicz:2009pj}.
The presence of the broad $f_0(500)/\sigma$ is signaled
by the interference with $f_0(980)$ (sharp drop in $|S_0|^2$ at
$m_{\pi\pi}\sim 1$ GeV) as in $\pi\pi$
elastic scattering. 
At higher energies other
exchange processes like $\rho$-exchange are possible \cite{Klempt:2007cp}.\\ 
(ii) {\it There is no phase coherence.} Contrary to the reaction
$\pi^-p\to \pi^0\pi^0 n$ \cite{Alde:1998mc} 
the phase differences $\varphi_{S_0}-\varphi_{D_0^-}$ 
and $\varphi_{S_0}-\varphi_{D_1^-}$ are very different. In particular,
while $|D_0^-|^2$ and $|D_1^-|^2$ both show a clear signal from $f_2(1270)$,
only $\varphi_{S_0}-\varphi_{D_0^-}$ reflects a related phase variation from 
its interference with the slowly moving $S$ wave, 
whereas $\varphi_{S_0}-\varphi_{D_1^-}$ is without resonance structure and
so ``phase coherence'' is
strongly violated.  

Therefore, one cannot construct a model based on formulae
\eref{ylmequations} with a resonance $f_2(1270)$ in $D_0^-$ and $D_1^-$
interfering with the same $S_0^-$. The violation of phase coherence 
as in figure \ref{fig:centralpi0pi0} is expected 
if $f_2(1270)$ couples to different partial waves $L_\mu^{(\pm)}$
with  different nucleon helicities.
As an example, let
the $S_0^-$ wave 
couple only to  $S_{0,++,++}^-$ but let $f_2(1270)$ couple to
$D^-_{0,++,++}$ and also to $D^-_{1,+-,+-}$ (the lower $(+)$ and $(-)$ 
refer to the nucleon helicities), then $f_2(1270)$ appears in both
$|D_0^-|^2,|D_1^-|^2$ moduli, $S_0\cdot D^-_0$ is resonant while
$S_0\cdot D^-_1=0$ similar to observations.

Such effects can occur if several exchange mechanisms
are at work 
with different spin composition and ``spin coherence'' is violated. 
Then,
in  \eref{ylmequations} it may be necessary to replace the single terms 
\begin{eqnarray}
|S|^2 & \to |S|^2+|\bar S|^2;\quad \nonumber\\ 
\Re (SD^{-*}) & \to \Re (SD^{-*}) + 
\Re (\bar S\bar D^{-*})
\label{incoherent}
\end{eqnarray}
where the terms $S$ and $\bar S$ correspond to contributions with different
nucleon helicities.
 Obviously, by addition of arbitrary 
incoherent contributions as in \eref{incoherent} one cannot in general
solve for all amplitudes in an energy independent analysis,
but possibly a model with selected
contributions to describe all moments can be constructed.

In a more general approach one may include a description for the transverse
momentum
dependences of the helicity amplitudes and in that way provide additional
information for a full separation of partial wave amplitudes. 
Transverse momentum
correlations of outgoing protons have been suggested as tool to discriminate
different intrinsic structures of the centrally produced object
(``glueball-filter'') by Close and Kirk \cite{Close:1997pj}.
Different effects 
have been found indeed
for the resonances in the $\pi\pi,K\bar K$ data obtained by the WA102 
Collaboration \cite{Barberis:1999cq}. 
Especially, the
azimuthal final $pp$ correlations behave differently for $f_2(1270)$ and
$f_0(1500)$ \cite{Close:2000dx} which implies again that a model with ``spin
coherence'' as incorporated in the amplitude analysis above is incomplete.
The $t$-dependence of helicity
amplitudes in the general framework of Regge theory and for specific models 
has been discussed by Kaidalov et al.
\cite{Kaidalov:2003fw}; and so this framework may be used for an extended
analysis including helicity dependent $t$-distributions. 

(iii) {\it Contributions from states $f_0(1370)$ and $f_0(1500)$}
have to be determined from a partial wave analysis, i.e. from
the complex amplitudes as discussed in subsection \ref{respipi}. In view of
the previous remarks this requires an extended analysis not relying on
spin-coherence.
Possibly at the higher LHC energies 
spin coherence is again a good approximation.
In any case, a determination of the resonance fractions
from $|S_0|^2$ alone without inclusion of phase movements does not yield
reliable results as was demonstrated in subsection \ref{respipi}.
For this reason the results from WA102 on $f_0(1370)\to\pi\pi,K\bar K$
\cite{Barberis:1999ap,Barberis:1999cq} are not considered here 
as evidence for the existence of this state.

\subsubsection{Production of $f_0(1370)$ and $f_0(1500)$ in the $4\pi$
system.}
The $4\pi$ decays of the scalar mesons have been investigated extensively 
in central
diffractive production $pp\to p (4\pi) p$ (WA102 \cite{Barberis:2000em,Barberis:1999wn}) and in 
$\bar p p\to 5\pi^0$ and $\bar p n\to \pi^- 4\pi^0$ annihilations 
(Crystal Barrel \cite{Abele:2001js}). 
A comparative
discussion is given by Klempt and Zaitsev \cite{Klempt:2007cp}.
They have collected the decay branching ratios of both resonances
into different $4\pi$ components as measured by central production and by
$p\bar p$ annihilation experiments (see table \ref{tab:4pi})

\begin{table}[h]
\caption{\label{tab:4pi} Decay branching ratios of isoscalar mesons into
$4\pi$ channels from central production (CP) and $\bar p p/\bar p n$ 
annihilation (from \cite{Klempt:2007cp}).\\
\# taken from PDG listing}
\begin{indented}
\item[]\begin{tabular}{llllll}
\br
 & $f_0(1370)$ &&& $f_0(1500)$ &\\
 & $\bar p  p/\bar p n$& CP &\qquad &  $\bar p  p/\bar p n$& CP\\
\mr
 $\Gamma_{4\pi}/\Gammatot$ & $0.80\pm 0.05$ && & $0.73\pm
 0.21$\#  & 
          $0.48 \pm 0.05$\\
 $\Gamma_{\rho\rho}/\Gamma_{4\pi}$ & $0.26\pm 0.07$ & $\sim 0.9$& & $0.13\pm
    0.08$ &  $0.74\pm 0.03$\\
 $\Gamma_{\sigma\sigma}/\Gamma_{4\pi}$ & $0.51\pm 0.09$ & $\sim 0$ &&
    $0.26\pm  0.07$ &  $0.26\pm 0.03$\\
\br
\end{tabular}
\end{indented}
\end{table}
Some discrepancy for $f_0(1500)$ disappears, if the different decay channels
included in the analysis
 are taken into account. In case of $f_0(1370)$ essential
discrepancies remain. Especially, in central production there is a clear
signal from $f_0(1500)$ in the $\sigma\sigma$ channels (including 
$4\pi^0$) but no obvious signal from $f_0(1370)$
($\sigma$ stands here for low mass
$\pi\pi$ $S$ wave). On the other hand, 
in $\bar p p\ (\bar p n)$ there is a strong peak in the $4\pi^0$
mass spectrum at $\sim 1470$ MeV but with a larger width 
than expected for $f_0(1500)$
and this effect 
has been related to $f_0(1370)$. The analysis has to take into
account the contributions from 
different processes and decay chains in $5\pi$ processes and 
combinatorial backgrounds in $5\pi^0$
not present in central production. 
An attempt to analyze
the phase motion of the $\sigma\sigma$ decay has lead to the conclusion
\cite{Klempt:2007cp} that there is only one resonant state with a motion
consistent with $f_0(1500)$; the strong $\rho\rho$ enhancement at threshold
in central production which has been related to $f_0(1370)$ 
and its complete absence in $\sigma\sigma$ in this view 
could  be a consequence of $\rho$-Regge exchange in $\pom\pom$
scattering, similar to the
threshold enhancement in $\pi\pi$ which has been attributed to pion 
exchange as discussed before. This mechanism
would explain the absence of $f_0(1370)\to \sigma\sigma$ in central production and
similarly the absence of the $\eta\eta$ production where a suitable Regge
exchange is lacking. There is also the interesting phenomenon of a dip in
the $\rho\rho$ mass spectrum which moves depending on the kinematic
selection for the outgoing protons \cite{Barberis:2000em}. For a particular
selection (azimuthal angle between protons $135^\circ < \phi < 180^\circ$) 
the deep sharp dip is just at the mass of 1500 MeV which suggests a negative
interference of $f_0(1500)$ with a broad background reminiscent of  the 
similar phenomenon 
in $\pi\pi$ scattering (figure \ref{fig:S00mass}).

While the signature of $f_0(1500)$ is very clear in $(4\pi)$, especially in
$(\sigma\sigma)$, there is no compelling evidence of additional
contributions which would require a new resonance 
at lower mass around 1370 MeV. As
learned in the 2 particle channels, instead of the broad $f_0(1370)$
($\Gamma\gtrsim 300$ MeV) there may also be a contribution from the long tail of
$f_0(500-1000)$ with $\Gamma\sim 1000$ MeV.  

\subsection{Flavour structure of $f_0(1500)$ from its decays}
This state has been considered at first as a glueball candidate following a
prediction from quenched lattice QCD. The study
of decay branching ratios suggested a mixed state of $q\bar q$
(mainly octet) and $gg$ as discussed in route 1 above
\cite{Amsler:1995td,Close:1996yc}. A pure $q\bar q$ composition was found
incompatible with certain decay properties \cite{Amsler:2002ey}.
On the other hand, the hint towards a negative phase between $K^+K^-$ and $\pi^+\pi^-$
decay amplitudes \cite{Minkowski:1998mf,Estabrooks:1978de} limits the contribution from a glueball component
which would prefer a singlet type $q\bar q$ state.

A direct measurement of the mixing angle $\alpha$ ($=\phi_{sc}$ in route 2) 
independent of the
gluonic component is possible through the measurement of $B_d$ decays
(see subsection \ref{Bddecay}). The negative relative sign in $D_s$ decay
between $s\bar s\to f_0(980)\to \pi\pi$ and $s\bar s\to f_0(1500)\to \pi\pi$ 
(see subsection \ref{Dsmesons}) requires an octet type mixing angle
$\alpha=\phi_{sc}$ in \eref{mixingf0pr} with
\beq
\tan \phi_{sc}>0 \label{mixingpositiv}
\eeq
once the singlet type mixing for
$f_0(980)$ is established. 
In the following we will investigate in more detail the
consequences of the measured decay branching ratios for the
constituent structure of $f_0(1500)$.   

\begin{figure*}[t]
\begin{center}
\includegraphics*[width=6cm]{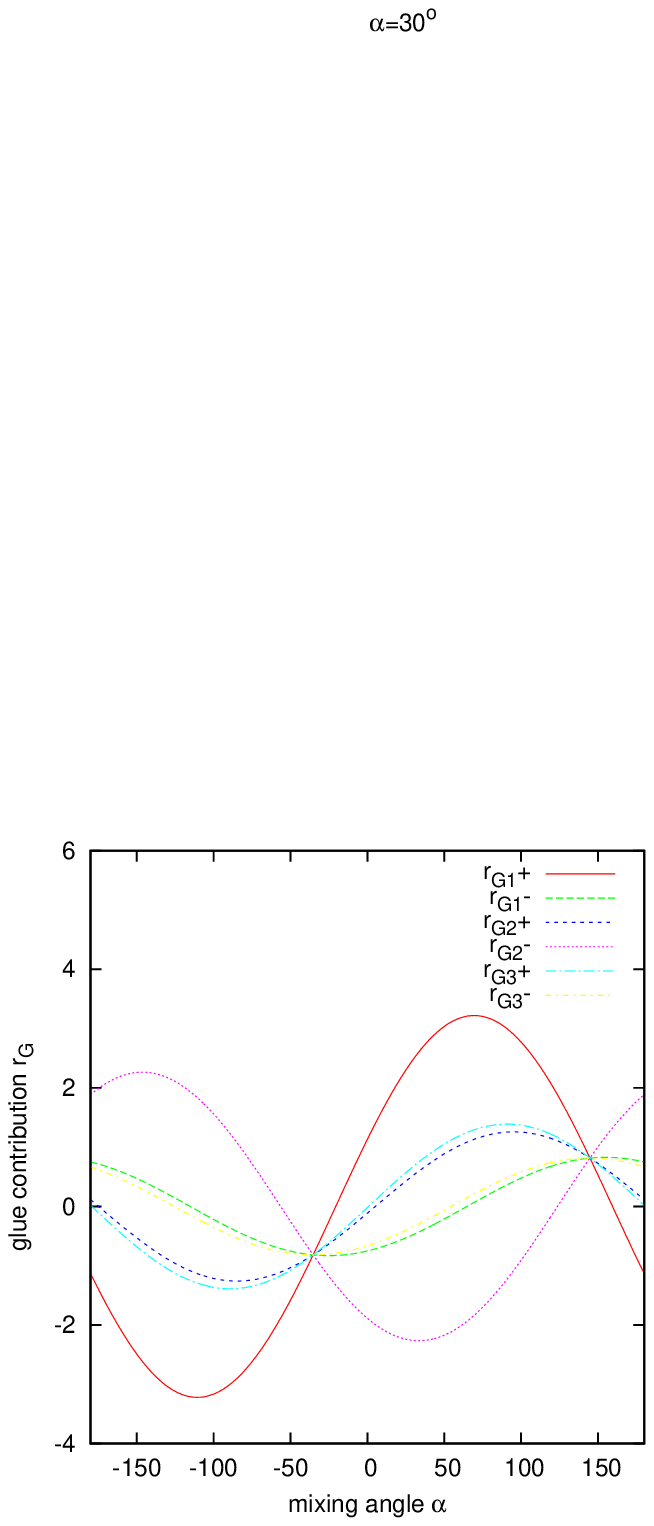}
 \includegraphics*[width=8.7cm]{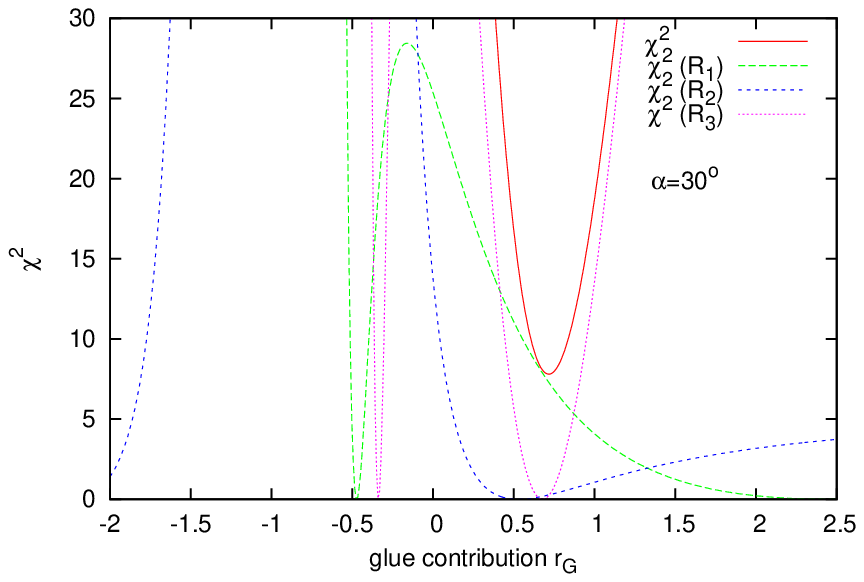}
\end{center}
\caption{\label{fig:ratiosra} 
Glue contribution $r_G$ in the $f_0(1500)$ state as function of
mixing angle $\alpha=\phi_{sc}$ for fixed ratios $R_i$ of branching ratios
as in \eref{R11500}-\eref{R31500} (left);
$\chi^2$ vs. $r_G$ for observed ratios $R_i$ and full $\chi^2$ at mixing angle
$\alpha=30^o$ (right).} 
\end{figure*}

The decay couplings $\gamma_{ij}^2$ of $f_0(1500)$ depend on the $q\bar q$
and $gg$ components through the mixing angle $\alpha$ (see  
\eref{mixing}, \eref{mixingf0pr}) and the parameter  $r_G\sim G/g$, 
as shown in table \ref{tab:decay}.
Now we consider ratios $R_i$ of branching ratios $B_k$ as in
\cite{Amsler:1995td}.
From the PDG fits \cite{beringer2012pdg} using the phase space 
correction $\gamma_k^2=B_k/q_k$ we derive the ratios
\bea
R_1&=\frac{\gamma^2[f_0(1500)\to \eta\eta]}{\gamma^2[f_0(1500)\to
\pi\pi]}&=0.208\pm 0.039\label{R11500}\\
R_2&=\frac{\gamma^2[f_0(1500)\to \eta\eta']}{\gamma^2[f_0(1500)\to
\pi\pi]}&=0.21\pm 0.09\\
R_3&=\frac{\gamma^2[f_0(1500)\to K\bar K]}{\gamma^2[f_0(1500)\to
\pi\pi]}&=0.321\pm 0.034 
\label{R31500}
\end{eqnarray}
(using also $q_{\eta\eta'}=195$ MeV for $R_2$ \cite{Amsler:1995td}).
For a given experimental ratio $R_i$ we can insert in
\eref{R11500}-\eref{R31500} 
the results for
$\gamma^2_{ij}$ from table~\ref{tab:decay}. Then,
for each value of the mixing angle
$\alpha$ there are two corresponding solutions for the glue contribution 
$r_{Gi}^\pm$.
Solving for $r_{Gi}^\pm$ we find (with $\phi\equiv \phi_{ps}$ and $\rho=R=1$)
\bea
r_{G1}^\pm\,(\sqrt{3R_1}\mp1)&=
    -\sqrt{3R_1}\cos\alpha\pm
    (\cos^2\phi_{}\cos\alpha-\sqrt{2}\sin^2\phi_{}\sin\alpha)\\
r_{G2}^\pm\,\sqrt{3R_2}&=
    -\sqrt{3R_2}\cos\alpha\pm 
      \sqrt{2} \sin\phi_{}\cos\phi_{}(\cos\alpha+\sqrt{2}\sin\alpha)\\
r_{G3}^\pm\,(\sqrt{3R_3}\mp2)&=
    -\sqrt{3R_3}\cos\alpha\pm (\cos\alpha-\sqrt{2}\sin\alpha)
\label{rpmsol}
\end{eqnarray}
In figure \ref{fig:ratiosra} (left panel) these solutions $r_{Gi}^\pm$ for every $R_i$
are shown for the given $\alpha$. There are two special points for $\tan \alpha=-1/\sqrt{2}$
where all solutions coincide in a trivial way
if the numerator $N_i$ and the denominator $D_i$ for $R_i D_i=N_i$ 
vanish: 
\beq
\alpha_1\approx-35^\circ,\ \alpha_2\approx145^\circ;\quad \text{with}\quad 
r_{G1}\approx -0.82,\ r_{G2}\approx0.82.   
\eeq
Otherwise, there is no crossing of the solutions from the three ratios $R_i$
expected for an exact solution for $r_G$, but rather some pairs of ratios
come close to each other ($r_{G2}^+,r_{G3}^+$) and ($r_{G1}^-,r_{G3}^-$). 

In the right panel of figure \ref{fig:ratiosra} we show, how the $\chi^2$ 
varies with the glue component $r_G$ for a mixing angle $30^\circ$; 
these two
parameters determine the ratios $R_i$ according to table~\ref{tab:decay}.
Also shown are the contributions to $\chi^2$ from $R_i$  separately
which vanish at the exact solutions for $r_{Gi}^\pm$ seen in figure 
\ref{fig:ratiosra} (left).
There is a $\chi^2$ minimum 
\beq
\text{for}\ \alpha=30^\circ: \qquad r_G=+0.7\pm 0.1.
\label{rsolution}
\eeq
For this solution there is a good agreement with $R_2,R_3$ but a 
discrepancy with the $\eta\eta$ channel ($R_1$)
by $\gtrsim 2\sigma$. There is another solution for $r_G\sim -0.4$ but with an
unacceptable $\chi^2>30$. The $\chi^2$ minimum is found to be practically
independent of $\alpha$ if $r_G$ moves along the solutions ($r_{G2}^+,r_{G3}^+$).
On that path there is also the solution ($\alpha=0^\circ,\ r_G\approx 0$) 
for $f_0(1500)$ without glue. Therefore, it needs further information
to determine the mixing angle $\alpha$ uniquely. Some information comes from
the $2\gamma$ decay rate (see, for example \cite{Amsler:2002ey}) which needs 
some model
assumptions (see next subsection). If $f_0(980)$ and $f_0(1500)$ belong to
the same multiplet as in route 2 then there are further relations as well as
discussed above. A direct determination of the mixing angle 
is possible by comparing $B_d$ and
$B_s$ decays as in \eref{f0prdirect}. Further information is available
from the interference with other states like $f_0(980)$ or $f_0(500-1000)$
in different channels.

According to Anisovich \cite{Anisovich:1997zw} the decay coupling $g_0$ of the
$q\bar q$ state into mesons and the coupling $G_0$ for the gluonium 
decay occur in the same leading order of a $1/N_c$ expansion. 
Then the couplings in \eref{couplings_g} are of the same order
\beq
g_0\approx G_0.
\eeq
For the allowed solution of $r_G$ following $r_{G2}^+,r_{G3}^+$ we deduce
from the figure \ref{fig:ratiosra} the bounds $-1 \lesssim r_G \lesssim 1$ and with
condition \eref{mixingpositiv} from the $f_0(980)-f_0(1500)$ interference
\beq
0 \lesssim r_G \lesssim 1
\eeq
so for $f_0(1500)$ 
the gluonium contribution $\simeq G$ does not exceed the
quarkonium contribution $\simeq g$.
As an example, if we assume the mixing $\alpha=\phi_{sc}\approx 30^\circ$
as in route 2 ($\phi_{sc}\approx \phi_{ps}$) then we find 
the glue component in \eref{rsolution}
\beq
G/g\approx \tan\phi_G
\approx 0.7;\quad \phi_G\approx 35^\circ. \label{glumix1500}
\eeq 
The alternative solutions $r_{G1}^-,r_{G3}^-$ have typically 
opposite sign and are disfavoured by $\chi^2$.

\subsection{Production in $\gamma\gamma$ scattering}
The Belle Collaboration \cite{Uehara:2008ep} has measured the angular
distribution of the process 
$$\gamma\gamma\to \pi^0\pi^0$$
with high statistics. A partial wave analysis of the
$\pi^0\pi^0$ system in the $cms$ energy range $0.6\ \GeV \leq W_{\gamma\gamma}\leq 1.6\
\GeV$ has been carried out with $S$,  $D_0$ and $D_2$ waves included. As the 
photons are unpolarized some
ambiguities remain. A parameterization
in terms of resonances and backgrounds 
yields parameters for $f_2(1270)$ and in the $S$ wave 
for $f_0(980)$ and an additional resonance $Y$, either
$f_0(1370)$ or $f_0(1500)$. They obtain the
mass $m=1470^{+6+72}_{-7-255}$ MeV  and the total 
width $\Gamma =90^{+2+50}_{-1-22}$ MeV
and partial width $\Gamma_{\gamma\gamma}B(f_0(Y)\to\pi^0\pi^0) =
11^{+4+603}_{-2-7}\ \eV $.
The large systematic errors are related to uncertainties in the coupling of
helicity $\mu=0$ of $f_2(1270)$ and the $K\bar K/\pi\pi$ branching ratio of
$f_0(980)$. 
Mass and width are nicely compatible with PDG values for $f_0(1500)$
while for the alternative $f_0(1370)$ the width is smaller than
the PDG estimate $300-500$ MeV.

With the identification of the extra state as $f_0(1500)$ we can correct the above 
radiative width for other hadronic decay modes and obtain
\begin{equation}
f_0(1500):\qquad\Gamma_{\gamma\gamma}/\Gamma_{{\rm tot}} = 
95^{+34+519}_{-17-60}\ \eV, \quad
\Gamma^{f_0(1500)}_{\gamma\gamma}/\Gamma^{a_0(980)}_{\gamma\gamma}=
0.38^{+0.14+2.08}_{-0.07-0.24}
\label{radwidth1500}
\end{equation}
We may also compare with constituent models. Assuming a mixing as in $\eta$
according to route 2 with 
$f_0(1500)\sim (u \bar u +d\bar d-s\bar s)/\sqrt{3}$ we obtain the
ratio, following \cite{Minkowski:1998mf} 
\begin{eqnarray}
\Gamma^{f_0(1500)}_{\gamma\gamma}/\Gamma^{a_0(980)}_{\gamma\gamma}&=
\tfrac{32}{27}\times (m_{f_0}/m_{a_0})^p\ \eV \nonumber \\
   &=  0.53\ /\ 1.8 \ / \ 4.0\ \eV \quad \text{for}\quad
        p= -3\ / 1\ / 3 
\label{radwidth1500c}
\end{eqnarray}
There is some uncertainty in the extrapolation in mass from the $a_0(980)$ to
$f_0(1500)$. In Born approximation which is good for light pseudoscalars
the power in \eref{radwidth1500c} is $p=3$; such a 
power is also consistent with the radiative widths of the lightest tensor
meson nonet if the mixing angle $\alpha_T=81^\circ$ is used
\cite{Amslerquark}. Taking
only phase space with $p=1$ or a form factor with $p=-3$ 
we find the lower values in \eref{radwidth1500c}.
Given the large upper 
systematic experimental error there is no real contradiction
with the pure quark composition models, on the other hand, there is 
enough room for
gluonic contributions which would lower the $q\bar q$ predictions
accordingly.

\section{Gluon rich processes in comparison}
So far, the different routes for classification of scalar mesons have
not yet led to a generally accepted identification of a
light glueball. In route 1 and 3 there is the need for $f_0(1370)$ 
where a clear evidence is missing: there is no established
2-body decay and their total decay rates 
are strongly limited (10-20\%); the decay rates into higher multiplicities
are controversial. A good argument why 2-body decays should be strongly
suppressed in comparison to  $f_0(1500)$, for example, is lacking as well.
New experiments, for example COMPASS \cite{Abbon:2007pq} could reach higher
sensitivities in such studies.

In route 2, the asymmetry between $\sigma$ and $\kappa$ is a bit puzzling,
although not fatal; $f_0(500)$ may not be purely gluonic and
the mixing properties of isoscalars have to be better
understood, possibly from $B_d,B_s$ decays.

The existence of glueballs 
as a basic prediction of QCD is now dependent
on the existence of a supernumerous state in the classification 
which is found 
controversial. Therefore there was the hope to get an additional signal
from the enhanced production of glueballs in ``gluon enriched'' reactions.
Looking over the discussions of the previous sections, there is not such a
really striking predominance of a particular isoscalar state in the
gluon rich processes (i)-(iii) listed in subsection \eref{gluon-rich}.

The reason could be that the relevant energies are of the order of the
glueball mass in question itself in the range 1-2 GeV which may be too low 
for a gluon to appear as a well distinguished object. Rather, because of the
running strong coupling the presence of intermediate $q\bar q$ components 
at low energies may be
important as well. As an example, we consider
the double Pomeron process (ii) 
and then we investigate processes where the presence of a gluon as primary
agent is better established. 

\subsection{The Pomeron as an example for mixed gluon and quark components
\label{pomeronpartons}} 
\begin{figure}[h]
\begin{center}
\vspace{-5.5cm}
\includegraphics*[angle=0,width=8cm]{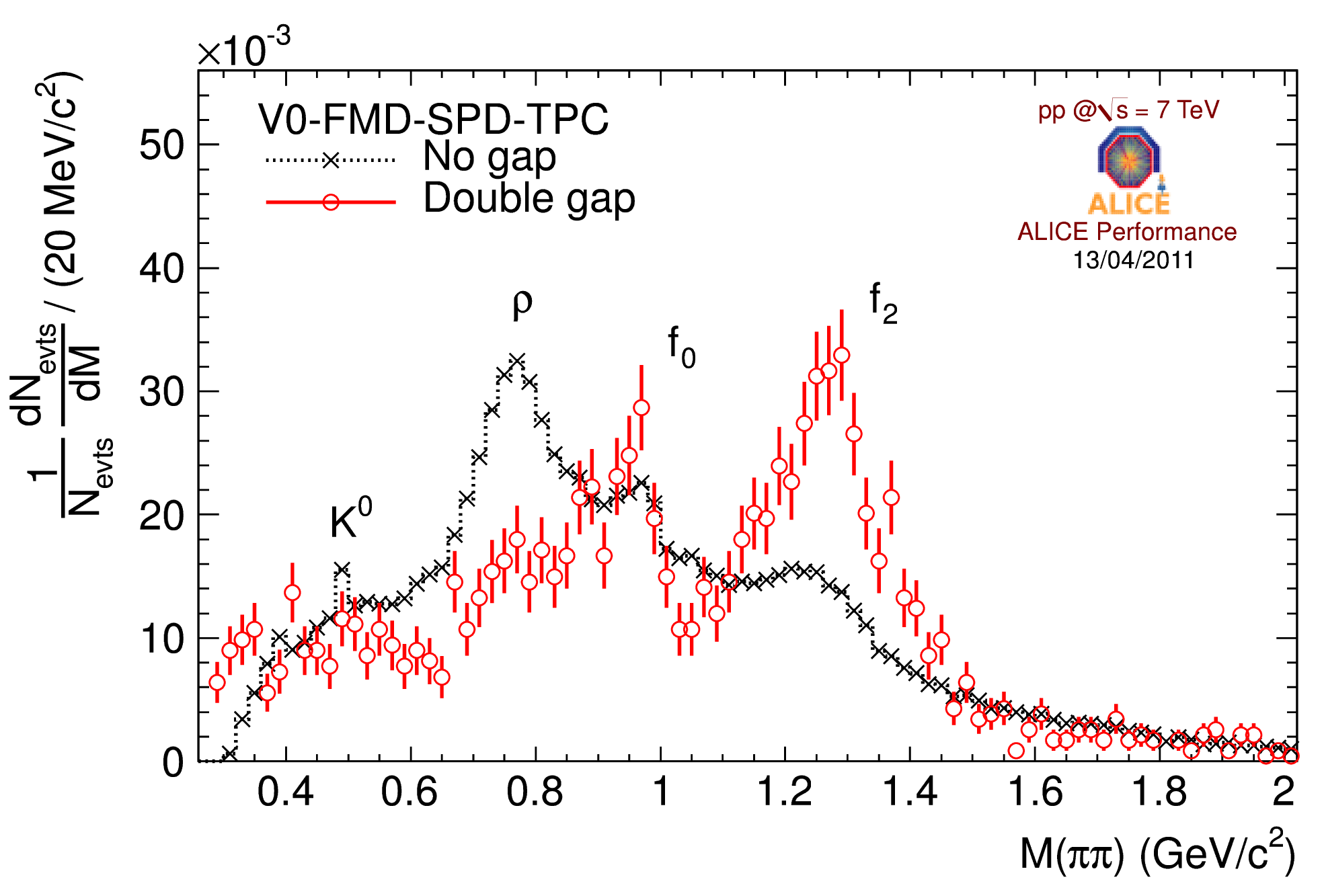}
\caption{
Central production of pion pairs at the LHC at 
7~TeV with and without gaps in pseudorapidity 
$-3.7<\eta<-0.9$ and $0.9<\eta<5.1$ (not corrected for acceptance),
figure from \cite{Schicker:2011ug} (ALICE Collaboration).}
\label{fig:centralalice}
\end{center}        
\end{figure}

At LHC energies of 7~TeV the double Pomeron process should dominate 
over Regge exchange processes for
sufficiently large rapidity gaps. The Pomeron has isospin $I=0$. If it is 
primarily a gluonic
object we expect the emergence of gluonic mesons in such processes.
A first result from the 
ALICE Collaboration with gap selections is shown in figure
\ref{fig:centralalice}.
While the $\pi\pi$ spectrum without gap selection 
shows the $\rho$ meson as dominant
structure, this state disappears if double gap events are selected. 
This is expected for the double Pomeron exchange process which cannot 
yield a final $\rho$ meson with isospin $I=1$.
Now, for the double gap events, 
the previously small structures related to $f_0(980)$ and $f_2(1270)$ become 
dominant. Certainly, the $f_2(1270)$ is a ``classical'' $q\bar q$ state
as is nicely confirmed by the 2-photon decay rates according to the quark
model (see, for example, \cite{Amslerquark}). 
The $f_0(980)$ is not typically considered as gluonic, though a
small gluonic component cannot be excluded (see section \eref{f0section}). 
Therefore, this 
result confirms that the double Pomeron process 
selects isoscalar states, but there is no evidence for the selection of 
gluonic states in particular. Such a behaviour may be related to a Pomeron,
which is constructed from mixed glueball and $q\bar q$ trajectories as 
it is realized within a non-perturbative QCD approach 
\cite{Kaidalov:1999de}.

In this context, it is interesting to recall the results from the studies of
deep inelastic scattering at
the HERA $ep$ collider by the 
H1 \cite{Aktas:2007bv} and ZEUS \cite{Chekanov:2009aa}
Collaborations on the partonic structure of the Pomeron 
as obtained from events with a rapidity gap adjacent to the
target proton.
At large momentum fraction $z$ of
partons inside the Pomeron the valence component is expected to dominate.
It is found, that for
$z\gtrsim 0.8$ (for all scales presented
with $Q^2\gtrsim 6$ GeV$^2$) the flavour singlet
quark density dominates over the density of gluons. In a perturbative QCD
treatment such dominant 
quark densities are initiated by Regge-like contributions
and evolve according to the QCD equations for 
the quark and gluon densities \cite{Martin:2006td}. Only in a global view, the
gluonic component dominates: 
the integrated gluon density is much bigger than the quark density and the
gluons carry a larger fraction of the total momentum as compared to the
quarks. 

\subsection{Processes with identified gluons}
In the gluon-rich processes (i)-(iv) the existence of gluons as intermediate partons is
required by theoretical arguments but at the low energy we cannot be sure
about the competing role of quark processes. 
We are therefore looking after processes involving higher gluon 
energies where the primary identity of gluons is better known. 

This is the case in the high energy production of quarks and gluons which
evolve into hadronic jets.
The identity of quark jets became obvious
in $e^+e^-$ annihilations above around $\sqrt{s}\sim 7$ GeV at SPEAR
\cite{Hanson:1975fe} whereas gluon jets have been identified
around 30 GeV at PETRA \cite{Brandelik:1979bd}. With increasing jet
energies, the quark and gluon jets can be better distinguished 
(see, for example \cite{Khoze:1996dn}, for a review).
At LHC the knowledge of parton distribution functions allows the calculation
of rates for quark and gluon jets and the selection of jets with a high
purity ($\gtrsim 80\%$) of either quark or gluon jets
\cite{Gallicchio:2011xc}. Then the fragmentation into particles of high
momentum fraction $z$ may yield information about gluonic mesons (see below in
section \ref{sectionleading}). 

Of special interest are exclusive 2-body
channels ($z\sim 1$) which can be more easily identified also at lower
energy and may carry a
primary quark or gluon. An example is the weak (``penguin'') 
decay of the heavy $b$ quark (iv)
$b\to sg$ which may contribute to $B\to K f_0 \to K \pi\pi$. This processes
have been studied in \cite{Minkowski:2004xf}. There are interesting tests
for the members of the scalar meson multiplets but the isolation of the
gluonic process has not been achieved so far.

A promising study concerns the $P$ wave charmonium (or bottomium) $\chi_c$ (or
$\chi_b$) decays (v) which proceed through $\chi\to gg$ with primary gluons 
(or other $C=+1$ heavy quarkonia) 
to which we turn in section \ref{symmetry_relations}.       
 
\section{Leading systems in gluon jets}
\label{sectionleading}

\begin{figure}[h]
\begin{center} 
\includegraphics*[angle=1.,width=5.0cm]{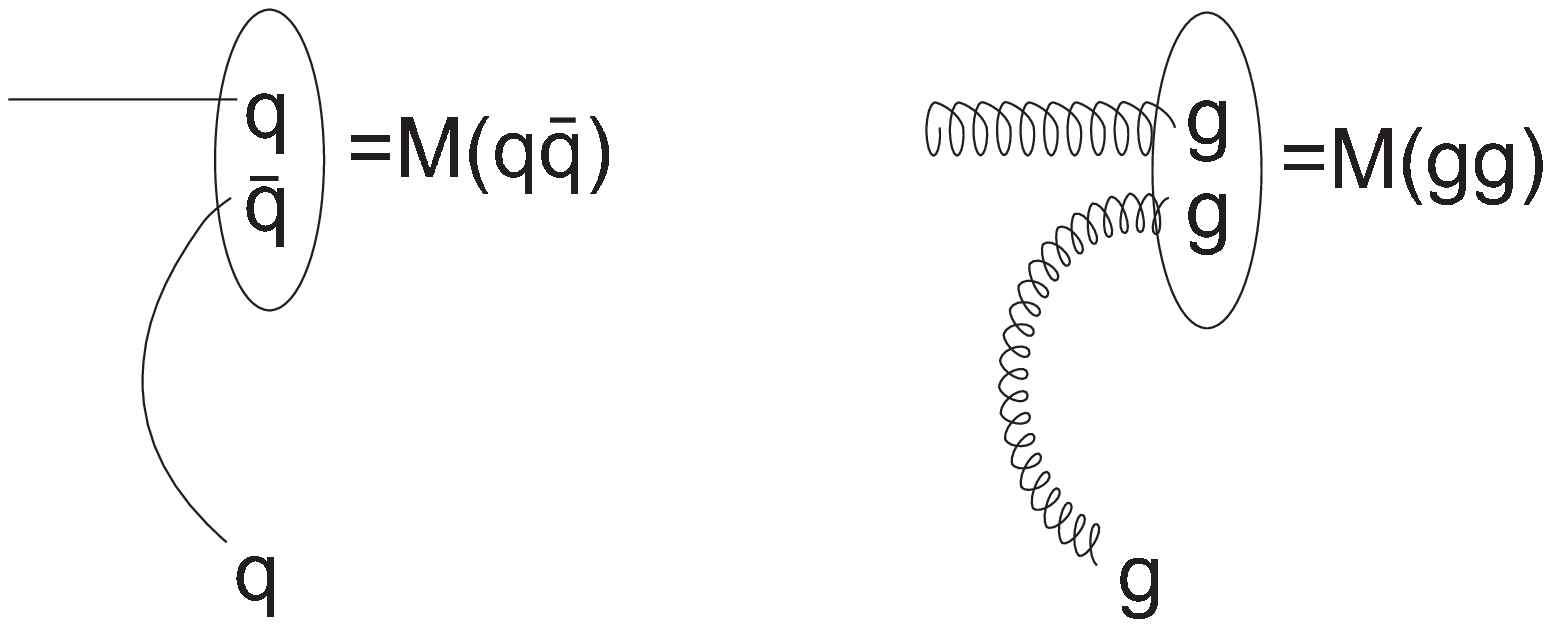}\\
\caption{Leading $q\bar q$ and gluonic particles in quark and gluon jets.}
\label{fig:leading}
\end{center}        
\end{figure}

A promising possibility to identify gluonic mesons is provided
by the comparative study
of leading particle systems in quark and gluon jets 
(see \cite{Ochs:2011cw} for a recent discussion).
According to the well known concept of quark fragmentation the leading 
particles at large momentum fraction ``Feynman $x$'' are those which carry
the primary quark of the jet as a valence quark, see figure \ref{fig:leading}.
For example, leading
particles in a $u$ quark jet are $\pi^+(u\bar d)$ or 
$\pi^0(\{u\bar u+d\bar d\}/\sqrt{2})$ with half strength 
whereas $\pi^-$ or $K^-$ without valence $u$ quark are suppressed
at large $x$. In analogy, one can consider the fragmentation of a gluon and 
hypothesize that the leading particle  has a gluonic valence component.

Models of this kind with leading glueballs in jets, and also with leading isoscalars
like $\eta,\eta',\omega$ at large $x$ have been suggested already long
ago \cite{Roy:1978pz}, $x$-distributions have been considered in 
\cite{Roy:1999bu}. The model for isoscalar $q\bar q$ meson production
has been studied at LEP with quark and gluon jets \cite{Acciarri:1995yp}, 
but no clear experimental support for the model has been 
established \cite{Ochs:2011cw}. This means that the isoscalars $\eta$ and
$\eta'$ behave like other $q\bar q$ mesons in fragmentation.

\begin{figure}[t]
\begin{center} 
\includegraphics*[width=6.0cm]{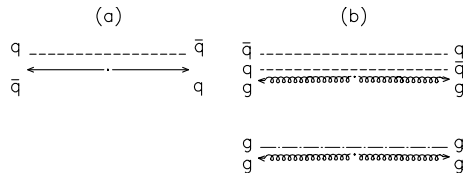}\\
\caption{Colour neutralization in quark and gluon jets (figure from
\cite{Minkowski:2000qp}).}
\label{fig:colour-neutralisation} 
\end{center}        
\end{figure}

A critical test for the presence of gluonic objects in the leading system of
a jet is the measurement of charge and mass of the leading cluster
$Q_{leading}$ and $M_{leading}$ beyond a rapidity gap.
Their distribution is expected to reflect the mechanism for colour 
neutralization~\cite{Minkowski:2000qp}. In a non-perturbative approach
an initial pair of $q\bar q$ with colour triplet charges 
creates another $q\bar q$ pair to neutralism the colour charges beyond the
confinement radius $R_c$. For an initial $gg$ pair the colour may be
neutralized either by another $gg$ pair (``colour octet neutralization'')
or by creation of two $q\bar q$ pairs (``colour triplet neutralization''),
see figure~\ref{fig:colour-neutralisation}. 
In a high energy jet one expects the charge distribution of the leading
cluster to approach with increasing rapidity gap a limiting behaviour with
charge $Q_{leading}=0$ for colour octet ($gg$)  
and charge $Q_{leading}=0,\pm1$ for
color triplet ($q\bar q g$) neutralization according to the minimal configuration of
partons beyond the gap. The mass distribution of the leading
system with $Q_{leading}=0$ should signal the presence of glueballs if any.

\subsection{Rate of neutral clusters at LEP above Monte Carlo expectations}

\begin{figure}[h]
\begin{center} 
\includegraphics*[width=7.5cm,bbllx=0.0cm,bblly=0.0cm,bburx=17.0cm,%
bbury=18.5cm]{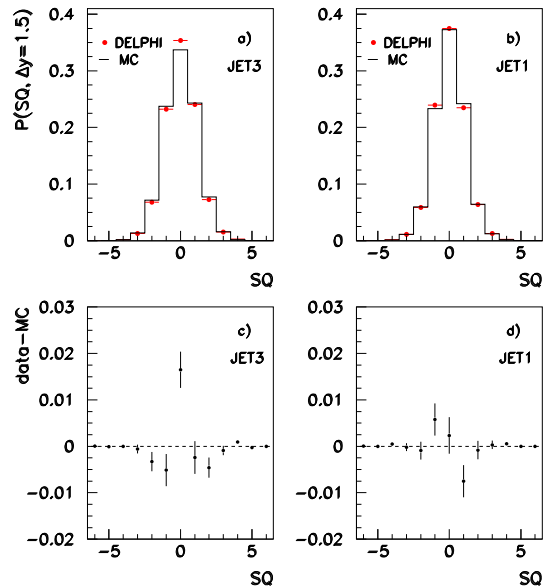}%
\hspace{0.1cm}
\\
\caption{Distribution of the sum of charges ``SQ'' (``	leading charge'')
 beyond a rapidity gap
of $\Delta y=1.5$ in gluon dominated (jet3) and quark dominated (jet1) jets
in comparison with Monte Carlo 
(JETSET); analysis by DELPHI, showing the surplus
of $SQ=0$ events in gluon jets (figure from \cite{Abdallah:2006ve}).
}
\label{fig:leadingcharge}
\end{center}        
\end{figure}

The distributions of charge and mass 
for the leading clusters in quark and gluon jets have been measured in
$e^+e^-$ annihilation at LEP by the DELPHI \cite{Abdallah:2006ve},
OPAL \cite{Abbiendi:2003ri} and ALEPH
\cite{Schael:2006ns} Collaborations. The results by DELPHI
on the distribution of the leading charge
beyond a rapidity interval $\Delta y=1.5$ are presented in 
figure~\ref{fig:leadingcharge}. For the most energetic jet in a 3-jet event
(``jet1'', in 90\% of cases a quark jet) the distribution of the leading
charge (sum of charges ``SQ'') follows closely the Monte Carlo expectation
(JETSET~\cite{Sjostrand:1993yb}). This numerical calculation 
is based on the QCD approach
to parton cascade evolution and string hadronization after tuning free
parameters by experimental data. 
On the other hand, for the jet of lowest
energy (``jet3'', a gluon jet with purity 70\%) there is a significant
excess of events with $Q_{leading}=0$, clearly visible in the lower figures
for the difference ``data-MC''. 
This excess is also shown in \cite{Abdallah:2006ve} 
to be built up by increasing the rapidity gap $\Delta y$.

\begin{figure}[h]
\begin{center} 
\includegraphics*[width=4.7cm,bbllx=0.0cm,bblly=0.0cm,bburx=19.0cm,%
bbury=20.0cm]{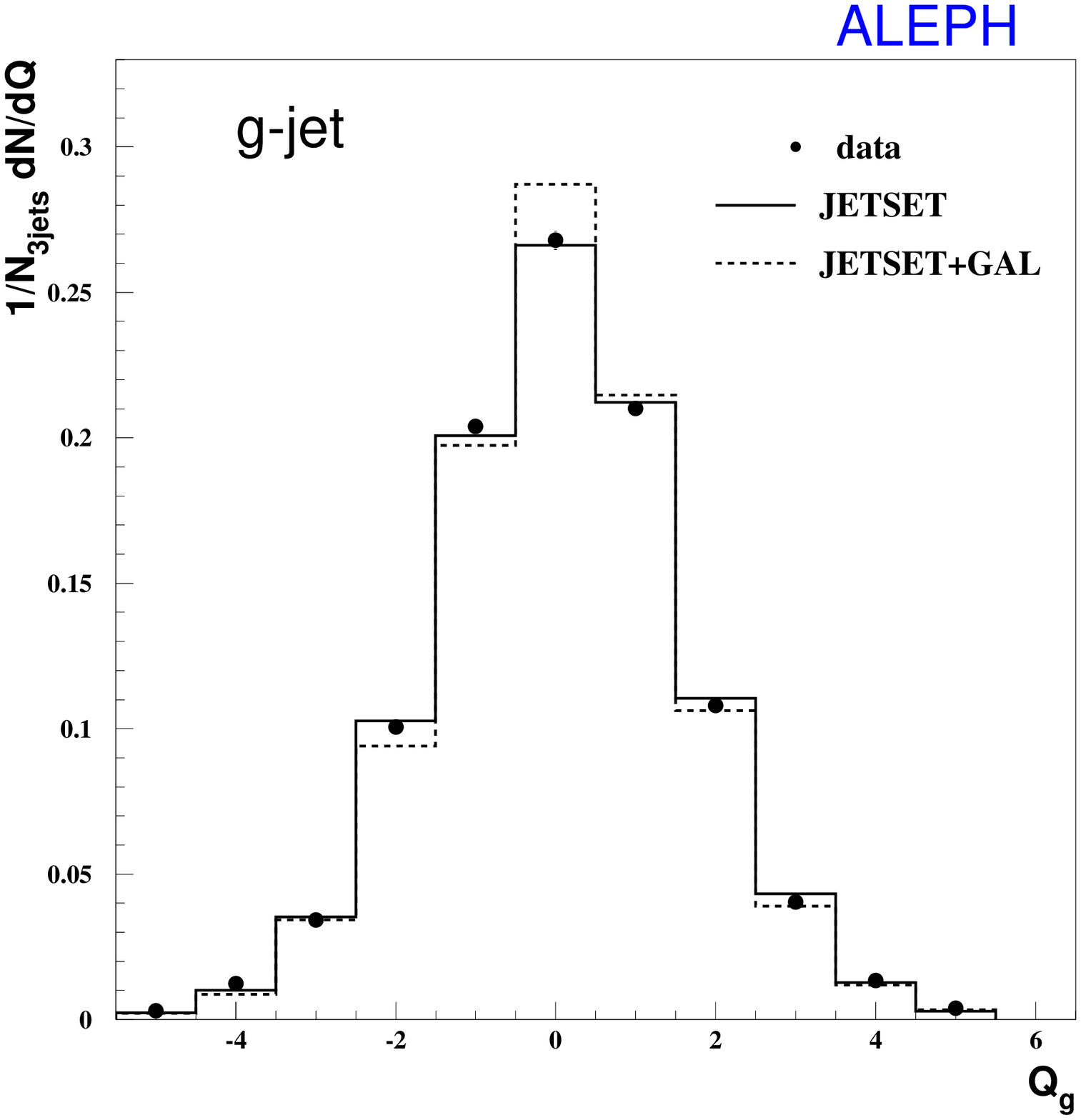}%
\hspace{1cm}
\includegraphics*[width=4.7cm,bbllx=0.0cm,bblly=0.0cm,bburx=19.0cm,%
bbury=20.0cm]{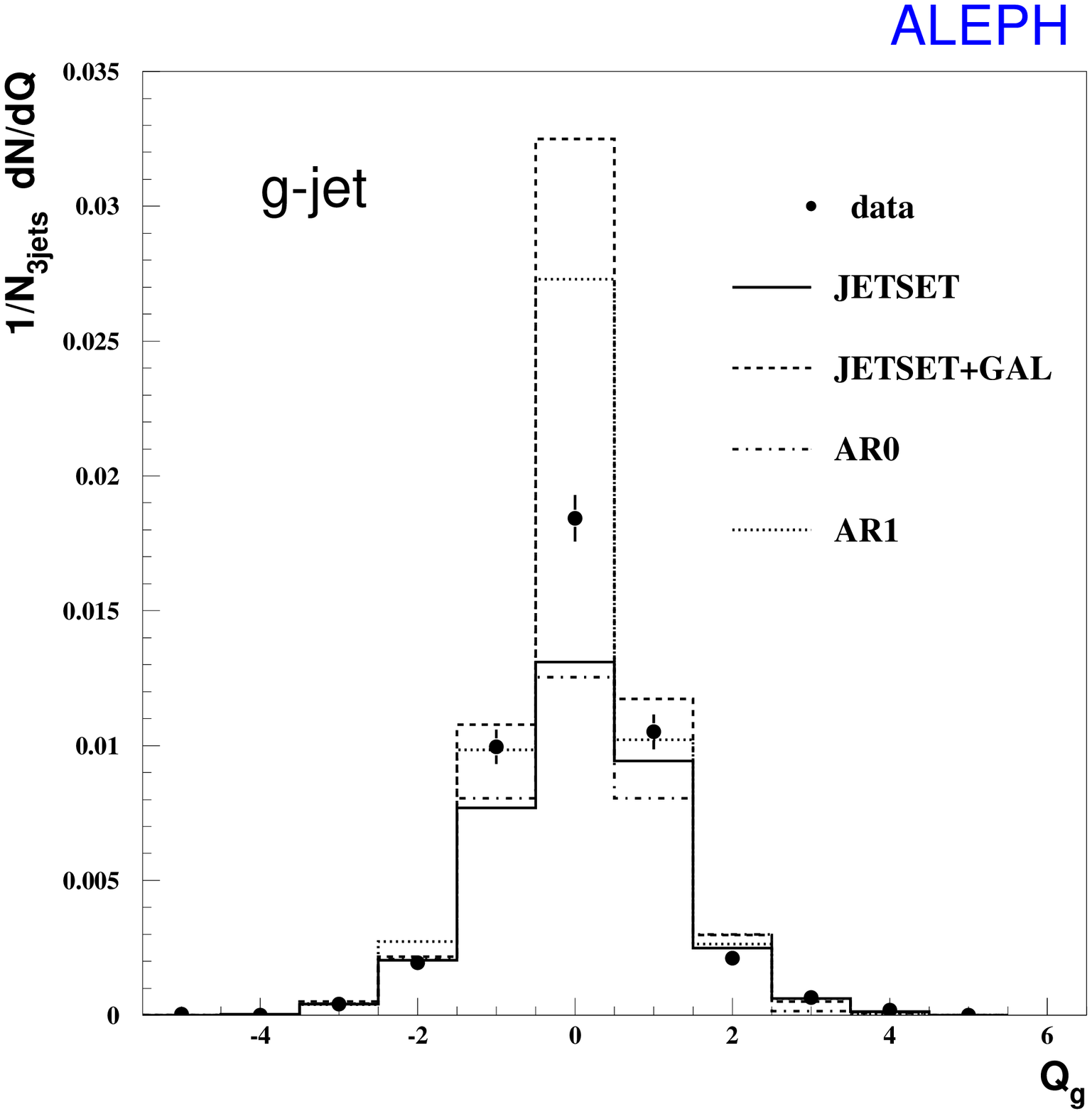}\\
\caption{\label{fig:gapaleph}Left: distribution of the leading charge $Q_{leading}$
in a gluon jet of high purity without gap; right: the 
same for gluon jets with rapidity gap $\Delta y=1.5$ showing the surplus
of events for $Q_{leading}=0$ over the JETSET Monte Carlo;
other models (GAL, AR) with ``colour reconnection'', figure from
\cite{Schael:2006ns} (ALEPH Collaboration).}
\end{center}        
\end{figure}

In the analysis by ALEPH \cite{Schael:2006ns} gluon jets of high purity have
been selected from the 3-jet events $b\bar b g$ with identified $b$-quark
jets.
The distribution of charge
has been studied in gluon jets without gap and in gluon jets with a
rapidity gap. As can be observed in figure \ref{fig:gapaleph}, the charge
distribution in the gluon jet without gap follows the MC calculation 
(JETSET) whereas
there is a large excess by about 40\% in the jet with a rapidity gap.
Models with non-standard colour connections along the parton cascade
fail to reproduce the data as well. 

\begin{figure}[t]
\begin{center} 
\hspace{1.5cm}
\includegraphics*[width=7.0cm,bbllx=0.0cm,bblly=0.0cm,bburx=18.5cm,%
bbury=18.5cm]{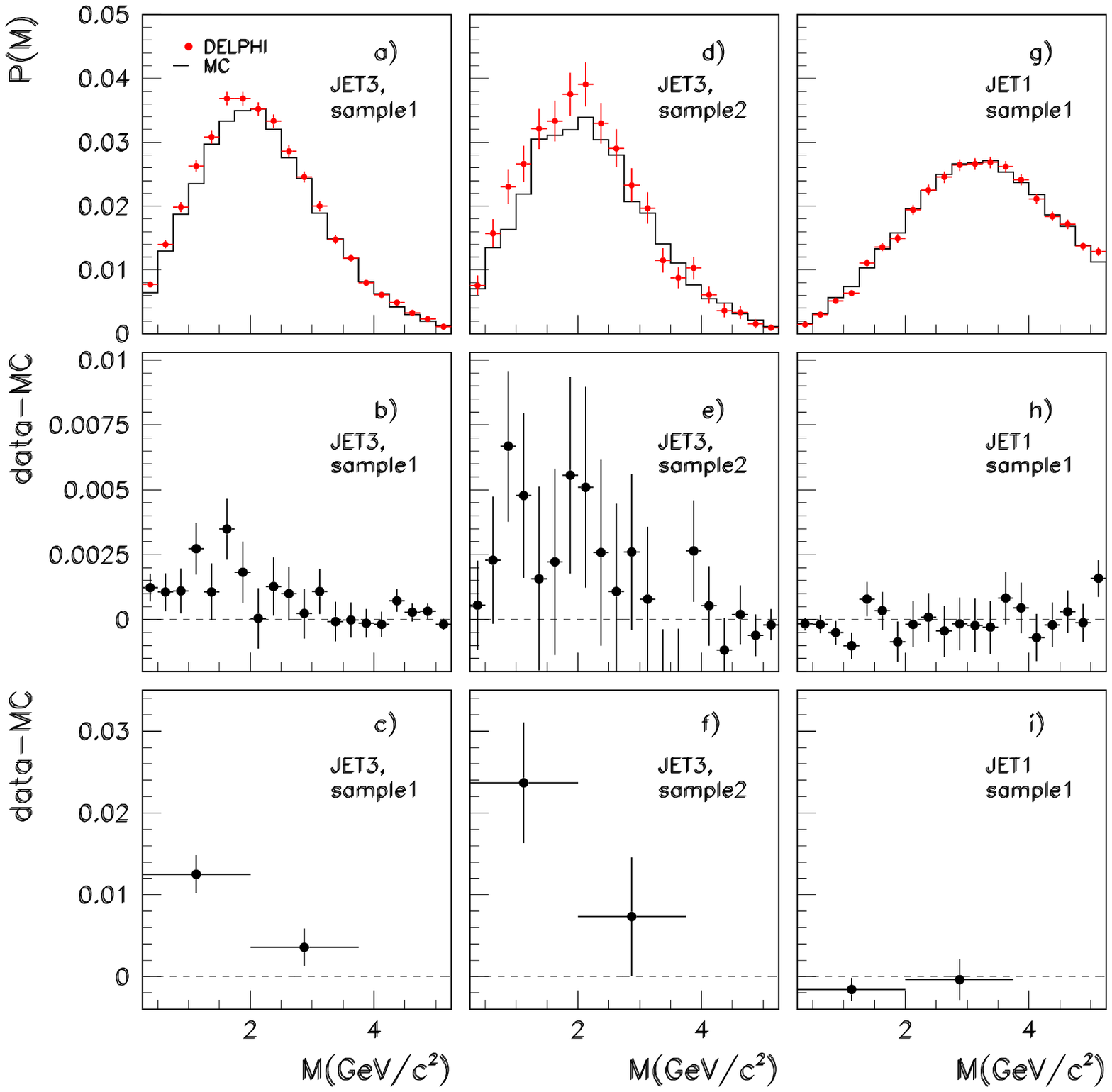}
\includegraphics*[width=6.5cm,bbllx=0.0cm,bblly=3.5cm,bburx=19.0cm,%
bbury=25.0cm]{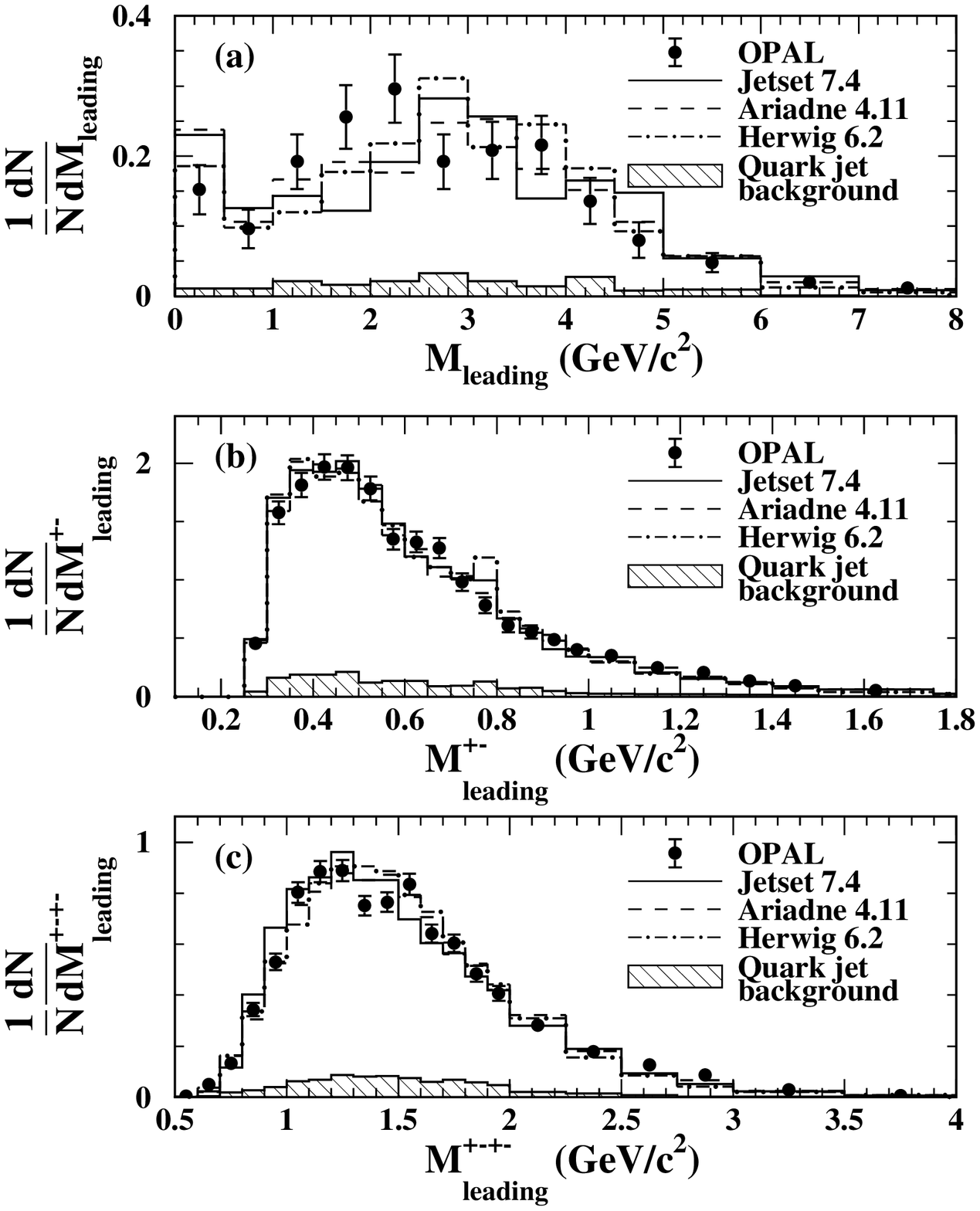}\\
\caption{Mass density distributions for the leading cluster $M_{leading}$
of charged and neutral particles;
left: DELPHI results for gluon jets (JET3 of lowest energy in 3 jet
events (sample 1) and in $b\bar b g$ events (sample 2)) and for quark jets
(JET1of highest energy), all with charge $Q_{leading}=0$ (figure from
\cite{Abdallah:2006ve}); 
right: OPAL results for all mass combinations
and for clusters with charges ($+-$) and ($+-+-$), 
figure from~\cite{Abbiendi:2003ri}. }
\label{fig:leading-mass}
\end{center}
\end{figure}

Finally, the distribution of masses of the leading cluster in figure
\ref{fig:leading-mass}, left side, shows again a good agreement with the
predictions from the JETSET Monte Carlo for quark jets (right column), 
whereas for gluon
jets the excess of events with leading charge $Q_{leading}=0$ is 
observed at small masses, mainly for $M\lesssim 2$ GeV. An excess of events 
at small masses is also observed by OPAL (figure \ref{fig:leading-mass},
right hand side). The mass spectra for selected charges with $Q_{leading}=0$ 
don't indicate striking resonant structures as expected for narrow glueballs,
a small enhancement around 1500 MeV in the ($+-+-$) spectrum is
indicated while for $(+-)$ the mass distribution is smooth. Interestingly,
there is no $\rho$ signal which the models seems to predict; this could
indicate a special feature of gluon jets which may not just be built from a
long $q\bar q$ colour triplet string.

While the study of spectroscopy did not provide a definitive evidence for a
gluonic component in the spectrum, the inclusive production of the leading
clusters in a gluon jet appears to be different from what is expected within
a model which reproduces otherwise 
the main feature of the multiparticle final states
in $e^+e^-$ annihilations. The lack of better signals from resonances may
indicate that in the selection procedure for rapidity gaps  particles
from a resonance have been lost. Also the rapidity gap with $\Delta
y\simeq 1.5$ may be too small for significant results.  

\subsection{Proposal for further measurements at the LHC}
There are some advantages to follow this line of studies at $pp$ colliders,
especially at the LHC \cite{Ochs:2011cw}:
\begin{itemize}
\item there is access to jets of much higher energies in comparison to LEP,
therefore larger rapidity gaps are available with better separation of the
leading cluster;
\item quark and gluon jets can be analyzed at comparable energies in the
same experiment;
\item the gluon jet sample could be much larger than before.
\end{itemize}
At $pp$ colliders quark and gluon jets can be produced in $2\to 2$ processes
of leading order: quark jets in $pp\to \gamma+jet+X$ with subprocess 
$qg\to \gamma q$ dominant at the lower $p_T$ and
gluon jets in di-jet events also at the smaller $p_T$. The rates can be
calculated from parton distribution functions
and parton-parton cross sections, the purities depend on
primary energy and $p_T$, for some sample calculations see table~\ref{tab:jet-rates}.
For quark jets a purity of 80\% seems sufficient for the study of leading
systems (quarks fragment harder than gluons), for gluons a higher purity
of $> 90\%$ is desirable.
\begin{table}[t]
\caption{\label{tab:jet-rates}
Purities (in\%) of quark and gluon jets at Tevatron \cite{Acosta:2004js}
and at LHC \cite{Gallicchio:2011xc}, taken from \cite{Ochs:2011cw}.}
\begin{indented}
\item[]\begin{tabular}{@{}lrcccc}
\br
           &       & $p_T$ & $x_T$ & g in di-jet & q in $\gamma + $ jet\\
\hline
Tevatron  & 1.8 TeV & 50 & 0.056 & 60\% & 75 \% \\
LHC    & 7 TeV & 200 & 0.057 & 60\% & 80 \% \\
                     &       & 50  & 0.014 & 75\% & 90 \% \\
                     &       & 800 & 0.229 &  25\% &75 \%  \\
\br
\end{tabular}
\end{indented}
\end{table}
This can be achieved by triggering on large total transverse energy and the
selection of 3-jet events. Similar to the case of $e^+e^-\to 3\ jets$ the
lowest momentum jet is most likely a gluon jet from QCD Bremsstrahlung
($qqg,\ qgg$ or $ggg$). Estimates suggest that purities of $95 \%$ for gluon
jets are within reach \cite{Ochs:2011cw}. 
Studies at LHC could be important in two directions:
\subsubsection{Leading clusters with larger rapidity gaps.}
The energies of identified quark and gluon jets at the LHC 
could be larger by an order
of magnitude as compared to LEP (typically $<25$ GeV) and the rapidity gaps
could extend up to $\Delta y\sim 4$. Then, the leading charges are expected
closer to their asymptotic values $Q_{leading}=0,\pm 1$ with a better
sensitivity to an extra component with  $Q_{leading}=0$.

\subsubsection{Identifying gluonic mesons through 
$[q\bar q]$-reference $x$-spectra.}
The observed mass spectra of the leading cluster beyond the rapidity gap 
did not show prominent resonance structures. As the rapidity gap selection
cuts into the angular distribution of the cluster decay particles 
the resonance signal
may be reduced. In an alternative approach, the mass spectra of particle
clusters $M(\pi\pi),\ M(K\bar K),\ M(4\pi)\ldots$ are measured
at a given total energy fraction $x$ of the cluster and the 
$x$-distribution of resonances is determined. Such resonance $x$-spectra 
have been determined in the past \cite{Boehrer:1996pr},
but not separately for quark and gluon jets.

The identification of gluonic mesons $gb$ by the classification as
 ``leading'' and
``non-leading'' 
appears feasible by the comparison of 
the $x$-spectra $D^{gb}(x)$ with the spectra of reference $[q\bar q]$
mesons of similar
mass $D^{[q\bar q]}(x)$. 
In comparison with the ``leading'' spectrum like $u\to \pi^+$ a
gluonic meson should be suppressed
\beq
D_u^{gb}\ll D_u^{\pi^+}\ \text{for large}\ x. 
\eeq
Similarly, a gluonic resonance $f_0(gb)$ should be suppressed in a quark jet
as compared to a
reference $[q\bar q]$ resonance and, vice versa, 
a $[q\bar q]$ meson should be suppressed in a gluon jet, as summarized in table
\ref{tab:resonanceprod} (for tetraquark mesons we assume here a steeper
fragmentation as for $[q\bar q]$ mesons according to the counting rules
\cite{Brodsky:1973kr}). 
\begin{table}[h]
\caption{\label{tab:resonanceprod}
Mesons are leading or suppressed in fragmentation $x$-spectra depending on
whether they carry the jet initiating parton ($q$ or $g$) 
as a valence constituent; gluonic mesons are detected by comparison to
reference $q\bar q$ mesons}
\begin{indented}
\item[]\begin{tabular}{@{}lllll}
\br
meson&          & quark jet & gluon jet      &  \\
&               &           & triplet neutr. & octet neutr.\\
\hline
$[q\bar q]:$ &$ f_0\ \{ref:\rho,f_2\}$ & {\underline{leading}} & 
                            suppressed & suppressed\\
$gb: $ & $    f_0\ \{ref:\rho,f_2\}$   & suppressed&suppressed&{\underline{leading}}  \\
$[4q]:$&$ \sigma, f_0(980)$ (?)  & suppressed & suppressed & suppressed\\
\br
\end{tabular}\\
\end{indented}
\end{table}
As an example, for a gluonic scalar meson $f_0(gb)$ decaying to $\pi\pi$ 
(candidates $f_0(500),\ f_0(980),\ f_01500)$) we can take
$\rho(770)$ and $f_2(1270)$ as reference $[q\bar q]$ states. Then, 
\beq
\text{at large}\ x:\quad D^{f_0(gb)}_q(x)\ll D^{\rho,f_2}_q(x);
\quad D_g^{f_0(gb)}\gg D^{\rho,f_2}_g(x).
\eeq
As found by the OPAL Collaboration \cite{Ackerstaff:1998ue} 
the $x$-distribution of $f_0(980)$ and  $f_2(1270)$ 
almost coincide, then there is 
no evidence so far for any 
constituent structure beyond $q\bar q$ of $f_0(980)$ from this study; a similar 
measurement of the corresponding $x$ spectra in gluon jets 
could reveal a gluonic admixture of $f_0(980)$ if there appeared a hard
component at large $x$. In this way, also mixed $gb-q\bar q$ states could be
recognized. 

Gluonic components could appear in the spectra of $ (\pi\pi)^0$ from
$f_0(500)/\sigma,\ f_0(980),$ $f_0(1500)$; in $(4\pi)^0$ from
$f_0(1370) (?), f_0(1500)$ and in $(K\bar K)^0$ from $f_0(980),\ f_0(1500)$
and $f_0(1710)$. As an example, if $f_0(1500)$ has a large glueball
component, it should fragment harder than $\rho$ or $f_2$ in a gluon jet
and softer in a quark jet. 

\section{$[q\bar q]$ multiplets and glueballs from symmetry relations: \\
first results from $\chi_c$ decays}
\label{symmetry_relations}

In this section we investigate the exclusive production of hadronic states
$H_1,H_2$ in decays of heavy Quarkonia with $C=+1$ 
through the intermediate process
\beq
[Q\bar Q]\to gg \to H_1H_2,
\eeq
in particular in decays of  $\chi_{c}$ or $\eta_c$ ($\chi_{b}$ or $\eta_b$). 
This decay  could be viewed
as fragmentation of gluons $g\to H_1,H_2$  with energy fraction $z=1$
with sensitivity to glueball production assuming that the mass
of $\chi_c$ is high enough for the $gg$ intermediate state to dominate.
Here, we explore in particular the symmetry relations for the set of 
decay rates if all $H_1$ and
$H_2$ are taken from $[q\bar q]$ multiplets $M_1$ and $M_2$ respectively. 
These relations follow
from the quark states being produced as flavour singlet. 

A nice example is given by the results in table
\ref{tab:chidecay} on $\chi_{c}$ decays 
where the properly weighted decay rates into pairs of
pseudoscalars follow a simple equipartition pattern. The same pattern 
should
be followed by all pairs of mesons $H_1,H_2$ taken from the same multiplets
$M_1,M_2$ each, i.e.
for the branching ratios $B$ into the members of two nonets 
after summing over charges
\beq
B:\quad [Q\bar Q]\to \pi_1\pi_2: K_1 \bar K_2: \bar K_1 K_2: \eta_1 \eta_2 : 
  \eta'_1 \eta'_2
 = 3: 2:2:1:1
\eeq
with obvious notation.
In this way, the classification of hadrons into multiplets can be 
tested. At the
same time any additional gluonic component should become visible by a
deviation from the symmetry relations. This will be discussed next for the
scalar multiplet. 
 
\subsection{Scalar multiplet according to route 2} 
In this case, the scalars $a_0(980),\ f_0(980),\
K^*_0(1430)$ and $f_0(1500)$ belong to the same multiplet. For three
channels results on $\chi_{c0}$ decays 
have been obtained by BES2 \cite{Ablikim:2004cg,Ablikim:2005kp} which we
list in table \ref{tab:chiscalar} (taking the values from PDG 
\cite{beringer2012pdg}) and we correct for unseen channels.
We take into account furthermore only corrections to the rates
due to phase space, but no corrections from quark masses or additional
dynamical form factors according to our findings on $\chi_c$ decays above
within the accuracy of around 10\%.
\begin{table}[h]
\caption{\label{tab:chiscalar} Branching ratios $B$ of $\chi_{c0}(3415)$ 
($0^{++}$)
into pairs of scalars $K^*_0(1430)^0$, $f_0\equiv f_0(980)$,
$f_0'\equiv f_0(1500)$ and baryons for comparison (data from PDG  
\cite{beringer2012pdg})
(i)~in observed channels, (ii) corrected for unseen decays (isospin, also
$B(f_0(980)\to \pi\pi)=0.75\pm 0.12$ \cite{Ablikim:2005kp} and $B(f_0(1500)\to \pi\pi\&K\bar K$
from PDG); (iii) $B/(cq)$ with charge 
weight $c$ and
momentum $q$ after rescaling to the value for $K^*_0(1430)^0
\bar{K^*_0}(1430)^0$,
together with the resp. theoretical
expectations for process $gg$ as in table \ref{tab:chidecay}.}
\begin{indented}
\item[]\begin{tabular}{@{}lccccc}
\br
 &$K^*_0 \bar{K^*_0}$ & $f_0f_0$ & $f_0'f_0'$ & $p\bar p$ 
     &$ \Lambda\bar \Lambda$ \\
weight $c$ &2&1&1&& \\ 
\mr
 & ($\pi^\pm K^\mp$) & ($\pi^\pm\pi^\mp$) & ($\pi^\pm K^\mp$) &
 & \\
\vspace{0.1cm}
(i)\ \ \ $B\times 10^3$ obs & $0.99^{+0.40}_{-0.29}$ & $0.67\pm0.21$ &$
<0.05$ &  \\
\vspace{0.1cm}
(ii)\ \ $B\times 10^3$ corr &$ 2.23^{+0.90}_{-0.65}$ &$2.68\pm1.21$ & $<5.0 $&$
0.223\pm0.013$ &  $0.33\pm 0.04$\\
(iii)\ $B/(cq)$ resc& $1.0^{+0.4}_{-0.3}$ &$ 1.64\pm0.73$ &$ <5.2$ & &  \\
\mr
expect (R=1.0) & 1. & 1.& 1.&   \\
\br
\end{tabular}
\end{indented}
\end{table}

The normalized decay rate for $f_0f_0$ is found slightly higher but
compatible with the rescaled $K^*_0 \bar K^*_0$ rate ($1.64\pm0.73$ vs. 1.0), as expected for members of the
same multiplet; $K^*_0(1430)$ is generally accepted as $q\bar q$ state.
An increased rate for decay into pairs of $f_0$'s could be indicative of 
an additional glueball component. 
The experimental uncertainties are still rather large for definitive
conclusions.
It is interesting to note the similarity of the two decay rates
\begin{eqnarray}
B(\chi_{c0}\to f_0(980)f_0(980))&= (2.68\pm 1.21) \times 10^{-3}
 \nonumber\\
    B(\chi_{c0}\to \eta'(958)\eta'(958))&= (2.02\pm0.22) \times 10^{-3}
\label{etaprimef0}
\end{eqnarray}
which again is in line with the two states being parity partners as in route 2
with the same $q\bar q$ mixing; the similar decay rates in 
$J/\psi\to (\eta',f_0(980)+V$) with
$V=(\phi,\omega)$ have been noted already in \cite{Minkowski:1998mf}
(see section \ref{route2}). 
The observation \eref{etaprimef0} for a gluonic decay indicates that not
only the quark structure but also the gluonic components (if any) 
are similar in $\eta'$ and
$f_0(980)$. A higher accuracy of the $f_0$ rates are certainly 
desirable.

A critical test for the existence of 
a scalar $q\bar q$ multiplet along route 2 would be
the observation 
\beq
B(\chi_{c0}\to a_0(980)a_0(980)):
B(\chi_{c0}\to K^*_0(1430) \bar K^*_0(1430))\ = 3:4
\eeq
if all charge states are counted, or 1:1 for single charge states.
The rates for $f_0f_0$ and $f_0'f_0'$ 
could possibly deviate from the symmetry expectations because of 
gluonic components. 

The glueball should appear through
a signal from  $\chi_{c0}\to f_0(500)f_0(500)$ where $f_0(500)/\sigma$ is
the broad object peaking at 1 GeV. The data in $\chi_{c0}\to 4\pi$ by BES 
\cite{Ablikim:2004cg} show indeed a considerable broad $\pi\pi$ background. 
Besides
$f_0(980)$ there is some enhancement between 1200 and 1500 MeV which could
get contributions from $f_2(1270)$ and $f_0(1500)$ interfering destructively
with the background so as to yield a drop of the mass spectrum around 1500
MeV. Whether this background corresponds to the broad
$f_0(500)$ could possibly be found out by observing the interferences with the
narrow states $f_0(980)$ and $f_0(1500)$.

In the decay of $\chi_{c2}$ \cite{Ablikim:2004cg} a strong signal of
decay into $\rho\rho$ shows up. In addition, there appear some bands around
$m_{\pi\pi}\sim 500$ MeV; this could be a signal from 
$\chi_{c2}\to f_0(500)f_0(500)$
where the mass spectrum is squeezed towards low masses because of the
threshold factor $p^5$ in cms momentum $p$. This could be a signal from
the decay into glueballs.  

If $f_0(500)$ has a high gluon component its pairwise production 
should be much favoured over the pairwise 
production of $K^*_0(800)/\kappa$ in $\chi_{c0}$ and $\chi_{c2}$
decays.  

\subsection{Scalar multiplet according to route 1 - tetraquark
expectations}
In this case
$\sigma,\kappa, f_0(980)$ and $a_0(980)$ are included 
in the lowest mass multiplet. 
If they are built from diquarks the decay would proceed according to the
diagram in figure \ref{fig:gbdecay}a
but with production of diquarks instead of quarks (a contribution from
figure \ref{fig:gbdecay}b has been found negligible from the study of table 
\ref{tab:chidecay}). In this process at cms energy of 3.5 GeV 
a formfactor for the diquark should play a role. This is actually observed
for the production of baryon pairs which are produced from a primary 
diquark and a primary
quark pair. These rates ($p\bar p,\Lambda\bar \Lambda$) 
are seen in table \ref{tab:chiscalar} to be an order
of magnitude smaller than the rates for $[q\bar q]$ scalars like $K^*_0\bar
K^*_0$. One would therefore expect
that scalars made up from two diquarks would be produced with a smaller rate
yet. This appears to contradict the large rate of $f_0f_0$ vs. $K^*_0\bar
K^*_0$ in table \ref{tab:chiscalar}. 

Furthermore for route 1, one should observe 
the decays $\sigma\sigma$ and $\kappa\kappa$ with the same
rate as $f_0f_0$ and $a_0a_0$ for particular charge states. In their
analyses the BES Collaboration finds indeed a signal from $\kappa\kappa$
of the order of $K^*_0\bar K^*_0$; they see signals with $\sigma$ 
but no signal from $\sigma\sigma$ where
they take $\sigma$ according to the observation in $J/\psi\to \omega \pi\pi$ 
with a peak at low mass. The results on $\sigma$ and $\kappa$ are considered
finally as too uncertain to be transformed into decay rates.
In $\chi_{c2}$ decays there seems to be the channel $\sigma\sigma$ as
discussed above, but there is no evidence for the channel $f_0(980)f_0(980)$.
It will be interesting to see, whether the branching ratios follow 
\beq
B(\chi_c\to \kappa\kappa): B(\chi_c\to \sigma\sigma)= 4:1 
\eeq
summed over all charge states, which is a result opposite to route 2 above.

We conclude that the further study of $\chi_c$ decays with higher statistics
(or even of $\chi_b$ decays) 
has a high potential
to clarify the structure of the lowest scalar multiplet and the potential
presence of gluonic mesons. The same, of course, is also true for 
multiplets with other quantum numbers, for example heavier pseudoscalars. 

\section{Summary and conclusions}
In a summary of our results we try to answer some questions and we provide
a minimal mixing scenario for the scalar mesons with glueball which takes
into account some relevant findings.

\subsection{Is there any experimental evidence for the existence of glueballs?}
We have studied different scenarios for a scalar meson spectroscopy
which includes the lightest glueball of QCD. None of the schemes 
considered is without
problem. It remains unsatisfactory, if the existence of the glueball as a
fundamental prediction of QCD, at the end depends on whether such a
controversial state like $f_0(1370)$ does exist. We have therefore looked
after other criteria as well.

There appears to be one observation which can be considered as a direct hint
for the existence of glueballs. That is the observation of several LEP
experiments that the leading system in a gluon jet has an excess of neutral
charge, in one measurement by 40\%, if compared with a standard Monte
Carlo program. There is a new chance at LHC to follow this line of
investigation further by obtaining a larger sample of gluon jets 
with higher energy to study jets with a larger rapidity gap. This should
make the effect clearer and could possibly give a hint towards the relevant
glueball mass. Future studies are in a ``win-win'' situation: either they
provide better evidence for gluonic mesons, or else they show a deficiency
of the ``standard Monte Carlo's'' with their hadronization models 
which would be important to understand as
well, as these Monte Carlos are the backbone of the phenomenological 
studies with jets.

\subsection{Are there supernumerous states in the scalar sector?}
This question is related to the selection of scalar nonets.\\
i) Extra state $f_0(1370)$. With the additional states $f_0(1500)$ and
$f_0(1710)$ nearby there is one state too much (discussed as route 1 and 3). 
Here our analysis has given a clear answer. We have not found 
convincing evidence for $f_0(1370)$ in any of the 2-body decay channels
in the study of a large variety of reactions. 
In particular,
elastic $\pi\pi$ scattering phase shift data put an 
upper limit of $\lesssim 10\%$ for its branching ratio into $\pi\pi$. There is
room for further studies in the $4\pi$ channel, but so far the results are
controversial. Further experimental investigations are desirable.\\
ii) Extra state $f_0(500)$. In this scenario the $K^{*0}(800)/\kappa$
is not needed. There is some enhanced interaction at low $K\pi$ 
energy but the
$K\pi$ phase shifts saturate  below $40^\circ$, very unusual for a resonance.
A possible clue could come from the study of symmetry relations for decay
rates in $\chi_c$ decays.

\subsection{Are there hints towards a gluonic component in isoscalar
mesons?}
Our analysis of the production and decay pattern has shown that 
$f_0(1500)$ can hardly be described as a pure $q\bar q$ state, a
particular solution with a $gg$ contribution is suggested depending on the
mixing angle.
For $f_0(980)$ there
is no hint for a strong gluonic component from the $\gamma\gamma$ decay
width. The state $f_0(500)$ couples to strange and non-strange quarks
but is not ``flavour blind'' so it could be a mixed gluonic-quarkonic state.
A promising tool for further clarification of the flavour structure of
isoscalars and their mixing angles 
are the study of $D, D_s$ and especially of $B,B_s$ decays.

\subsection{Future studies of symmetry relations in $\chi_c$ decays}
Besides the study of leading clusters in gluon jets we find the study
of decays of heavy quarkonia like $\chi_{c0},\chi_{c2}$ into pairs 
of scalar particles particularly interesting. This follows from the
success of the symmetry relations for pairs of pseudoscalar particles.
Such relations could establish which particles form a multiplet. Deviations
from symmetry for the isoscalars would be a hint for glueball.   
The present results are consistent with $f_0(980)$ and $K^{*0}(1430)$
to be in the same multiplet, $f_0(980)$ as tetraquark state is disfavoured.
Although the treatment of particles with a very large width is difficult,
there may be some clues on $\sigma$ and $\kappa$ and their multiplets 
or gluonic component from such studies.

\subsection{A minimal scenario for scalar quarkonium-glueball spectroscopy}
Finally we present a solution which takes into account the main
results on the flavour structure of the $f_0$ mesons discussed in 
this report. 
It is based on the $q\bar q$ nonet of route~2, i.e. it includes
$f_0(980),\ a_0(980),\ K^{*0}(1430),\ f_0(1500)$ with a flavour mixing
similar to the mixing of $\eta,\eta'$. The near singlet flavour structure of
$f_0(980)$ has been strongly supported with a mixing angle around
30$^\circ$ (as compared to $42^\circ$ for the pseudoscalars). Also the
relative negative sign of the $s\bar s$ component within both $f_0$ states
is established. In that case, 
$f_0(1500)$ needs a contribution from $gg$
to explain the decay branching ratios ($\pi\pi,\
K\bar K,\ \eta\eta$) with their small errors.
At the same time, $f_0(500)$ does not look like 
a pure glue state, as it doesn't seem to realize flavour symmetry in the decays. 
The interpretation of $\gamma\gamma$ decays, sometimes with large errors,
is not without controversy because of the model dependence.
Our findings on the flavour structure 
suggest as a minimal solution the existence of a glueball 
which is strongly mixed into
$f_0(500-1000)$ and $f_0(1500)$ whereas $f_0(980)$ remains a quarkonium
state.
We recall that in the physical
region the maximal strength of $\pi\pi$ scattering is near 
1000 MeV (and not 500 MeV) with a comparable width. In this sense
the broad $f_0(500-1000)$ and the narrow $f_0(1500)$ overlap in mass, a situation 
reminiscent of the $\rho-\omega$ system, apart for the very different widths. 
This suggests a mixing 
\bea
|f_0(500-1000)\ran&= \sin \phi_G |q\bar q\ran - \cos\phi_G |gg\ran\\
|f_0(1500)\ran&= \cos \phi_G |q\bar q\ran + \sin\phi_G |gg\ran\\
|q\bar q\ran &=  \cos \phi_{sc} |n\bar n\ran - \sin\phi_{sc} |s\bar s\ran
\end{eqnarray}
We prefer for $f_0(1500)$ the solution $\tan\phi_G>0$
and especially with our value of $\phi_{sc}$ 
\beq
\phi_{sc}= (30\pm 3)^\circ, \quad \phi_{G}\sim 35^\circ.
\eeq
This may be compared with  maximal $q\bar q - gg$ mixing at $\phi_{G}\sim
45^\circ$ while the model \cite{Minkowski:1998mf} corresponds to $\phi_G=0$.
The result on $\phi_G$ assumes a relation between the decay couplings
($g_0\approx G_0$); more generally, one may introduce an extra parameter
$r_{G0}=G_0/g_0$, for now $r_{G0}\approx 1$. 
The flavour structure of $f_0(1500)$
with dominant $gg$ and $q\bar q$ octet components is also comparable to 
the models 
\cite{Amsler:1995td,Close:1996yc} (see \eref{mixingclose}) but with a very
different multiplet structure.

Specific experimental tests of the presented scenario 
include the better determination of mixing angles of the $f_0$ mesons and
their gluonic components. Also the amplitude signs lead to observable
effects. A special property of $f_0(500-1000)$ in this scheme is the
negative coupling of the $s\bar s$ component with respect to the one of
$f_0(980)$ which is indicated in the Argand diagram of figure
\ref{fig:dsdec}. 

A new possibility for glueball spectroscopy is opened up by the gluonic
coupling of $f_0(500)/\sigma$ and its dominant decay into $\pi\pi$.
Because of this coupling a higher mass glueball should have a favoured decay 
into $\sigma\sigma$. In case of $f_0(1500)$, which has a probability 1/3
for a gluonic component according to the above simple estimate, 
there is a decay BR of around 13\%
into $\sigma\sigma$ according to table \ref{tab:4pi}; the evidence
is particularly clear and unique for the decay channel $\sigma\to
\pi^0\pi^0$. 

As an interesting aspect of the scenario presented 
for scalar spectroscopy we note the similarity to the
recent solution from QCD sum rules for the scalar sector which
suggests two gluonic contributions  
at 1000 and 1500 MeV, but with strong correlation. Also there is one group
with a similar result from lattice calculations, but alternative views exist as
well.

\section*{Acknowledgments}
I would like to express my gratitude to Peter Minkowski 
for the long term collaboration on
gluonic mesons, for his inspiration and the theoretical insights 
he shared with me. I am much obliged to Alan Martin for suggesting 
and supporting this report and for his helpful comments on the manuscript.
Also I would like to thank many colleages for exchanges
on the topics of this report at various stages, especially I appreciated
the collaboration with
Gerard Mennessier and Stephan Narison, and the interest and the 
stimulating discussions
with V. V. Anisovich, D. Bugg, B. Buschbeck, S. U. Chung, F.
Close, B. Gary, R. Kaminski, E. Klempt, F. Mandl, W. M\"anner, B. Meadows, 
J.R. Pelaez, M. Pennington, A. Sarantsev, J. Wosiek and B.S. Zou.
Thanks also to Rainer Schicker for inviting me to the 
WE-Heraeus Summer school in Heidelberg 2011, which initiated this study.

\section*{List of reprinted figures}
\addcontentsline{toc}{section}{\numberline{}List of reprinted figures}
\footnotesize
The following figures are reprinted with kind permission of the
copyright holders:

{\it American Physical Society:}\\
 ``Readers may view, browse, and/or download material for temporary copying
purposes only, provided these uses are for noncommercial personal purposes.
Except as provided by law, this material may not be further reproduced,
distributed, transmitted, modified, adapted, performed, displayed,
published, or sold in whole or part, without prior written permission from
the American Physical Society.''\\
Fig. 1, from \cite{Morningstar:1999rf},  C.~J.~Morningstar and
M.~J.~Peardon, 
``The Glueball spectrum from an
anisotropic lattice study,'' 
http://link.aps.org/abstract/PRD/v60/e034509;\\
Figs. 2,3, from \cite{Amsler:1995td},  C.~Amsler and F.~E.~Close, 
``Is $f_0 (1500)$ a scalar glueball?,''
http://link.aps.org/abstract/PRD/v53/p295;\\
Fig. 5c, from
\cite{delAmoSanchez:2010fd},  P.~del Amo Sanchez {\it et al.}, 
``Analysis of the $D^+ \to K^- \pi^+ e^+ \nu_e$
decay channel,'' http://link.aps.org/abstract/PRD/v83/e072001;\\
Fig. 16, from \cite{LHCb:2012ae}, R.~Aaij {\it et al.},  
``Analysis of the resonant components in
$B_s \to J/\psi\pi^+\pi^-$, '' DOI: 10.1103/PhysRevD.86.052006.

{\it Elsevier B. V.:}\\
Fig. 6, from \cite{Klempt:2007cp},  E.~Klempt and A.~Zaitsev,
``Glueballs, Hybrids, Multiquarks.
Experimental facts versus QCD inspired concepts;'' \\
Fig. 7c, from \cite{Armstrong:1993fh}, T.~A.~Armstrong {\it et al.},
``Evidence for $\eta \eta$ resonances in $\bar{p}p$ annihilations 
at $2950 < \sqrt{s} < 3620$ MeV;''\\
Figs. 10, 11, from  \cite{Ochs:2006rb}, W.~Ochs,
``Scalar mesons: In search of the lightest glueball;'' \\
Fig. 18, from  \cite{Barberis:1999ap}, D.~Barberis {\it et al.},
``A Partial wave analysis of the
centrally produced $\pi^0 \pi^0$ system in p p interactions at 450-GeV/c;''\\
Fig. 22, from  \cite{Minkowski:2000qp},  P.~Minkowski and W.~Ochs,
``Gluon fragmentation into glueballs and hybrid mesons,''\\
Figs. 23, 25a, from \cite{Abdallah:2006ve}, J.~Abdallah {\it et al.}, 
 ``Study of Leading Hadrons in Gluon and Quark Fragmentation;''

 {\it Springer Science + Business Media B.V.:}\\
Figs. 7a,b, from  \cite{Amsler:2002qq},  C.~Amsler {\it et al.}, 
``Proton anti-proton annihilation at 900-MeV/c into $\pi^0 \pi^0 \pi^0, 
\pi^0 \pi^0 \eta$ and $\pi^0 \eta \eta$;'' \\
Fig. 9, from \cite{Alde:1998mc}, D.~Alde {\it et al.},
``Study of the $\pi^0 \pi^0$ system with the GAMS-4000 spectrometer at
100-GeV;''\\
Fig. 12, from \cite{Buettiker:2003pp},  P.~Buettiker {\it et al.}, 
``A new analysis of $\pi K$ scattering from Roy and Steiner type
  equations;''\\
Fig. 14, from \cite{Minkowski:2004xf},  P.~Minkowski and W.~Ochs, 
``B decays into light scalar particles and glueball'';\\
Fig. 24, from \cite{Schael:2006ns},  S.~Schael {\it et al.}, 
``Test of Colour Reconnection Models using Three-Jet Events in Hadronic Z
Decays;''\\
Fig. 25b, from \cite{Abbiendi:2003ri},  G.~Abbiendi {\it et al.}, 
``Tests of models of color reconnection and a search for glueballs using
  gluon jets with a rapidity gap.''


\end{document}